\documentclass[preprint]{ptptex}

\usepackage{graphicx}
\usepackage{wrapft}
\usepackage{amssymb}
\usepackage{amsmath}

\usepackage{epsf}
\usepackage{graphics}
\usepackage{amssymb}
\usepackage{amsmath}
\usepackage{enumerate}
\usepackage{hhline}
\usepackage{array}
\usepackage{tabularx}

\renewcommand{\theenumi}{(\arabic{enumi})}

\newcommand{\bear}{\begin{array}}  \newcommand{\eear}{\end{array}}
\newcommand{\bea}{\begin{eqnarray}}  \newcommand{\eea}{\end{eqnarray}}
\newcommand{\beq}{\begin{equation}}  \newcommand{\eeq}{\end{equation}}
\newcommand{\bef}{\begin{figure}}  \newcommand{\eef}{\end{figure}}
\newcommand{\bec}{\begin{center}}  \newcommand{\eec}{\end{center}}

\def\Vec#1{\mbox{\boldmath $#1$}}
\def\Vecs#1{\mbox{\boldmath\tiny $#1$}}

\notypesetlogo                       
\preprintnumber[3cm]{
OU-TAP-262}

\title{
Curvaton Scenario in the Presence of Two Dilatons \\
Coupled to the Scalar Curvature
}

\author{
Kazuharu \textsc{Bamba}$^{1}$
and Motohiko \textsc{Yoshimura}$^{2}$
}

\inst{
$^1$Department of Earth and Space Science, Graduate School of 
Science, \\
Osaka University, Toyonaka 560-0043, Japan
\\
$^2$Department of Physics, Okayama University, \\
Okayama 700-8530, Japan\\ 
}

\abst{
The possibility of the realization of a curvaton scenario is 
studied in a theory in which two dilatons are introduced along 
with coupling to the scalar curvature.  
It is shown that 
when the two dilatons have an approximate $O(2)$ symmetric coupling, 
a scalar field playing the role of the curvaton may exist 
in the framework of this theory 
without introducing any other scalar field for the curvaton.  
Thus the curvaton scenario can be realized, 
and 
in the simple version of the curvaton scenario, 
in which the curvaton potential is quadratic, 
the curvature perturbation 
with a sufficient 
amplitude and a nearly scale-invariant spectrum suggested by 
observations obtained from WMAP can be generated.  
}

\begin{document}

\maketitle

\section{Introduction}
The hierarchy problem exists 
in both particle physics and cosmology.  
In the former case, this is the hierarchy problem between gravity and 
standard particle physics mass scales, 
while in the latter case, 
it is the presence of a very small but finite cosmological constant 
or the dark energy indicated by the analysis of type-Ia supernovae 
observations \cite{Perlmutter}, 
whose magnitude is as small as 120 orders smaller than 
its theoretically natural value, i.e., the Planck scale 
(or a similar scale).\cite{Weinberg} \ 
The origin of this small cosmological constant 
is not well understood yet.  
There has yet been no definite convection established 
in theoretical studies.  

Recently, one of the present authors attempted to construct a model 
\cite{Yoshimura} which simultaneously solves the above two hierarchy 
problems by radically changing cosmology in the same spirit as the 
ideas of 
Dirac \cite{Dirac} and Brans and Dicke.\cite{B-D,F-M} \ 
In this model, a theory with two dilaton fields 
coupled to the scalar curvature 
considered.  The existence of many dilatons is conceivable 
in light of higher-dimensional theories.   
Moreover, 
it has been argued that an effective gravity mass scale 
(the square root of the inverse of Newton's constant) 
increases in the inflationary stage \cite{Guth}
(for a review of inflation, see Refs.~\citen{Linde1} and \citen{Kolb}), 
and that 
the small cosmological constant or 
the dark energy density in the present universe could be dynamically 
realized in the case that two dilatons have approximately 
an $O(2)$ symmetry, 
taking the fundamental mass scale to be of TeV.  
In such a TeV scale model, where the potential of the dilatons is 
of order $ \sim (\mathrm{TeV})^4$, 
however, 
it is conjectured that 
a naive estimate of the curvature perturbation gives a 
magnitude which is much smaller 
than the recent observational results of 
the anisotropy of the cosmic microwave background (CMB) 
radiation obtained from Wilkinson microwave background probe (WMAP).
\cite{Bennet} \ 
Moreover, as is shown in \S 3, power-law inflation could be 
realized in this model.  
In the standard inflation models, however, 
inflation is driven by the potential energy of 
a scalar field, called an ``inflaton", as it slowly 
rolls down the potential hill.  
This slow roll over in the quasi-de Sitter stage is 
necessary to account for the nearly scale-invariant 
spectrum of the primordial curvature perturbation out of the quantum 
fluctuation of the inflaton, which is also suggested by the observational 
results obtained from WMAP.\cite{Spergel}  

However, 
in recent years a new mechanism for generating the primordial 
curvature perturbation has been proposed, in which a late-decaying 
massive scalar field provides the dominant source of the 
curvature perturbation.\cite{Moroi,Enqvist,Lyth,Lyth2} \ 
In this scenario, the dominant part of the curvature perturbation 
originates from the quantum fluctuations of a new scalar field, called the 
``curvaton", which is different from the inflaton.  
In contrast to the situation with the usual mechanism, 
with this new mechanism, 
inflation need not be of the slow-roll variety.  
Hence, even in the case of power-law inflation, 
the nearly scale-invariant spectrum of the curvature perturbation 
demanded by the observations can be realized.  
Instead, 
it is required that the curvaton potential be sufficiently 
flat during inflation.  

The purpose of the present paper is to argue that 
the curvaton scenario can be realized in the above theory, 
in which two dilatons are introduced along with 
the coupling to the scalar curvature.\cite{Yoshimura} \ 
In particular, 
we show that there exists a scalar field corresponding to the curvaton 
in the framework of this theory without introducing any other scalar field 
that 
plays the role of the curvaton, 
and discuss the case in which the curvaton scenario 
is realized.  
Also, we are able to construct 
the curvature perturbation with sufficiently large amplitude and 
nearly scale-invariant spectrum suggested by the 
observational results obtained from WMAP.   

This paper is organized as follows.  
In \S 2 we describe our model and derive field equations 
from its action.  
In \S 3 we show that power-law inflation can be 
realized in this model.  
Next, 
in the framework of this theory 
the existence of a scalar field 
corresponding to the curvaton is demonstrated in \S 4. 
Then, in \S 5, we study the case in which the curvaton scenario can be 
realized in this model.  
Finally, \S 6 is devoted to a conclusion.  
Throughout the present paper 
we use units in which $k_\mathrm{B} = c = \hbar = 1$ 
and denote Newton's constant by 
$G = {m_{\mathrm{Pl}}}^{-2}$, where 
$m_{\mathrm{Pl}} = 1.2 \times 10^{19}$GeV is the Planck mass.

\section{Model}

\subsection{Action}
We introduce two scalar fields, $\varphi_i \hspace{1mm} (i=1,2)$.  
Furthermore, we introduce dilatonic coupling of these scalars to 
the scalar curvature.  
Our model action is the following:\cite{Yoshimura}  
\begin{eqnarray}
S &=& \int d^{4}x \sqrt{-g} \left[ 
-f(\varphi_i) R + 
\frac{1}{2}(\partial \varphi_i)^2
- V[\varphi_i] + {\cal L}_{\mathrm{m}} 
\right], \label{eq:1} \\[5mm]
f(\varphi_i) &=& \epsilon_1 \varphi_1^2 + \epsilon_2 \varphi_2^2, 
\label{eq:2} \\[5mm]
V[\varphi_i] &=& V_{0} \cos \frac{\varphi_{\mathrm{r}}}{M} + \Lambda, 
\label{eq:3}
\end{eqnarray}
where 
$R$ is the scalar curvature arising from the spacetime 
metric tensor $g_{\mu\nu}$, 
$g$ is the determinant of $g_{\mu\nu}$, 
$f(\varphi_i)$ is the coupling between the dilaton and the 
scalar curvature [with $\epsilon_i \hspace{0.5mm} 
(\hspace{0.5mm} > 0 \hspace{0.5mm})$
being dimensionless constants], 
$V[\varphi_i]$ is the potential of the dilaton (with 
$\varphi_{\mathrm{r}} = \sqrt{\varphi_1^2 + \varphi_2^2}$ 
and $V_0 > \Lambda > 0$), 
and 
$M$ denotes a mass scale.  
Here $\Lambda$ is the collection of all the constants in the Standard Model 
Lagrangian ${\cal L}_{\mathrm{m}}$ such that the potential of the Standard 
Model Lagrangian vanishes at its minimum.  
We assume $O(2)$ symmetry for the potential $V[\varphi_i]$, 
while we allow its violation by choosing 
$\epsilon_1\not=\epsilon_2$.  
Moreover, we use the simplified notation 
$
(\partial \varphi_i)^2 
\equiv 
g^{\mu\nu}{\partial}_{\mu}{\varphi_i}{\partial}_{\nu}{\varphi_i}. 
$

The dilatonic coupling of the scalar field 
to the scalar curvature given by $f(\varphi_i) R$ is 
taken from the Brans-Dicke theory.\cite{B-D} \ 
However, with regard to 
the choice of the potential $V[\varphi_i]$, 
we part with 
the Brans-Dicke theory, 
in which a single dilaton, $\varphi$, is introduced along with the null 
potential and $f(\varphi) = (1/2) \xi \varphi^2$, 
where $\xi$ is 
a dimensionless constant.\cite{F-M} \ 
For both simplicity and naturalness, 
we assume that all mass parameters are of the same order, 
and thus 
$V_{0} \approx \Lambda = O[M^4]$ for the choice (\ref{eq:3}).  
In the present model, 
we use a common mass scale $M$ of order $m_{\mathrm{Pl}}$, 
although in the scenario proposed in Ref.~\citen{Yoshimura}, 
a common mass scale $M$ of order $\mathrm{TeV}$, 
favored by the present small value of the dark energy, 
is used.  
\if
The reason why we have chosen the form of the dilaton potential in 
Eq.\ (\ref{eq:3}) is as follows \cite{Yoshimura}.  
We do not assume any fine tuning of the potential $V[\varphi_i]$ 
except that it is a bounded function allowing infinitely many negative values 
and infinitely many local minima.  
In this way a large mass hierarchy and 
dynamical relaxation towards a small cosmological constant might be realized.  
The simplest choice realizing these is a periodic 
potential of minimum numbers of parameters.  

Important features of our assertions below are valid irrespective of the
precise form of the potential.  
The essential requirement on the potential for a successful scenario is
that (1) boundedness, (2) infinitely many local
minima, and (3) infinitely many regions of negative values 
between minima and maxima.  
Nevertheless, it would be useful to have
a simple realization such as Eq.\ (\ref{eq:3}) 
of our idea and to discuss a model explicitly.  
\fi

\subsection{Field equations}
The field equations can be derived by taking variations of the 
above action in Eq.\ (\ref{eq:1}) with respect to the 
metric $g_{\mu\nu}$ and the dilatons $\varphi_i$ as follows:\cite{Zee}  
\begin{eqnarray}
R_{\mu \nu} - \frac{1}{2}g_{\mu \nu}R 
= 
\frac{1}{2f} 
\left[ T^{(\mathrm{m})}_{\mu \nu} + T^{(\varphi_i)}_{\mu \nu} \right] 
+ \frac{1}{f} 
\left( {\nabla}_{\mu} {\nabla}_{\nu} f - g_{\mu \nu} \Box f \right),
\label{eq:4}
\end{eqnarray}
with
\begin{eqnarray}
T^{(\varphi_i)}_{\mu \nu}
= {\partial}_{\mu}{\varphi_i}{\partial}_{\nu}{\varphi_i} 
- g_{\mu\nu} 
\left[ \frac{1}{2} (\partial \varphi_i)^2 - V[\varphi_i] 
\right], 
\label{eq:5}
\end{eqnarray}
and 
\begin{eqnarray}
\Box {\varphi}_{i}
=
-\frac{\partial V}{\partial \varphi_i} - \frac{\partial f}
{\partial \varphi_i}R,  
\label{eq:6}
\end{eqnarray}
where 
${\nabla}_{\mu}$ is the covariant derivative operator associated with 
$g_{\mu \nu}$, and $\Box \equiv g^{\mu \nu} {\nabla}_{\mu} {\nabla}_{\nu}$ 
is the covariant d'Alembertian for a scalar field.  
In addition, $R_{\mu \nu}$ is the Ricci curvature tensor, 
while 
$T^{(\varphi_i)}_{\mu \nu}$ is the contribution to 
the energy-momentum tensor from the scalars $\varphi_i$, and 
$T^{(\mathrm{m})}_{\mu \nu}$ is the usual contribution of radiation, matter 
and other fields.  
Taking the trace of Eq.\ (\ref{eq:4}), we obtain 
\begin{eqnarray}
- R = \frac{1}{2f} 
\left[ T^{(\mathrm{m})} - \left( \partial \varphi_i \right)^2 + 4V(\varphi_i)
- 6 \Box f \right], 
\label{eq:7}
\end{eqnarray}
where $T^{(\mathrm{m})}$ is the trace of $T^{(\mathrm{m})}_{\mu \nu}$.

We now assume the spatially flat 
Friedmann-Robertson-Walker (FRW) spacetime with the metric
\begin{eqnarray}
 {ds}^2 = g_{\mu\nu}dx^{\mu}dx^{\nu} =  {dt}^2 - a^2(t)d{\Vec{x}}^2,
\label{eq:8}
\end{eqnarray} 
where $a(t)$ is the scale factor.  
In the FRW metric (\ref{eq:8}), 
the equations of motion for the background homogeneous scalar fields read 
\begin{eqnarray}
  \ddot{\varphi_i} + 3H\dot{\varphi_i} = 
-\frac{\varphi_i}{\varphi_{\mathrm{r}}}V^{\prime} -2\epsilon_i \varphi_i R,
\label{eq:9}
\end{eqnarray} 
where the dot denotes differentiation with respect to time, 
and the prime denotes differentiation with respect to $\varphi_{\mathrm{r}}$.  
Here $H$ is the Hubble parameter.  
Using the equations of motion for the dilaton fields 
(\ref{eq:9}) with Eq.\ (\ref{eq:7}), we find the dynamical equations 
in terms of the two field variables 
$f=\epsilon_1 \varphi_1^2 + \epsilon_2 \varphi_2^2$ and 
$\tilde{f}=\epsilon_1 \varphi_1^2 - \epsilon_2 \varphi_2^2$ as 
\begin{eqnarray}
\hspace{-10mm}
\ddot{f} + 3H\dot{f} &=& 2F^{-1} 
\biggl\{  
(\epsilon_1 \dot{\varphi_1}^2 + \epsilon_2 \dot{\varphi_2}^2)
-\frac{f}{\varphi_{\mathrm{r}}} V^{\prime} 
\nonumber \\[3mm]
&& \hspace{12mm}
{}
+\frac{\epsilon_1^2 \varphi_1^2 + \epsilon_2^2 \varphi_2^2}{f} 
\left[ T^{(\mathrm{m})} + 4V -(\dot{\varphi_1}^2 + \dot{\varphi_2}^2)  
\right] 
\biggr\},
\label{eq:10} \\[5mm]
\hspace{-10mm}
\ddot{\tilde{f}} + 3H\dot{\tilde{f}} &=& 
2F^{-1} 
\Biggl\{
F (\epsilon_1 \dot{\varphi_1}^2 - \epsilon_2 \dot{\varphi_2}^2) 
-
\left[
\frac{\tilde{f}}{f}
- 24 \epsilon_1 \epsilon_2 (\epsilon_1 - \epsilon_2)
\left( \frac{\varphi_1 \varphi_2}{f} \right)^2 
\right] 
\frac{f}{\varphi_{\mathrm{r}}} V^{\prime} 
\nonumber \\[3mm]
&&\hspace{-5mm}
{} + 
\left(
\frac{\epsilon_1^2 \varphi_1^2 - \epsilon_2^2 \varphi_2^2}{f} 
\right)
\left[ T^{(\mathrm{m})} + 4V -(\dot{\varphi_1}^2 + \dot{\varphi_2}^2) 
-12(\epsilon_1 \dot{\varphi_1}^2 + \epsilon_2 \dot{\varphi_2}^2)
\right]
\Biggr\}
, 
\label{eq:11} \\[5mm]
F &=& 1+12 \hspace{0.5mm}
\frac{\epsilon_1^2 \varphi_1^2 + \epsilon_2^2 \varphi_2^2}{f}. 
\label{eq:12}
\end{eqnarray}
(The reason we have chosen the field variable 
$\tilde{f}$ as the partner of the coupling $f$ is stated in \S A.2.) 
Furthermore, the gravitational field equations read 
\begin{eqnarray}
H^2 &=&  \left( \frac{\dot{a}}{a} \right)^2 = 
\frac{1}{6f} 
\left[ T^{(\mathrm{m})}_{00} + 
\frac{1}{2} (\dot{\varphi_1}^2 + \dot{\varphi_2}^2) + V \right] 
- H \frac{\dot{f}}{f}, 
\label{eq:13} \\[5mm] 
\hspace{0mm}
\dot{H} - 2H\frac{\dot{f}}{f} &=& 
-\frac{F^{-1}}{f}  
\biggl[  
\frac{1}{12} \left( 4FT^{(\mathrm{m})}_{00} - T^{(\mathrm{m})} \right) 
+ (\epsilon_1 \dot{\varphi_1}^2 + \epsilon_2 \dot{\varphi_2}^2)
\nonumber \\[3mm] 
&& \hspace{12.5mm}
{}+ \frac{1}{4} 
\left(
1+ 8 \hspace{0.5mm}
\frac{\epsilon_1^2 \varphi_1^2 + \epsilon_2^2 \varphi_2^2}{f}
\right)
(\dot{\varphi_1}^2 + \dot{\varphi_2}^2) 
\nonumber \\[3mm]
&& \hspace{12.5mm}
{}
-\frac{f}{\varphi_{\mathrm{r}}} V^{\prime} 
+ 4 \hspace{0.5mm}
\frac{\epsilon_1^2 \varphi_1^2 + \epsilon_2^2 \varphi_2^2}{f} 
\hspace{0.5mm} V 
\biggr],
\label{eq:14}
\end{eqnarray} 
where in deriving the expression (\ref{eq:14}) 
we have used Eq.\ (\ref{eq:10}). 

\if
and $T^{(\mathrm{m})}_{00}$ is the $0-0$ component of 
$T^{(\mathrm{m})}_{\mu \nu}$.  
\fi

\section{Power-law inflation}
In this section, we show that power-law inflation can be 
realized in the presently considered model.  
Here we seek solutions with the ansatz 
\begin{eqnarray}
 f=At^2, \hspace{5mm} 
 L=Bt,   \hspace{5mm}  
 \varphi_{\mathrm{r}}=Ct,  \hspace{5mm} 
 a \propto t^{\omega},  
\label{eq:15}
\end{eqnarray}
which is valid for large $t$.  Here, 
$L = \varphi_1 \dot{\varphi_2}- \varphi_2 \dot{\varphi_1}$ 
is the angular momentum.  The existence of the angular momentum 
is 
one of the important features of this model.  
We now assume that 
in the inflationary stage, the cosmic energy density is 
dominated by the dilatons, and hence 
$T^{(\mathrm{m})}_{\mu \nu}$ 
is negligible.  
Moreover, for the moment we ignore 
variation of the potential 
and 
hence we replace $V$ by its average value, $\Lambda$.  

Substituting the ansatz (\ref{eq:15}) into 
Eqs.\ (\ref{eq:10}), (\ref{eq:13}) and (\ref{eq:14}), 
we obtain 
\begin{eqnarray}
&& \hspace{-1mm}
1+3\omega = 
F^{-1} 
\biggl[
4
\left(
\frac{\epsilon_1^2 \varphi_1^2 + \epsilon_2^2 \varphi_2^2}{f} 
\right)
\frac{\Lambda}{A} \nonumber \\[3mm]  
&& \hspace{23.5mm}
{}-
\left( 
\frac{\epsilon_1^2 \varphi_1^2 + \epsilon_2^2 \varphi_2^2}{f}
- 
\frac{\epsilon_1 \dot{\varphi_1}^2 + \epsilon_2 \dot{\varphi_2}^2}
{\dot{\varphi_1}^2 + \dot{\varphi_2}^2}
\right)
\frac{1}{A} \left( C^2 + \frac{B^2}{C^2} \right)
\biggr], 
\label{eq:16} \\[5mm] 
&& \hspace{-3mm}
\omega^2 + 2\omega = \frac{1}{12} 
\left[ 2 \hspace{0.5mm}  \frac{\Lambda}{A} 
+ \frac{1}{A} \left( C^2 + \frac{B^2}{C^2} \right)
\right],
\label{eq:17} \\[5mm] 
&& \hspace{8mm}
\omega = 
F^{-1}
\biggl[
\frac{4}{5} 
\left(
\frac{\epsilon_1^2 \varphi_1^2 + \epsilon_2^2 \varphi_2^2}{f} 
\right)
\frac{\Lambda}{A} \nonumber \\[3mm]  
&& \hspace{14.5mm}
{}+\frac{1}{20} 
\left(
1+ 8 \hspace{0.5mm}
\frac{\epsilon_1^2 \varphi_1^2 + \epsilon_2^2 \varphi_2^2}{f} 
+4 \hspace{0.5mm}
\frac{\epsilon_1 \dot{\varphi_1}^2 + \epsilon_2 \dot{\varphi_2}^2}
{\dot{\varphi_1}^2 + \dot{\varphi_2}^2}
\right)
\frac{1}{A} \left( C^2 + \frac{B^2}{C^2} \right)
\biggr], 
\label{eq:18}
\end{eqnarray}
where in deriving the expressions (\ref{eq:16})$-$(\ref{eq:18}) 
we have used the following relation: 
\begin{eqnarray}
\dot{\varphi_1}^2 + \dot{\varphi_2}^2 = 
\dot{\varphi_{\mathrm{r}}}^2 + \frac{L^2}{\varphi_{\mathrm{r}}^2}
= C^2 + \frac{B^2}{C^2}.  
\label{eq:19}
\end{eqnarray}
From Eqs.\ (\ref{eq:17}) and (\ref{eq:18}), we find 
\begin{eqnarray}
&&\hspace{0mm}
\frac{\Lambda}{A} = 
2
\left(
1+ 4 \hspace{0.5mm}
\frac{\epsilon_1 \dot{\varphi_1}^2 + \epsilon_2 \dot{\varphi_2}^2}
{\dot{\varphi_1}^2 + \dot{\varphi_2}^2}
\right)^{-1} \nonumber \\[3mm]  
&&\hspace{7.5mm}{}\times 
\left[
3(\omega^2 + 2\omega) 
\left(
1+ 8 \hspace{0.5mm}
\frac{\epsilon_1^2 \varphi_1^2 + \epsilon_2^2 \varphi_2^2}{f} 
+4 \hspace{0.5mm}
\frac{\epsilon_1 \dot{\varphi_1}^2 + \epsilon_2 \dot{\varphi_2}^2}
{\dot{\varphi_1}^2 + \dot{\varphi_2}^2}
\right)
-5 \omega F
\right], 
\label{eq:20} \\[5mm] 
&&\hspace{0mm}
\frac{1}{A} \left( C^2 + \frac{B^2}{C^2} \right)  =
4 
\left(
1+ 4 \hspace{0.5mm}
\frac{\epsilon_1 \dot{\varphi_1}^2 + \epsilon_2 \dot{\varphi_2}^2}
{\dot{\varphi_1}^2 + \dot{\varphi_2}^2}
\right)^{-1} \nonumber \\[3mm]  
&&\hspace{29mm}{}\times 
\left[
-24(\omega^2 + 2\omega) 
\left(
\frac{\epsilon_1^2 \varphi_1^2 + \epsilon_2^2 \varphi_2^2}{f} 
\right)
+5 \omega F
\right].  
\label{eq:21} 
\end{eqnarray}
Eliminating $\Lambda/A$ and 
$\left( C^2 + B^2/C^2 \right)/A$ from 
Eqs.\ (\ref{eq:16}), (\ref{eq:20}) and (\ref{eq:21}), we obtain 
\begin{eqnarray}
&&\hspace{-7.5mm}
24 
\left(
\frac{\epsilon_1^2 \varphi_1^2 + \epsilon_2^2 \varphi_2^2}{f} 
\right)
\omega^2 
-\left[ 
12 
\left(
\frac{\epsilon_1^2 \varphi_1^2 + \epsilon_2^2 \varphi_2^2}{f} 
\right)
-8 \hspace{0.5mm}
\frac{\epsilon_1 \dot{\varphi_1}^2 + \epsilon_2 \dot{\varphi_2}^2}
{\dot{\varphi_1}^2 + \dot{\varphi_2}^2}
+3
\right] 
\omega \nonumber \\[3mm] 
&& \hspace{65mm} 
{}-\left(
1+ 4 \hspace{0.5mm}
\frac{\epsilon_1 \dot{\varphi_1}^2 + \epsilon_2 \dot{\varphi_2}^2}
{\dot{\varphi_1}^2 + \dot{\varphi_2}^2}
\right)
=0.  
\label{eq:22} 
\end{eqnarray}
We now assume $\varphi_1^2 \approx \varphi_2^2$ and 
$\dot{\varphi_1}^2 \approx \dot{\varphi_2}^2$ and 
approximately determine $\omega$ in terms of $\epsilon_i$.  
It follows from these approximate relations that 
Eq.\ (\ref{eq:22}) can be rewritten in the form 
\begin{eqnarray}
24 \beta \omega^2 -(12\beta -4\alpha^2 +3\alpha)\omega 
-\alpha (1+2\alpha) =0, 
\label{eq:23} 
\end{eqnarray}
with 
\begin{eqnarray}
\alpha &\equiv& \epsilon_1 + \epsilon_2, 
\label{eq:24} 
\\[3mm]
\beta &\equiv& \epsilon_1^2 + \epsilon_2^2.  
\label{eq:25} 
\end{eqnarray}
The solution of Eq.\ (\ref{eq:23}) is given by 
\begin{eqnarray}
\omega &=& \frac{1}{48\beta} 
\left[ (12\beta -4\alpha^2 +3\alpha) + 
\sqrt{D}
\right], 
\label{eq:26} \\[5mm]
D &=&
9\alpha^2-24\alpha^3+168\alpha\beta+16\alpha^4
+96\alpha^2\beta+144\beta^2, 
\label{eq:27}
\end{eqnarray}
where we have taken the positive solution of $\omega$.  
In the case $\epsilon_i \ll 1$, 
considering only 
the terms of leading order in $\epsilon_i$ in 
expressions (\ref{eq:26}) and (\ref{eq:27}), we find 
\begin{eqnarray}
\omega  \approx \frac{\alpha}{8\beta} = 
\frac{\epsilon_1 + \epsilon_2}{8(\epsilon_1^2 + \epsilon_2^2)}.  
\label{eq:28}
\end{eqnarray}
Hence, for $\epsilon_i \ll 1$, 
the index of the power $\omega$ 
can be much larger than unity, and 
thus the power-law inflation can certainly be realized.  
From this point, we consider only the case $\epsilon_i \ll 1$.  

A large value of $\omega$ is favored to approximately mimic 
the exponential expansion of the cosmological scale factor. 
The gravity mass scale increases as 
$
f \propto a^{2/\omega}
$ 
during inflation.  
Thus, there is no difficulty in obtaining a large enough number of 
\textit{e}-folds 
during inflation, and simultaneously resolving the mass hierarchy
problem.  This is a feature that also appears 
in the model of extended inflation 
\cite{E-I} (see also Ref.~\citen{Mathiazhagan}).  
The number of \textit{e}-folds for a particular case in this model 
is estimated in \S 4.3.  
The requirement for power-law inflation in the curvaton scenario is that 
the power-law exponent $\omega$ be much larger than unity, 
as is discussed in \S 5.  
Furthermore, we note the following point.  
The discussion in this section corresponds to analysis in 
the Jordan frame, in which 
there exist scalar fields nonminimally coupled to the scalar curvature.  
It is not obvious, however, that power-law inflation can be realized 
not only in the Jordan frame but also 
in the Einstein frame, in which there ordinarily exist scalar fields 
minimally coupled to the scalar curvature.  
For this reason, in Appendix B, we show that 
power-law inflation can also be realized in the Einstein frame.  

%
Here we present approximate expressions of $A$, $B$, and $C$.  
Their detailed derivations are given 
in \S A.1.  
These approximate expressions are as follows: 
\begin{eqnarray} 
A &\approx&
\frac{32}{3} \left( \frac{\beta}{\alpha} \right)^2 \Lambda,  
\label{eq:39} \\[3mm]
B &\approx&
\pm 
\frac{128\beta^2}{3\alpha^3} 
\sqrt{\frac{\alpha^2}{2\beta-\alpha^2}} \hspace{0.5mm} \Lambda, 
\label{eq:44} \\[3mm]
C 
&\approx&
\sqrt{
\frac{64\beta^2}{3\alpha^3} 
\hspace{0.5mm} \Lambda}.  
\label{eq:42}
\end{eqnarray}

Finally, we discuss the following points.  
We first consider the allowed case in which we can 
ignore the variation of the potential 
$V[\varphi_i] 
= V_{0} \cos \left( \varphi_{\mathrm{r}}/M \right) + \Lambda$ in 
Eq.\ (\ref{eq:3}), 
and therefore replace $V$ by 
its average value, $\Lambda$, as done in deriving the 
solutions of power-law inflation at the beginning of this section.  
As explained in \S 4.2, in this model, we consider the following case.  
In the inflationary stage, 
the field amplitude of the dilatons is given by 
$
\varphi_1 = \gamma_1 M,   
\varphi_2 = \gamma_2 M, 
$
where $\gamma_1$ and $\gamma_2$ are dimensionless parameters with time 
dependence.  In addition, we assume 
$\gamma_1 \approx \gamma_2 \approx \gamma = \gamma(t)$; 
these relations are 
consistent with the assumption $\varphi_1^2 \approx \varphi_2^2$, 
used 
in deriving the approximate expressions of $A$, $B$, $C$ and $\omega$ above.  
In this case, we have $\varphi_{\mathrm{r}} = \sqrt{\varphi_1^2 + \varphi_2^2} 
\approx \gamma M$.  
Hence, if $\gamma (t) \ll 1$ during inflation, we can 
ignore variation of the potential.  
In this case, 
we find $V \approx V_0 + \Lambda$.  
Thus it follows from the assumption 
$V_{0} \approx \Lambda = O[M^4]$ made in \S 2.1 that 
it is valid to recognize $V$ as being approximately equal to $\Lambda$.  
As shown in \S 4.2, the case in which 
we are interested corresponds to just this case.  
\if
Finally, we discuss the following points: 
We first consider the allowed case that we can 
ignore the variation of the potential 
$V[\varphi_i] 
= V_{0} \cos \left( \varphi_{\mathrm{r}}/M \right) + \Lambda$ in 
Eq.\ (\ref{eq:3})
and then replace $V$ by 
its averaged value $\Lambda$ as done in deriving the 
solutions of power-law inflation at the beginning of this section.  
As is explained in Sec.\ 4.2, in this model we consider the following case:  
In the inflationary stage, 
the field amplitude of the dilatons is given by 
$
\varphi_1 = \gamma_1 M,   
\varphi_2 = \gamma_2 M, 
$
where $\gamma_1$ and $\gamma_2$ are dimensionless parameters with time 
dependence.  In addition, we suppose 
$\gamma_1 \approx \gamma_2 \approx \gamma = \gamma(t)$; these relation is 
consistent with the assumption $\varphi_1^2 \approx \varphi_2^2$ 
taken 
in deriving the approximate expressions of $A$, $B$, $C$, and $\omega$ above.  
In this case, $\varphi_{\mathrm{r}} = \sqrt{\varphi_1^2 + \varphi_2^2} 
\approx \gamma M$.  Hence, if $\gamma (t) \gtrsim 1$ and 
$\gamma (t_{\mathrm{f}}) - \gamma (t_{\mathrm{i}}) \gg 1$, 
where $t_{\mathrm{f}}$ and $t_{\mathrm{i}}$ are the time at the end and 
at the beginning of inflation, respectively, 
or $\gamma (t) \ll 1$ during inflation, we can 
neglect potential variation.  In the former case, we can 
replace $V$ by its averaged value $\Lambda$.  On the other hand, 
in the latter case, we find $V \approx V_0 + \Lambda$.  
Thus it follows from the assumption 
$V_{0} \approx \Lambda = O[M^4]$ in Sec.\ 2.1 that 
it is valid to recognize $V$ as to be approximately $\Lambda$.  
In particular, as is shown in Sec.\ 4.2, the case in which 
we are interested corresponds to the latter.  
\fi
In fact, for example, 
in the case $\epsilon_1 = 3.1 \times 10^{-3}$, 
$\epsilon_2 = 3.0 \times 10^{-3}$ and 
$\gamma = 3.5 \times 10^{-4}$, which is the case (v) in Table I of 
\S 4.2, the solution derived by 
partially taking the variation of the potential $V$ 
into consideration 
is $\omega=4.15 \times 10^{1} \gg 1$.  Hence power-law inflation 
can be practically realized.  Furthermore, 
the value $\omega=4.10 \times 10^{1}$, which is estimated by 
ignoring the potential variation and 
replacing $V$ by its average value $\Lambda$, as shown in Table I, 
is very close to that in the above solution.  Thus, if 
$\gamma (t) \ll 1$ during inflation, 
the above approximation in which we ignore the variation of the 
potential and 
regard $V$ as being approximately $\Lambda$ is valid.  
Here we note that in deriving the above solution 
by partially 
taking the variation of the potential $V$ into consideration, 
we have used 
$\varphi_{\mathrm{r}} \approx \gamma M$ and the 
relation 
$t \approx \omega \sqrt{6 \alpha \gamma^2} /M$, which is derived by using 
Eqs.\ (\ref{eq:74}) and (\ref{eq:75}), which are given in 
\S 4.2, and the relation $H=\omega/t$.  
\if
Thus it follows from the assumption 
$V_{0} \approx \Lambda = O[M^4]$ in Sec.\ 2.1 that 
we can make use of the approximate relation $V \approx \Lambda$, 
where we have dropped numerical factor because it does not significantly 
affect final results.  
In particular, as is shown in Sec.\ 4.2, the case in which 
we are interested is corresponding to the latter.  
\fi

\if
in the latter case, we find $V \approx V_0 + \Lambda \approx \Lambda$, 
where in deriving the second approximate equality we have dropped numerical 
factor because it does not significantly affect final results.  
In particular, as is shown in Sec.\ 4.2, the case in which 
we are interested is corresponding to the latter.   
\fi

From the above consideration, we see that 
if $\gamma (t) \ll 1$ during inflation, 
then this case is just the case considered 
in the later discussion appearing in \S 4 and 5, as stated above, 
and power-law inflation can be practically realized in this model.  
In contrast to our scenario, 
in the scenario proposed 
in Ref.~\citen{Yoshimura}, which simultaneously solves  
the hierarchy problem between gravity and 
particle physics mass scales and the small cosmological constant 
or the dark energy problem, 
the following case is considered.  
In the inflationary stage, 
$\varphi_{\mathrm{r}}$ goes over the many local maxima of the potential and 
then changes for many periods, i.e., 
$\gamma (t) \gtrsim 1$ and 
$\gamma (t_{\mathrm{f}}) - \gamma (t_{\mathrm{i}}) \gg 1$, 
where $t_{\mathrm{f}}$ and $t_{\mathrm{i}}$ are the time at the end and 
at the beginning of inflation, respectively.  
Thus, the situation for the evolution of $\varphi_{\mathrm{r}}$ 
during inflation considered in the present model is different from 
that in the scenario proposed in Ref.~\citen{Yoshimura}.  
\if
Thus it is exceedingly difficult for the present model to be 
compatible with the scenario proposed in Ref.~\citen{Yoshimura}.  
Here we indicate this point clearly, and then advance our 
consideration and discussion further.  
\fi

Second, we consider the reason why power-law inflation occurs.  
In the inflationary stage, 
from Eq.\ (\ref{eq:13}) we find 
$
H^2 
\approx
1/(6f) 
\left[ \left( \dot{\varphi_1}^2 + \dot{\varphi_2}^2 \right)/2 + V \right] 
\approx
\Lambda/(6f)
$.  
Here, in deriving the first approximate equality, 
we have used $H^2 \gg H \dot{f}/f$.  This relation follows from 
$H/(\dot{f}/f)=\omega/2 \gg 1$, where 
we have used $H=\omega/t$ and $f=At^2$.  Moreover, in deriving 
the second approximate equality, 
we have used 
$
(\dot{\varphi_1}^2 + \dot{\varphi_2}^2)/2 + V 
\approx \left[16\beta/(3\alpha)\right] \Lambda + \Lambda 
\approx \left[2/(3\omega)\right] \Lambda + \Lambda \approx \Lambda, 
$
where in deriving the first approximate equality 
we have used Eqs.\ (\ref{eq:19}) and (\ref{eq:40}), 
in deriving the second approximate equality we have used Eq.\ (\ref{eq:28}), 
and 
in deriving the last approximate equality 
we have used $\omega \gg 1$.  
Furthermore, we have ignored variation of the potential and hence replaced 
$V$ by its average value, $\Lambda$.  
From the above consideration, we understand 
that the contribution of the cosmological constant $\Lambda$ 
to the power-law inflation is much larger than that of 
the kinetic term of the dilatons.  
Thus, the contribution to the power-law inflation 
comes mainly from the effect of cosmological constant, $\Lambda$.

\section{Extremely large mass hierarchy} 
In this section, 
we show that there exists a scalar field corresponding to the curvaton 
in the framework of this theory without introducing any other scalar field 
that plays the role of the curvaton.  
First, we discuss the fluctuation equation 
in terms of 
$f=\epsilon_1 \varphi_1^2 + \epsilon_2 \varphi_2^2$ and 
$\tilde{f}=\epsilon_1 \varphi_1^2 - \epsilon_2 \varphi_2^2$ 
in order to derive
these mass values in the inflationary stage.  
Next, we consider the case in which an extremely large hierarchy of 
these mass values in the inflationary stage can be realized.  
The realization of this extremely large hierarchy implies that 
the potential of 
the scalar field with the lighter mass 
is much flatter than that of the scalar field with the heavier mass.  
Furthermore, 
if the lighter mass is much smaller than the Hubble parameter 
in the inflationary stage, the scalar field with the lighter mass can 
be the curvaton, as discussed in \S 5.2.

\if
To begin with, 
as the preliminary stage to discussing 
the fluctuation equation in terms of $f$ and $\tilde{f}$, 
we consider the terms in the right hand side of 
Eqs.\ (\ref{eq:10}) and (\ref{eq:11}).  
From $f = \epsilon_1 \varphi_1^2 + \epsilon_2 \varphi_2^2$ 
and ${\varphi_{\mathrm{r}}}^2 = \varphi_1^2 + \varphi_2^2$, we find 
\begin{eqnarray}
\varphi_1^2 = \frac{f - \epsilon_2 {\varphi_{\mathrm{r}}}^2}
{\epsilon_1 - \epsilon_2}, \hspace{5mm} 
\varphi_2^2 = \frac{-f + \epsilon_1 {\varphi_{\mathrm{r}}}^2}
{\epsilon_1 - \epsilon_2}.  
\label{eq:29}
\end{eqnarray}
Using equations (\ref{eq:29}), we obtain 
\begin{eqnarray}
\frac{\epsilon_1^2 \varphi_1^2 + \epsilon_2^2 \varphi_2^2}{f} 
&=&
(\epsilon_1 + \epsilon_2) - 
\epsilon_1 \epsilon_2 \frac{{\varphi_{\mathrm{r}}}^2}{f}, 
\label{eq:30} \\[5mm]
\frac{\varphi_1 \varphi_2}{f} 
&=&
\pm \frac{1}{\epsilon_1 - \epsilon_2} 
\sqrt{
\left(-1+ \epsilon_1 \frac{{\varphi_{\mathrm{r}}}^2}{f} \right) 
\left(1 - \epsilon_2 \frac{{\varphi_{\mathrm{r}}}^2}{f} \right)
}, 
\label{eq:31} \\[5mm]
\tilde{f} = \epsilon_1 \varphi_1^2 - \epsilon_2 \varphi_2^2
&=&
\frac{f}{\epsilon_1 - \epsilon_2}
\left[ (\epsilon_1 + \epsilon_2) - 
2 \epsilon_1 \epsilon_2 \frac{{\varphi_{\mathrm{r}}}^2}{f} \right]. 
\label{eq:32}
\end{eqnarray}
Furthermore, from 
$\dot{f} = 2 \left( \epsilon_1 \varphi_1 \dot{\varphi_1} 
+ \epsilon_2 \varphi_2 \dot{\varphi_2} \right)$ 
and 
$L = \varphi_1 \dot{\varphi_2}- \varphi_2 \dot{\varphi_1}$, we find 
\begin{eqnarray}
\dot{\varphi_1} = 
\frac{\varphi_1 \dot{f} - 2 \epsilon_2 \varphi_2 L}{2f}, \hspace{5mm} 
\dot{\varphi_2} = 
\frac{\varphi_2 \dot{f} + 2 \epsilon_1 \varphi_1 L}{2f}.  
\label{eq:33}
\end{eqnarray}
Using equations (\ref{eq:33}), we obtain 
\begin{eqnarray}
&& \hspace{-10mm}
\epsilon_1 \dot{\varphi_1}^2 + \epsilon_2 \dot{\varphi_2}^2
= 
\frac{1}{4f} \left( \dot{f}^2 + 4 \epsilon_1 \epsilon_2 L^2 \right),
\label{eq:34} \\[5mm]
&& \hspace{-10mm}
\epsilon_1 \dot{\varphi_1}^2 - \epsilon_2 \dot{\varphi_2}^2
=
\frac{1}{4f} \left[ 
\frac{\tilde{f}}{f} \left( \dot{f}^2 - 4 \epsilon_1 \epsilon_2 L^2  \right)
- 8 \epsilon_1 \epsilon_2 \dot{f} L 
\left( \frac{\varphi_1 \varphi_2}{f} \right) 
\right],
\label{eq:35} \\[5mm]
&& \hspace{-3.5mm}
\dot{\varphi_1}^2 + \dot{\varphi_2}^2 
= 
\dot{\varphi_{\mathrm{r}}}^2 + \frac{L^2}{\varphi_{\mathrm{r}}^2} 
\nonumber \\[3mm]
&& \hspace{14mm}
=
\frac{1}{4f^2} \left[ 
{\varphi_{\mathrm{r}}}^2 \dot{f}^2 + 
4 (\epsilon_1 - \epsilon_2) \varphi_1 \varphi_2 \dot{f} L + 
4 (\epsilon_1^2 \varphi_1^2 + \epsilon_2^2 \varphi_2^2) L^2
\right].  
\label{eq:36}
\end{eqnarray}
It follows from Eq.\ (\ref{eq:36}) that the quadric equation in terms of 
$L$ reads 
\begin{eqnarray}
&& \hspace{-15mm}
\left[ 
\left( \frac{\epsilon_1^2 \varphi_1^2 + \epsilon_2^2 \varphi_2^2}{f} 
\right) \left( \frac{{\varphi_{\mathrm{r}}}^2}{f} \right) -1
\right] L^2 
+ (\epsilon_1 - \epsilon_2) 
\left( \frac{{\varphi_{\mathrm{r}}}^2}{f} \right)
\left( \frac{\varphi_1 \varphi_2}{f} \right) \dot{f} L \nonumber \\[3mm]
&& \hspace{55mm} {}+
\frac{1}{4} \left( \frac{{\varphi_{\mathrm{r}}}^2}{f} \right)^2 \dot{f}^2 
-{\varphi_{\mathrm{r}}}^2 {\dot{\varphi_{\mathrm{r}}}}^2
=0.  
\label{eq:37}
\end{eqnarray}
The solution of Eq.\ (\ref{eq:37}) is given by 
\begin{eqnarray}
L = 
\left[
2 (\epsilon_1 - \epsilon_2) 
\left( \frac{\varphi_1 \varphi_2}{f} \right)
\right]^{-1}
\left[ - \left( \frac{{\varphi_{\mathrm{r}}}^2}{f} \right) \dot{f} 
\pm 2 \varphi_{\mathrm{r}} \dot{\varphi_{\mathrm{r}}}
\right], 
\label{eq:38}
\end{eqnarray}
where in deriving this expression we have used Eq.\ (\ref{eq:31}).  
Thus, from Eqs.\ (\ref{eq:30})$-$(\ref{eq:32}), 
(\ref{eq:34})$-$(\ref{eq:36}), and (\ref{eq:38}), 
we see that 
all the terms in the right-hand side of 
Eqs.\ (\ref{eq:10}) and (\ref{eq:11}) can 
be represented in terms of $f$ and $\varphi_{\mathrm{r}}$.  

Next, 
we consider the approximate expressions of $A$, $B$, and $C$.  
We again assume $\varphi_1^2 \approx \varphi_2^2$ and 
$\dot{\varphi_1}^2 \approx \dot{\varphi_2}^2$ in the same way 
as we approximately determined $\omega$ in terms of $\epsilon_i$ 
in \S 3.   
Applying these approximate relations to Eq.\ (\ref{eq:20}) and 
using Eqs.\ (\ref{eq:24}), (\ref{eq:25}) and (\ref{eq:28}), we find 
\begin{eqnarray} 
A \approx \frac{32}{3} \left( \frac{\beta}{\alpha} \right)^2 \Lambda,  
\label{eq:39}
\end{eqnarray}
where in deriving this approximate expression we have taken account of 
the small value limit $\epsilon_i \ll 1$, that is, $\alpha/\beta \gg 1$ 
and hence $\omega \gg 1$.  
In the same way, from Eqs.\ (\ref{eq:21}) and (\ref{eq:39}) we find 
\begin{eqnarray} 
C^2 + \frac{B^2}{C^2} 
\approx 
\frac{32\beta}{3\alpha} \hspace{0.5mm} 
\Lambda.  
\label{eq:40}
\end{eqnarray}
Furthermore, applying one of the above approximate relations 
$\varphi_1^2 \approx \varphi_2^2$ to ${\varphi_{\mathrm{r}}}^2/f$ and 
using the ansatz (\ref{eq:15}), we find 
\begin{eqnarray} 
\frac{{\varphi_{\mathrm{r}}}^2}{f} = \frac{C^2}{A} 
\approx \frac{2}{\alpha}.  
\label{eq:41}
\end{eqnarray}
Thus, from Eqs.\ (\ref{eq:39}) and (\ref{eq:41}) we find 
\begin{eqnarray} 
C \approx 
\sqrt{
\frac{64\beta^2}{3\alpha^3} 
\hspace{0.5mm} \Lambda}, 
\label{eq:42}
\end{eqnarray}
where we have taken the positive expression of $C$. 

On the other hand, substituting the ansatz (\ref{eq:15}) into 
the solution of the quadric equation in terms of $L$ (\ref{eq:38}), 
we find 
\begin{eqnarray} 
B &=& \pm 2C^2
\left[
\left(-1+ \epsilon_1 \hspace{0.5mm} \frac{C^2}{A} \right) 
\left(1 - \epsilon_2 \hspace{0.5mm} \frac{C^2}{A} \right)
\right]^{-1/2}
\label{eq:43} \\[5mm]
&\approx&
\pm 
\frac{128\beta^2}{3\alpha^3} 
\sqrt{\frac{\alpha^2}{2\beta-\alpha^2}} \hspace{0.5mm} \Lambda, 
\label{eq:44}
\end{eqnarray} 
where the second approximate equality follows from 
Eqs.\ (\ref{eq:39}) and (\ref{eq:41}).  
\fi


\if
Finally, we note the following point:  
From Eqs.\ (\ref{eq:39}) and (\ref{eq:40}), we find 
\begin{eqnarray}
\frac{1}{A} \left( C^2 + \frac{B^2}{C^2} \right)  
\approx
\frac{\alpha}{\beta}.
\label{eq:45}
\end{eqnarray} 
Eliminating $B$ from Eqs.\ (\ref{eq:43}) and (\ref{eq:45}), 
we obtain 
\begin{eqnarray} 
\left(\frac{C^2}{A}\right)^3 
- \frac{\alpha(\alpha^2 + \beta)}{\beta(\alpha^2 - \beta)} 
\left(\frac{C^2}{A}\right)^2 
+ \frac{2(\alpha^2 - 3 \beta)}{\beta(\alpha^2 - \beta)} 
\hspace{0.5mm}
\frac{C^2}{A}
- \frac{2\alpha}{\beta(\alpha^2 - \beta)} = 0.
\label{eq:46}
\end{eqnarray}
Hence we could find the value of $C^2/A$ 
by numerically solving this equation.  However, we could not get 
the simple expression of the solution in terms of $\alpha$ and $\beta$.  
Thus 
we use the approximation in (\ref{eq:41}).  
Incidentally, for example, 
in the case $\epsilon_1 = 1.1 \times 10^{-2}$, and 
$\epsilon_2 = 1.0 \times 10^{-2}$, the real number solution 
of the numerical solution is 
$C^2/A = 322$, while the approximate solution is 
$C^2/A \approx 2/\alpha = 95.2$.  
The ratio of the former to the latter is therefore about 3.3, that is, 
the difference between the numerical solution and 
the approximate one is only about a factor.  
Consequently, we consider that the approximation in (\ref{eq:41}) is 
suitable to estimate $A$, $B$, and $C$.  
\fi


\subsection{Linearized equations} 
To begin with, 
we discuss the fluctuation equation 
in terms of 
$f=\epsilon_1 \varphi_1^2 + \epsilon_2 \varphi_2^2$ and 
$\tilde{f}=\epsilon_1 \varphi_1^2 - \epsilon_2 \varphi_2^2$ 
in order to derive
these mass values in the inflationary stage.  
With the ansatz  
\begin{eqnarray}
f &=& f_0 + \delta f, 
\hspace{5mm} \left| \delta f \right| \ll \left| f_0 \right|, 
\nonumber \\[3mm] 
\tilde{f} &=& {\tilde{f}}_0 + \delta \tilde{f}, 
\hspace{5mm} \left| \delta \tilde{f} \right| \ll \left| {\tilde{f}}_0 \right|,
\label{eq:45}
\end{eqnarray}
where $f_0$ and ${\tilde{f}}_0$ are the zeroth order quantities and 
satisfy Eqs.\ (\ref{eq:10}) and (\ref{eq:11}), respectively, 
we obtain the linearized equations 
\begin{eqnarray}
\left(
\begin{array}{c}
  \ddot{\delta f} + 3H \dot{\delta f} \\
  \ddot{\delta \tilde{f}} + 3H \dot{\delta \tilde{f}}
\end{array}
\right)
&=&
-{\cal M}^2
\left(
\begin{array}{c}
 \delta f \\  \delta \tilde{f}
\end{array}
\right),  
\label{eq:46} 
\end{eqnarray}
where
\begin{eqnarray}
{\cal M}^2 \equiv 
\left(
\begin{array}{cc}
  {{\cal M}^2}_{11} & {{\cal M}^2}_{12} \\
  {{\cal M}^2}_{21} & {{\cal M}^2}_{22} 
\end{array}
\right).  
\label{eq:48}
\end{eqnarray}
In deriving the expression (33), 
we have kept terms up to those of first order in 
$\delta f$ and $\delta \tilde{f}$.  
A detailed derivation of Eq.\ (33) and 
the expressions of the components of ${\cal M}^2$ 
are given in \S A.2. 

Here we emphasize the reason why we need to estimate 
the mass values of the scalar quantities $f$ and $\tilde{f}$.  
As stated at the beginning of this section, 
the final purpose of this section is to show that there exists a scalar field 
corresponding to the curvaton in the framework of this theory, without 
introducing any other scalar field that plays the role of the curvaton.  
The necessary condition for a scalar field to correspond to the curvaton 
is that its mass be much smaller than the Hubble parameter in the inflationary 
stage.  
Hence, in order to examine the existence of a scalar field 
corresponding to the curvaton, we consider not the mass scales of 
the dilatons $\varphi_i$ themselves, but those of the scalar quantities 
$f=\epsilon_1 \varphi_1^2 + \epsilon_2 \varphi_2^2$ and 
$\tilde{f}=\epsilon_1 \varphi_1^2 - \epsilon_2 \varphi_2^2$, 
which are expressed as linear combinations of $\varphi_1^2$ and 
$\varphi_2^2$.  
We cannot easily estimate these mass values from 
Eqs.\ (\ref{eq:10}) and (\ref{eq:11}) themselves, however, 
because the parts corresponding to the mass terms of these equations are too 
complicated.  
Therefore we consider the linearized equations (33)
by making use of the ansatz (32), and then we estimate 
the eigenmasses of $f$ and $\tilde{f}$ 
by carrying out the diagonalization of the matrix ${\cal M}^2$ in (34), 
as discussed in the next subsection.  

\if
where we have taken the terms up to the first order of 
$\delta f$ and $\delta \tilde{f}$.  
Moreover, we have neglected the quantities in terms of 
$\dot{\delta f}$ and $\dot{\delta \tilde{f}}$ 
derived from the right-hand side 
of Eqs.\ (\ref{eq:10}) and (\ref{eq:11}).  
This is because, 
as we see from the ansatz (\ref{eq:15}) and Eq.\ (\ref{eq:32}), 
both $f$ and $\tilde{f}$ are proportional to $t^2$ and 
thus in the large $t$ limit 
$\left| \dot{\delta f} \right| \ll \delta f$ and 
$\left| \dot{\delta \tilde{f}} \right| \ll \delta \tilde{f}$.  
Incidentally, the reason why we have chosen the field variable 
$\tilde{f}$ as the partner of the coupling $f$ is that 
$\tilde{f}$ is a simple quantity satisfying the condition that 
it should have the same dimension as $f$ and should be expressed by 
the linear combination of $\varphi_1^2$ and $\varphi_2^2$.

In deriving the linearized equations (\ref{eq:46}), we have paid 
attention to the following point:  
As noted in the previous subsection, 
from Eqs.\ (\ref{eq:30})$-$(\ref{eq:32}), 
(\ref{eq:34})$-$(\ref{eq:36}), and (\ref{eq:38}), 
we see that all the terms in the right-hand side of 
Eqs.\ (\ref{eq:10}) and (\ref{eq:11}) can 
be represented in terms of $f$ and $\varphi_{\mathrm{r}}$.  
Furthermore, 
from Eq.\ (\ref{eq:32}) we find 
$
{\varphi_{\mathrm{r}}}^2 
=
\left[ (\epsilon_1 + \epsilon_2)f 
-(\epsilon_1 - \epsilon_2) \tilde{f}
\right]/ (2 \epsilon_1 \epsilon_2). 
$
Hence the fluctuation $\delta \varphi_{\mathrm{r}}$ 
induced by $\delta f$ and $\delta \tilde{f}$ 
can be represented as follows: 
\begin{eqnarray}
\delta \varphi_{\mathrm{r}} 
=
\frac{1}{4\epsilon_1 \epsilon_2 \varphi_{\mathrm{r}}} 
\left[ (\epsilon_1 + \epsilon_2) \delta f
-(\epsilon_1 - \epsilon_2) \delta \tilde{f}
\right].  
\label{eq:47}
\end{eqnarray}
Consequently, all the fluctuations derived from the right-hand side of 
Eqs.\ (\ref{eq:10}) and (\ref{eq:11}) can be represented 
in terms of $\delta f$ and $\delta \tilde{f}$ 
by using Eqs.\ (\ref{eq:30})$-$(\ref{eq:32}), 
(\ref{eq:34})$-$(\ref{eq:36}), (\ref{eq:38}) and (\ref{eq:47}).  

As a result, ${\cal M}^2$ is given by 
\begin{eqnarray}
{\cal M}^2 \equiv 
\left(
\begin{array}{cc}
  {{\cal M}^2}_{11} & {{\cal M}^2}_{12} \\
  {{\cal M}^2}_{21} & {{\cal M}^2}_{22} 
\end{array}
\right),   
\label{eq:48}
\end{eqnarray}
with
\begin{eqnarray}
{{\cal M}^2}_{11} &=& 
-4 F^{-2} \left( 2F-1 \right) Q 
\hspace{0.5mm} 
\frac{1}{f}
\hspace{0.5mm} 
V
\nonumber \\[3mm]
&& \hspace{0mm}
{}-2 F^{-1} 
\left[ 
-1 + \frac{\epsilon_1 + \epsilon_2}{\epsilon_1 \epsilon_2} 
\left( \frac{1}{4} \frac{f}{{\varphi_{\mathrm{r}}}^2} + U \right)
-\frac{6Q}{F}
\right] 
\frac{1}{\varphi_{\mathrm{r}}}
\hspace{0.5mm} 
V^{\prime} 
\nonumber \\[3mm]
&& \hspace{0mm}
{}+F^{-1} 
\hspace{0.5mm}
\frac{\epsilon_1 + \epsilon_2}{2 \epsilon_1 \epsilon_2} 
\hspace{0.5mm}
\frac{f}{{\varphi_{\mathrm{r}}}^2} 
\hspace{0.5mm} 
V^{\prime \prime} 
\nonumber \\[3mm]
&& \hspace{0mm}
{}-F^{-1} 
\Biggl\{
\frac{Q}{F} \hspace{0.5mm}
\frac{\dot{\varphi_1}^2 + \dot{\varphi_2}^2}{f} 
+ \frac{\epsilon_1 + \epsilon_2}{\epsilon_1 \epsilon_2} 
\hspace{0.5mm} 
U 
\hspace{0.5mm} 
\frac{L^2}{{\varphi_{\mathrm{r}}}^4}
-\frac{1 + 6(\epsilon_1 + \epsilon_2)}{2F} J 
\nonumber \\[3mm]
&& \hspace{14mm}
{}+
\frac{Q}{\epsilon_1 \epsilon_2} 
\hspace{0.5mm} 
\frac{I}{{\varphi_{\mathrm{r}}}^2} 
\left[ 
Q^2 \frac{I}{f} - Q \frac{\dot{f}}{f} 
\pm (\epsilon_1 + \epsilon_2) 
\frac{\dot{\varphi_{\mathrm{r}}}}{\varphi_{\mathrm{r}}}
\right] 
\Biggr\} 
\label{eq:49} \\[5mm]
&\approx& 
-\frac{3\alpha(2\beta-\alpha^2)}{8\beta^2}
\frac{V}{\Lambda t^2}
+\frac{3\sqrt{3}\alpha^{3/2}(\alpha^2-4\beta)}{16\beta(\alpha^2-\beta)}
\frac{V^{\prime}}{\sqrt{\Lambda} t}
+\frac{\alpha^2}{2(\alpha^2-\beta)} 
V^{\prime \prime}
\nonumber \\[3mm]
&& \hspace{0mm}
{}
+\frac{C_{11}}{\beta(\alpha^2-\beta)(2\beta-\alpha^2)}
\hspace{0.5mm}
\frac{1}{t^2},  
\label{eq:50} \\[5mm]
{{\cal M}^2}_{12} &=& 
-4 F^{-2} (\epsilon_1 - \epsilon_2)
\hspace{0.5mm}
\frac{1}{f} \hspace{0.5mm} 
V
\nonumber \\[3mm] 
&& \hspace{0mm}
{}-2 F^{-1} (\epsilon_1 - \epsilon_2) 
\left[ 
-\frac{1}{\epsilon_1 \epsilon_2} 
\left( \frac{1}{4} \frac{f}{{\varphi_{\mathrm{r}}}^2} + U \right)
+ \frac{6}{F}
\right] 
\frac{1}{\varphi_{\mathrm{r}}}
\hspace{0.5mm} 
V^{\prime} 
\nonumber \\[3mm]
&& \hspace{0mm}
{}-F^{-1} 
\hspace{0.5mm}
\frac{\epsilon_1 - \epsilon_2}{2 \epsilon_1 \epsilon_2} 
\hspace{0.5mm}
\frac{f}{{\varphi_{\mathrm{r}}}^2} 
\hspace{0.5mm} 
V^{\prime \prime} 
\nonumber \\[3mm]
&& \hspace{0mm}
{}+F^{-1} 
\hspace{0.5mm}
(\epsilon_1 - \epsilon_2) 
\Biggl[
\frac{1}{F} \hspace{0.5mm}
\frac{\dot{\varphi_1}^2 + \dot{\varphi_2}^2}{f} 
+ \frac{U}{\epsilon_1 \epsilon_2} 
\hspace{0.5mm}
\frac{L^2}{{\varphi_{\mathrm{r}}}^4}
+\frac{3}{F} J
\nonumber \\[3mm]
&& \hspace{28mm}
{}+\frac{Q}{\epsilon_1 \epsilon_2} 
\hspace{0.5mm} 
\frac{I}{{\varphi_{\mathrm{r}}}^2} 
\left( 
Q \frac{I}{f} - \frac{\dot{f}}{f} 
\pm \frac{\dot{\varphi_{\mathrm{r}}}}{\varphi_{\mathrm{r}}}
\right)
\Biggr]
\label{eq:51} \\[5mm]
&\approx& 
\pm \sqrt{2\beta-\alpha^2}
\biggl\{
-\frac{3\alpha^2}{8\beta^2}
\frac{V}{\Lambda t^2}
+\frac{\sqrt{3}\alpha^{3/2}(\alpha^2+8\beta)}{16\alpha\beta(\alpha^2-\beta)}
\frac{V^{\prime}}{\sqrt{\Lambda} t}
-\frac{\alpha}{2(\alpha^2-\beta)} 
V^{\prime \prime}
\nonumber \\[3mm]
&& \hspace{22mm}
{}
+\frac{C_{12}}{\beta(\alpha^2-\beta)(2\beta-\alpha^2)}
\hspace{0.5mm}
\frac{1}{t^2}
\biggr\},
\label{eq:52} \\[5mm]
{{\cal M}^2}_{21} &=& 
4 F^{-2} 
\hspace{0.5mm}
\frac{Q}{\epsilon_1 - \epsilon_2} 
\hspace{0.5mm}
\left[ (\epsilon_1 + \epsilon_2) + 24 \epsilon_1 \epsilon_2 \right]
\hspace{0.5mm} 
\frac{1}{f}
\hspace{0.5mm} 
V
\nonumber \\[3mm]
&& \hspace{0mm}
{}-\frac{F^{-1}}{\epsilon_1 - \epsilon_2} 
\biggl\{ 
Q
\left(
-\frac{12Q}{F} 
+ \frac{\epsilon_1 + \epsilon_2}{2 \epsilon_1 \epsilon_2} 
\hspace{0.5mm} 
\frac{f}{{\varphi_{\mathrm{r}}}^2} 
\right) 
+ 48 \epsilon_1 \epsilon_2 
\left[ 
\left( 1 + \frac{6Q}{F} \right) P 
- Q \hspace{0.5mm} \frac{{\varphi_{\mathrm{r}}}^2}{f}
\right]
\nonumber \\[3mm]
&& \hspace{19mm}
{}+12(\epsilon_1 + \epsilon_2) 
\left( 
2 Q - P \hspace{0.5mm} \frac{f}{{\varphi_{\mathrm{r}}}^2}
\right)
+\frac{(\epsilon_1 + \epsilon_2)(\epsilon_1 - \epsilon_2)}
{\epsilon_1 \epsilon_2}
\hspace{0.5mm} 
W
\biggr\}
\hspace{0.5mm}
\frac{1}{\varphi_{\mathrm{r}}}
\hspace{0.5mm} 
V^{\prime} 
\nonumber \\[3mm]
&& \hspace{0mm}
{}-F^{-1} 
\hspace{0.5mm}
\frac{\epsilon_1 + \epsilon_2}{\epsilon_1 - \epsilon_2} 
\left( 
-\frac{Q}{2 \epsilon_1 \epsilon_2} + 12P
\right)
\frac{f}{{\varphi_{\mathrm{r}}}^2} 
\hspace{0.5mm} 
V^{\prime \prime} 
\nonumber \\[3mm]
&& \hspace{0mm}
{}+\frac{1}{\epsilon_1 - \epsilon_2} 
\hspace{0.5mm} 
\frac{1}{f^2}
\left[ 
Q \left( \dot{f}^2 - 4 \epsilon_1 \epsilon_2 L^2 \right) 
+ \left( 4 \epsilon_1 \epsilon_2 P - Q^2 \right) \dot{f} I 
\right]
\nonumber \\[3mm]
&& \hspace{0mm}
{}-F^{-1}
\biggl\{
\frac{Q}{(\epsilon_1 - \epsilon_2)F}
\hspace{0.5mm}
\left[ (\epsilon_1 + \epsilon_2) + 24 \epsilon_1 \epsilon_2 \right]
\left( \frac{\dot{\varphi_1}^2 + \dot{\varphi_2}^2}{f} + 3J \right)
\nonumber \\[3mm]
&& \hspace{14mm}
{}
+W 
\left( 
3J + \frac{\epsilon_1 + \epsilon_2}{2 \epsilon_1 \epsilon_2} 
\hspace{0.5mm} 
\frac{L^2}{{\varphi_{\mathrm{r}}}^4}
\right)
\biggr\}
\nonumber \\[3mm]
&& \hspace{0mm}
{}+
\frac{1}{\epsilon_1 - \epsilon_2} 
\left\{
\frac{\dot{f}}{f} 
-
\left[
Q + 
\frac{\epsilon_1 - \epsilon_2}{2 \epsilon_1 \epsilon_2}
\hspace{0.5mm} 
\frac{W}{F}
\left( 
\frac{f}{{\varphi_{\mathrm{r}}}^2} + 12 \epsilon_1 \epsilon_2
\right)
\right]
\frac{I}{f}
\right\}
\nonumber \\[3mm]
&& \hspace{55mm}
{}\times 
\left[ Q^2 \frac{I}{f} - Q \frac{\dot{f}}{f} 
\pm (\epsilon_1 + \epsilon_2) \frac{\dot{\varphi_{\mathrm{r}}}}
{\varphi_{\mathrm{r}}}
\right]
\label{eq:53} \\[5mm]
&\approx& 
\pm \sqrt{2\beta-\alpha^2}
\left\{
\frac{3\alpha^2}{8\beta^2}
\frac{V}{\Lambda t^2}
-\frac{9\sqrt{3}\alpha^{5/2}}{16\beta(\alpha^2-\beta)}
\frac{V^{\prime}}{\sqrt{\Lambda} t}
+\frac{\alpha}{2(\alpha^2-\beta)} 
V^{\prime \prime}
\right\}
\nonumber \\[3mm]
&& \hspace{22mm}
{}
+\frac{C_{21}}
{(\pm \sqrt{2\beta-\alpha^2})\beta(\alpha^2-\beta)}
\hspace{0.5mm}
\frac{1}{t^2},
\label{eq:54} \\[5mm] 
{{\cal M}^2}_{22} &=& 
-4 F^{-1} 
\left[ 
(\epsilon_1 + \epsilon_2) - 6(\epsilon_1 - \epsilon_2)
\hspace{0.5mm}
\frac{W}{F}
\right]
\frac{1}{f}
\hspace{0.5mm} 
V
\nonumber \\[3mm]
&& \hspace{0mm} 
{}+F^{-1} 
\biggl\{
2
\left[ 
1-\frac{6Q}{F} 
+ 6\left( 2Q - P \hspace{0.5mm} \frac{f}{{\varphi_{\mathrm{r}}}^2} \right) 
+ 144 \epsilon_1 \epsilon_2 \hspace{0.5mm} \frac{P}{F} 
\right] 
\nonumber \\[3mm]
&& \hspace{13mm}
{}+
\frac{1}{2 \epsilon_1 \epsilon_2} 
\left[ 
2(\epsilon_1 - \epsilon_2) W 
+ Q \hspace{0.5mm} \frac{f}{{\varphi_{\mathrm{r}}}^2} 
\right]
\biggr\}
\hspace{0.5mm}
\frac{1}{\varphi_{\mathrm{r}}}
\hspace{0.5mm} 
V^{\prime} 
\nonumber \\[3mm]
&& \hspace{0mm}
{}-F^{-1} 
\left( \frac{Q}{2 \epsilon_1 \epsilon_2} - 12P \right)
\frac{f}{{\varphi_{\mathrm{r}}}^2} 
\hspace{0.5mm} 
V^{\prime \prime}
\nonumber \\[3mm]
&& \hspace{0mm}
{}+\frac{1}{2f^2} 
\left( 
-\dot{f}^2 + 4 \epsilon_1 \epsilon_2 L^2 + 2 Q \dot{f} I 
\right)
\nonumber \\[3mm]
&& \hspace{0mm}
{}+F^{-1} 
\left\{
\left[ 
(\epsilon_1 + \epsilon_2) - 6(\epsilon_1 - \epsilon_2)
\hspace{0.5mm}
\frac{W}{F}
\right]
\left( 
\frac{\dot{\varphi_1}^2 + \dot{\varphi_2}^2}{f} + 3J
\right)
+ \frac{\epsilon_1 - \epsilon_2}{2 \epsilon_1 \epsilon_2}
\hspace{0.5mm}
W
\hspace{0.5mm}
\frac{L^2}{{\varphi_{\mathrm{r}}}^4}
\right\}
\nonumber \\[3mm]
&& \hspace{0mm}
{}-
\left\{
\frac{\dot{f}}{f}
- 
\left[ 
Q + \frac{\epsilon_1 - \epsilon_2}{2 \epsilon_1 \epsilon_2}
\hspace{0.5mm} 
\frac{W}{F}
\left( 
\frac{f}{{\varphi_{\mathrm{r}}}^2} + 12 \epsilon_1 \epsilon_2 
\right)
\right]
\frac{I}{f} 
\right\} 
\left( 
Q \frac{I}{f} - \frac{\dot{f}}{f} 
\pm \frac{\dot{\varphi_{\mathrm{r}}}}{\varphi_{\mathrm{r}}}
\right)
\label{eq:55} \\[5mm]
&\approx&
-\frac{3\alpha^3}{8\beta^2}
\frac{V}{\Lambda t^2}
+\frac{\sqrt{3}\alpha^{3/2}(14\beta-5\alpha^2)}{16\beta(\alpha^2-\beta)}
\frac{V^{\prime}}{\sqrt{\Lambda} t}
-\frac{2\beta-\alpha^2}{2(\alpha^2-\beta)} 
V^{\prime \prime}
\nonumber \\[3mm]
&& \hspace{0mm}
{}
+\frac{C_{22}}
{\beta(\alpha^2-\beta)(2\beta-\alpha^2)}
\hspace{0.5mm}
\frac{1}{t^2},
\label{eq:56} 
\end{eqnarray}
where 
\begin{eqnarray}
F &=& 
1 + 12 
\left[ 
(\epsilon_1 + \epsilon_2) 
- \epsilon_1 \epsilon_2 \frac{{\varphi_{\mathrm{r}}}^2}{f} 
\right] = 1+12U 
\hspace{0.5mm} \approx \hspace{0.5mm}
1+ 12 \frac{\beta}{\alpha}, 
\label{eq:57} \\[3mm]
Q &=& (\epsilon_1 + \epsilon_2) 
- 2 \epsilon_1 \epsilon_2 \frac{{\varphi_{\mathrm{r}}}^2}{f}
\hspace{0.5mm} \approx \hspace{0.5mm}
\frac{2\beta - \alpha^2}{\alpha}, 
\label{eq:58} \\[3mm]
U &=& (\epsilon_1 + \epsilon_2) 
- \epsilon_1 \epsilon_2 \frac{{\varphi_{\mathrm{r}}}^2}{f}
\hspace{0.5mm} \approx \hspace{0.5mm}
\frac{\beta}{\alpha}, 
\label{eq:59} \\[3mm]
J &=& 
\frac{1}{f^2} 
\left( \dot{f}^2 + 4 \epsilon_1 \epsilon_2 L^2 \right)
\hspace{0.5mm} \approx \hspace{0.5mm}
4 \left( \frac{7\alpha^2 - 6\beta}{2\beta - \alpha^2} \right) 
\frac{1}{t^2},
\label{eq:60} \\[3mm] 
I &=& -\frac{1}{2P} 
\left[ - \left( \frac{{\varphi_{\mathrm{r}}}^2}{f} \right) \dot{f} 
\pm 2 \varphi_{\mathrm{r}} \dot{\varphi_{\mathrm{r}}}
\right]
\hspace{0.5mm} \approx \hspace{0.5mm}
\frac{128}{3} \hspace{0.5mm} \frac{\beta^2}{\alpha (2\beta - \alpha^2)} 
\hspace{0.5mm} \Lambda \hspace{0.5mm} t, 
\label{eq:61} \\[3mm] 
P &=& - 
\left(1- \epsilon_1 \frac{{\varphi_{\mathrm{r}}}^2}{f} \right) 
\left(1 - \epsilon_2 \frac{{\varphi_{\mathrm{r}}}^2}{f} \right)
\hspace{0.5mm} \approx \hspace{0.5mm}
\frac{2\beta - \alpha^2}{\alpha^2}, 
\label{eq:62} \\[3mm] 
W &=& (\epsilon_1 - \epsilon_2) + 
\frac{\epsilon_1 + \epsilon_2}{\epsilon_1 - \epsilon_2} 
\hspace{0.5mm}  
Q
\hspace{0.5mm} \approx \hspace{0.5mm}
\pm 2 \sqrt{2\beta - \alpha^2},  
\label{eq:63} \\[3mm]
C_{11} &=& 
7\alpha^4\beta-16\beta^3-2\alpha^2\beta^2-\alpha^6
-\frac{96\beta(\alpha^2-\beta)^2(2\beta-\alpha^2)}{\alpha},
\label{eq:64}  \\[5mm]
C_{12} &=& 
14\alpha\beta^2-\alpha^3\beta-\alpha^5 
+\frac{48\beta(\alpha^2-\beta)
\left[
2\alpha(\alpha^2-\beta)-3\beta(7\alpha^2-6\beta)
\right]
}
{\alpha},
\label{eq:65}  \\[5mm] 
C_{21} &=& 
\alpha^5-8\alpha\beta^2-5\alpha^3\beta 
+12\beta(-7\alpha^4+\alpha^2\beta+10\beta^2),
\label{eq:66}  \\[5mm] 
C_{22} &=& 
\alpha
\left[
\alpha(14\beta^2-\alpha^4-\alpha^2\beta)
+12\beta(7\alpha^4-9\alpha^2\beta-2\beta^2)
\right].  
\label{eq:67}   
\end{eqnarray}
Here the expression of $F$ (\ref{eq:57}) follows from 
Eqs.\ (\ref{eq:12}) and (\ref{eq:30}).  

In deriving the last approximate equalities in each of the equations
(\ref{eq:50}), (\ref{eq:52}), (\ref{eq:54}), and 
(\ref{eq:56})$-$(\ref{eq:63}), 
we have used Eqs.\ (\ref{eq:24}), (\ref{eq:25}), 
(\ref{eq:39}), (\ref{eq:40}), (\ref{eq:42}), and (\ref{eq:44}).  
Moreover, 
in the last approximate equalities in each of the equations 
(\ref{eq:50}), (\ref{eq:52}), (\ref{eq:54}), and (\ref{eq:56}),
we have taken all the leading order terms and 
a part of the sub-leading order ones in $\epsilon_i$ and 
have neglected the rest of the sub-leading order ones which are 
explicitly unimportant.  
Furthermore, 
in deriving the last approximate equality in 
(\ref{eq:61}) we have taken the negative sign so that 
$I$ could have a finite value.  
If we take the positive sign, 
under the approximation in (\ref{eq:41}), $I \approx 0$.  
Since 
it follows from Eqs.\ (\ref{eq:38}), (\ref{eq:61}), and (\ref{eq:62}) 
that the angular momentum is given by $L = - \left( \pm P^{1/2} \right) I$
and here we consider the case $L\not= 0$, 
we have taken the negative sign so that the angular momentum could 
have a finite value.  
Hence, in deriving the approximate expressions 
(\ref{eq:50}), (\ref{eq:52}), (\ref{eq:54}), and (\ref{eq:56}), 
we have taken the negative sign 
in each last term in the right-hand side of 
Eqs.\ (\ref{eq:49}), (\ref{eq:51}), (\ref{eq:53}), and (\ref{eq:55}) 
because the sign $\pm$ in these terms derives from the angular 
momentum $L$ and $L$ is proportional to $I$ as shown above.  

We finally emphasize that all the terms in each of the 
components of ${\cal M}^2$ 
can be expressed in terms of the background quantities 
$f=At^2$, $L=Bt$, and $\varphi_{\mathrm{r}}=Ct$ in the inflationary stage 
and hence can be approximately represented in terms of 
$\alpha$ and $\beta$ by using expressions 
(\ref{eq:39}), (\ref{eq:40}), (\ref{eq:42}), and (\ref{eq:44}) 
as shown in expressions 
(\ref{eq:50}), (\ref{eq:52}), (\ref{eq:54}), and (\ref{eq:56}).  
\fi

\subsection{Mass diagonalization} 
\if
Next, in order to consider that a huge hierarchy between 
the mass values of $f$ and $\tilde{f}$ in the inflationary 
stage could be realized we investigate these eigenmasses by 
solving the following 
characteristic 
equation of ${\cal M}^2$:  
\fi
Next, we show that an extremely large hierarchy of 
the mass values of $f$ and $\tilde{f}$ in the inflationary 
stage can be realized.  For this purpose, 
we investigate these eigenmasses by 
solving the following 
characteristic 
equation of ${\cal M}^2$:  
\begin{eqnarray} 
\det \left({\cal M}^2 - \lambda E  \right)  = 0 
\hspace{1mm}
\Longleftrightarrow 
\hspace{1mm}
\lambda^2 - \mathrm{Tr} \hspace{0.5mm} {\cal M}^2  \hspace{0.5mm} \lambda 
+ \det {\cal M}^2  = 0, 
\label{eq:68}
\end{eqnarray}
where $\lambda$ is an eigenvalue of ${\cal M}^2$, and 
$E$ is the unit matrix.  The solutions of Eq.\ (35) 
are equivalent to the diagonal components of the diagonalized 
matrix of ${\cal M}^2$, namely, the square of the eigenmasses of $f$ and 
$\tilde{f}$.  
The solution of Eq.\ (35) is given by 
\begin{eqnarray} 
\lambda_{\pm} = \frac{\mathrm{Tr} \hspace{0.5mm} {\cal M}^2}{2} 
\left[
1 \pm 
\sqrt{1-
\frac{4\det {\cal M}^2}{(\mathrm{Tr} \hspace{0.5mm} {\cal M}^2)^2}
}
\hspace{0.5mm}
\right],
\label{eq:69}
\end{eqnarray}
where the signs $\pm$ in $\lambda_{\pm}$ correspond to those on 
the right-hand side of this equation.  

The necessary conditions for 
an extremely large hierarchy of 
the eigenmasses of $f$ and $\tilde{f}$ 
to be realized are the following three:  
(I) $\mathrm{Tr} \hspace{0.5mm} {\cal M}^2 > 0$,   
(II) $\det {\cal M}^2 > 0$,   
and 
(III) $ \Theta \equiv 
4\det {\cal M}^2 / (\mathrm{Tr} \hspace{0.5mm} {\cal M}^2)^2 \ll 1$.  
The first and second conditions are necessary in order that 
both the square of the eigenmasses can be positive, 
that is, $\lambda_{+} > 0$ and $\lambda_{-} > 0$.  
The third condition is necessary in order that 
an extremely large hierarchy of the eigenmasses can be realized, 
that is, $\lambda_{-}/\lambda_{+} \ll 1$.  
In the case that the above three conditions are satisfied, 
it follows from Eq.\ (36) that 
the square of the eigenmasses of $f$ and $\tilde{f}$ can be approximately 
expressed as 
\begin{eqnarray} 
\lambda_{+} &\approx& 
\mathrm{Tr} \hspace{0.5mm} {\cal M}^2
\left[
1-\frac{\det {\cal M}^2}{(\mathrm{Tr} \hspace{0.5mm} {\cal M}^2)^2 }
\right] > 0,
\label{eq:70} \\[3mm]
\lambda_{-} &\approx& 
\frac{\det {\cal M}^2}{\mathrm{Tr} \hspace{0.5mm} {\cal M}^2} > 0,
\label{eq:71} \\[3mm]
\frac{\lambda_{-}}{\lambda_{+}} &\approx& 
\frac{\det {\cal M}^2}{(\mathrm{Tr} \hspace{0.5mm} {\cal M}^2)^2} 
\left[
1+\frac{\det {\cal M}^2}{(\mathrm{Tr} \hspace{0.5mm} {\cal M}^2)^2 }
\right]
\ll 1, 
\label{eq:72}
\end{eqnarray}
where each approximate equality in these relations follows from 
the above third condition, 
$\Theta = 4\det {\cal M}^2 / 
(\mathrm{Tr} \hspace{0.5mm} {\cal M}^2)^2 \ll 1$.  
We therefore investigate the case in which the above three conditions are 
satisfied by estimating $\mathrm{Tr} \hspace{0.5mm} {\cal M}^2$ and 
$\det {\cal M}^2$.  

Here we consider the following case.  
As stated in \S 2, in regard to the dilaton potential 
appearing in 
Eq.\ (\ref{eq:3}), we assume $V_{0} \approx \Lambda \approx M^4$.  
Moreover, 
in the inflationary stage, 
the field amplitude of the dilatons is given by 
$
\varphi_1 = \gamma_1 M
$
and 
$
\varphi_2 = \gamma_2 M, 
$
where $\gamma_1$ and $\gamma_2$ are dimensionless parameters with time dependence. In addition, we assume 
$\gamma_1 \approx \gamma_2 \approx \gamma = \gamma(t)$.  (These relations 
are 
consistent with the assumption $\varphi_1^2 \approx \varphi_2^2$, 
made 
in deriving the approximate expressions of $A$, $B$, $C$ and $\omega$ 
in \S 3.)  
Furthermore, we consider the case in which the value of $\epsilon_1$ 
is close to that of $\epsilon_2$ and introduce a dimensionless 
constant 
$\bar{\epsilon} \equiv \left(\epsilon_1 + \epsilon_2 \right)/2 = 
\alpha/2 = O[\epsilon_i]$.  
In this case, 
from Eq.\ (\ref{eq:2}) 
we find
\begin{eqnarray}
f \approx 2 \bar{\epsilon} \gamma^2 M^2 
= \alpha \gamma^2 M^2.  
\label{eq:73}
\end{eqnarray}
Hence, it follows from Eqs.\ (\ref{eq:13}) and (\ref{eq:73}) 
that the Hubble parameter in the inflationary stage is expressed as 
\begin{eqnarray}
H &\approx& \sqrt{\frac{M^4}{6f}} \approx \Upsilon M,  
\label{eq:74} \\[3mm]
\Upsilon &\equiv&
\frac{1}{\sqrt{12 \bar{\epsilon} \gamma^2}}
= \frac{1}{\sqrt{6\alpha \gamma^2}}.  
\label{eq:75}
\end{eqnarray}
In deriving the first approximate equality in Eq.\ (\ref{eq:74}) 
we have used two relations.  The first is 
$
(\dot{\varphi_1}^2 + \dot{\varphi_2}^2)/2 + V 
\approx \left[16\beta/(3\alpha)\right] \Lambda + \Lambda 
\approx \left[2/(3\omega)\right] \Lambda + \Lambda \approx \Lambda, 
$
where, in deriving the first approximate equality, 
we have used Eqs.\ (\ref{eq:19}) and (\ref{eq:40}), 
in deriving the second approximate equality, we have used 
Eq.\ (\ref{eq:28}), 
and 
in deriving the last approximate equality, 
we have used $\omega \gg 1$.  
\if
$
(\dot{\varphi_1}^2 + \dot{\varphi_2}^2)/2 + V 
\approx \left[16\beta/(3\alpha)\right] \Lambda + \Lambda 
= 2/(3\omega) + \Lambda \approx \Lambda, 
$
where in deriving the first approximate equality 
we have used Eqs.\ (\ref{eq:19}) and (\ref{eq:40}), 
in deriving the second equality we have used Eq.\ (\ref{eq:28}), 
and 
in deriving the last approximate equality 
we have used $\omega \ll 1$.  
\fi
\if
$
(\dot{\varphi_1}^2 + \dot{\varphi_2}^2)/2 + V 
\approx \left[ 16\beta/(3\alpha) \right] \Lambda + \Lambda \approx \Lambda, 
$
where in deriving the first approximate equality 
we have used Eqs.\ (\ref{eq:19}) and (\ref{eq:40}). 
\fi 
Furthermore, we have ignored variation of the potential, and hence replaced 
$V$ by its average value, $\Lambda$.  
The second relation used here is $H^2 \gg H \dot{f}/f$.  
This relation follows from 
$H/(\dot{f}/f)=\omega/2 \gg 1$, where 
we have used $H=\omega/t$ and $f=At^2$ in the ansatz (\ref{eq:15}).  

From Eq.\ (\ref{eq:3}) and the expressions 
(\ref{eq:50}), (\ref{eq:52}), (\ref{eq:54}) and (\ref{eq:56}), 
we can obtain approximate expressions for 
$\mathrm{Tr} \hspace{0.5mm} {\cal M}^2 $ and $\det {\cal M}^2$.  
For convenience in describing these expressions, 
we represent 
$\mathrm{Tr} \hspace{0.5mm} {\cal M}^2$ and 
$\det {\cal M}^2$ as 
\begin{eqnarray}
\mathrm{Tr} \hspace{0.5mm} {\cal M}^2 
&=&
{{\cal M}^2}_{11} + {{\cal M}^2}_{22} 
\nonumber \\[5mm]
&\equiv& 
(\mathrm{Tr} \hspace{0.5mm} {\cal M}^2)^{(0)} 
+ (\mathrm{Tr} \hspace{0.5mm} {\cal M}^2)^{(1)}, 
\label{eq:76} 
\end{eqnarray}
and 
\begin{eqnarray}
\det {\cal M}^2
&=& 
{{\cal M}^2}_{11} {{\cal M}^2}_{22} - {{\cal M}^2}_{12} {{\cal M}^2}_{21}
\nonumber \\[5mm]
&\equiv&
(\det {\cal M}^2)^{(0)} 
+ (\det {\cal M}^2)^{(1)}
+ (\det {\cal M}^2)^{(2)}.  
\label{eq:79}   
\end{eqnarray}
Here, 
$(\mathrm{Tr} \hspace{0.5mm} {\cal M}^2)^{(0)}$ 
and $(\det {\cal M}^2)^{(0)}$ are the parts independent of the 
sine functions.  
Moreover,
$(\mathrm{Tr} \hspace{0.5mm} {\cal M}^2)^{(1)}$ 
and $(\det {\cal M}^2)^{(1)}$ are the parts proportional to 
$\sin [ \varphi_{\mathrm{r}}/M + 
\arctan \left( X_1/X_2 \right) ]$ and 
$\sin \left[ \varphi_{\mathrm{r}}/M + 
\arctan \left( Y_1/Y_2 \right) \right]$, respectively, 
and $(\det {\cal M}^2)^{(2)}$ is the part proportional to 
$\sin \left[ 2\varphi_{\mathrm{r}}/M + 
\arctan \left( Z_1/Z_2 \right) \right]$.  
\if
where 
$(\mathrm{Tr} \hspace{0.5mm} {\cal M}^2)^{(0)}$ 
and $(\det {\cal M}^2)^{(0)}$ are the parts independent of the 
sine functions, 
while 
$(\mathrm{Tr} \hspace{0.5mm} {\cal M}^2)^{(1)}$ 
and $(\det {\cal M}^2)^{(1)}$ are the parts proportional to 
$\sin \left[ \varphi_{\mathrm{r}}/M + 
\arctan \left( X_1/X_2 \right) \right]$ and 
$\sin \left[ \varphi_{\mathrm{r}}/M + 
\arctan \left( Y_1/Y_2 \right) \right]$, respectively, 
and $(\det {\cal M}^2)^{(2)}$ is the part proportional to 
$\sin \left[ 2\varphi_{\mathrm{r}}/M + 
\arctan \left( Z_1/Z_2 \right) \right]$.  
\fi
These approximate expressions for 
$\mathrm{Tr} \hspace{0.5mm} {\cal M}^2 $ and $\det {\cal M}^2$ and 
the expressions for $X_1$, $X_2$, $Y_1$, $Y_2$, $Z_1$ and $Z_2$ are 
presented in \S A.3.  
Here, as stated above, we have considered the case 
$V_{0} \approx \Lambda \approx M^4$.  
Moreover, from Eq.\ (\ref{eq:74}) we have used 
the relation 
\begin{eqnarray}
M \approx \frac{H}{\Upsilon} 
=
\frac{\omega}{\Upsilon} \hspace{0.5mm} \frac{1}{t} 
\approx 
\frac{\alpha}{8\beta}\frac{1}{\Upsilon} \hspace{0.5mm} \frac{1}{t}, 
\label{eq:89}
\end{eqnarray}
where the last approximate equality follows from Eq.\ (\ref{eq:28}).

\if
From Eq.\ (\ref{eq:3}) and expressions 
(\ref{eq:50}), (\ref{eq:52}), (\ref{eq:54}), and (\ref{eq:56}), 
we can obtain the approximate expressions of 
$\mathrm{Tr} \hspace{0.5mm} {\cal M}^2 $ and $\det {\cal M}^2$.  
For convenience in describing these expressions, 
we represent 
$
\mathrm{Tr} \hspace{0.5mm} {\cal M}^2 \equiv
(\mathrm{Tr} \hspace{0.5mm} {\cal M}^2)^{(0)} 
+ (\mathrm{Tr} \hspace{0.5mm} {\cal M}^2)^{(1)} 
$
and
$
\det {\cal M}^2 \equiv 
(\det {\cal M}^2)^{(0)} 
+ (\det {\cal M}^2)^{(1)}
+ (\det {\cal M}^2)^{(2)},
$
where 
$(\mathrm{Tr} \hspace{0.5mm} {\cal M}^2)^{(0)}$ 
and $(\det {\cal M}^2)^{(0)}$ are the parts independent of 
sine functions, 
while 
$(\mathrm{Tr} \hspace{0.5mm} {\cal M}^2)^{(1)}$ 
and $(\det {\cal M}^2)^{(1)}$ are the parts proportional to 
$\sin \left[ \varphi_{\mathrm{r}}/M + 
\arctan \left( X_1/X_2 \right) \right]$ and 
$\sin \left[ \varphi_{\mathrm{r}}/M + 
\arctan \left( Y_1/Y_2 \right) \right]$, respectively, 
and $(\det {\cal M}^2)^{(2)}$ is the part proportional to 
$\sin \left[ 2\varphi_{\mathrm{r}}/M + 
\arctan \left( Z_1/Z_2 \right) \right]$.  
($X_1$, $X_2$, $Y_1$, $Y_2$, $Z_1$, and $Z_2$ will be 
given by Eqs.\ (\ref{eq:83})$-$(\ref{eq:88}).) 
Consequently, 
the approximate expressions of 
$\mathrm{Tr} \hspace{0.5mm} {\cal M}^2 $ and $\det {\cal M}^2$ 
are given by  
\if
It follows from Eq.\ (\ref{eq:3}) and the expressions 
(\ref{eq:52}), (\ref{eq:54}), (\ref{eq:56}), and (\ref{eq:58}) 
that the approximate expressions of 
$\mathrm{Tr} \hspace{0.5mm} {\cal M}^2 $ and $\det {\cal M}^2$ 
are given by  
\fi
\begin{eqnarray}
\mathrm{Tr} \hspace{0.5mm} {\cal M}^2 &=& 
{{\cal M}^2}_{11} + {{\cal M}^2}_{22} 
\nonumber \\[5mm]
&\equiv& 
(\mathrm{Tr} \hspace{0.5mm} {\cal M}^2)^{(0)} 
+ (\mathrm{Tr} \hspace{0.5mm} {\cal M}^2)^{(1)}, 
\label{eq:76} \\[5mm]
(\mathrm{Tr} \hspace{0.5mm} {\cal M}^2)^{(0)} 
&\approx& \hspace{0mm}
\left[
-\frac{3\alpha}{4\beta} 
+ \frac{C_{11}+C_{22}}{\beta(\alpha^2-\beta)(2\beta-\alpha^2)}
\right] 
\hspace{0mm}
\frac{1}{t^2},
\label{eq:77} \\[5mm]
(\mathrm{Tr} \hspace{0.5mm} {\cal M}^2)^{(1)}
&\approx& \hspace{0mm}
\sqrt{X_1^2 +X_2^2} 
\sin \left( \frac{\varphi_{\mathrm{r}}}{M} + 
\arctan \frac{X_1}{X_2} \right)
\frac{1}{t^2},
\label{eq:78}  
\end{eqnarray}
and 
\begin{eqnarray}
\det {\cal M}^2
&=& 
{{\cal M}^2}_{11} {{\cal M}^2}_{22} - {{\cal M}^2}_{12} {{\cal M}^2}_{21}
\nonumber \\[5mm]
&\equiv& 
(\det {\cal M}^2)^{(0)} 
+ (\det {\cal M}^2)^{(1)}
+ (\det {\cal M}^2)^{(2)},
\label{eq:79} \\[5mm]
(\det {\cal M}^2)^{(0)} 
&\approx& \hspace{0mm}
\biggl\{
\frac{3\alpha}{8\beta^3(\alpha^2-\beta)}
\left[
-\frac{\alpha^2}{2\beta-\alpha^2}C_{11} -C_{22}
+ \alpha\left( C_{21} - C_{12} \right)
\right]
\nonumber \\[3mm]
&& \hspace{-5mm}
{}
+ \frac{C_{11}C_{22} - (2\beta-\alpha^2)C_{12}C_{21}}
{\beta^2(\alpha^2-\beta)^2(2\beta-\alpha^2)^2}
\nonumber \\[3mm]
&& \hspace{-5mm}
{}
+\frac{3\alpha^3}{64\beta^4}
\left[
9\alpha(2\beta-\alpha^2)
+\frac{\alpha^6-11\alpha^2\beta^2+2\alpha^4\beta+8\beta^3}
{64(\alpha^2-\beta)^2} \frac{1}{\Upsilon^2}
\right]
\biggr\} \hspace{0.5mm}
\frac{1}{t^4},
\label{eq:80} \\[5mm]
(\det {\cal M}^2)^{(1)}
&\approx& \hspace{0mm}
\sqrt{Y_1^2 +Y_2^2} 
\sin \left( \frac{\varphi_{\mathrm{r}}}{M} + 
\arctan \frac{Y_1}{Y_2} \right) 
\hspace{0mm}
\frac{1}{t^4},
\label{eq:81} \\[5mm]
(\det {\cal M}^2)^{(2)}
&\approx& \hspace{0mm}
\sqrt{Z_1^2 +Z_2^2} 
\sin \left( 2\frac{\varphi_{\mathrm{r}}}{M} + 
\arctan \frac{Z_1}{Z_2} \right)
\hspace{0mm}
\frac{1}{t^4},
\label{eq:82}  
\end{eqnarray}
where
\begin{eqnarray}
X_1 &=& 
-\frac{3\alpha}{4\beta}
\left(1+\frac{\alpha}{48\beta}\frac{1}{\Upsilon^2} \right),
\label{eq:83}  \\[5mm]
X_2 &=& 
\frac{\sqrt{3} \alpha^{5/2}}{64\beta^2} 
\frac{1}{\Upsilon},
\label{eq:84}  \\[5mm]
Y_1 &=& 
\frac{3\alpha}{8\beta^3(\alpha^2-\beta)}
\left[
-\frac{\alpha^2}{2\beta-\alpha^2}C_{11} -C_{22}
+ \alpha\left( C_{21} - C_{12} \right)
\right]
+\frac{9\alpha^4(2\beta-\alpha^2)}{16\beta^4}
\nonumber \\[3mm]
&& \hspace{0mm}
{}
-\frac{\alpha^2}{128\beta^3(\alpha^2-\beta)}
\biggl\{
\frac{1}{\alpha^2-\beta}
\left[
-C_{11}+\frac{\alpha^2}{2\beta-\alpha^2}C_{22} 
+ \alpha\left( C_{21} - C_{12} \right)
\right]
\nonumber \\[3mm]
&& \hspace{31.5mm}
{}
+\frac{3\alpha(2\beta^2-\alpha^4)}{4\beta}
\biggr\}
\frac{1}{\Upsilon^2},
\label{eq:85}  \\[5mm] 
Y_2 &=& 
-\frac{\sqrt{3}\alpha^{5/2}}{128\beta^3(\alpha^2-\beta)}
\biggl\{
\frac{1}{\alpha^2-\beta}
\biggl[ 
\frac{(14\beta-5\alpha^2)C_{11} + 3(\alpha^2-4\beta)C_{22}}
{2\beta-\alpha^2}
+9 \alpha C_{12}
\nonumber \\[3mm]
&& \hspace{28.5mm}
{}
-\frac{(\alpha^2+8\beta)C_{21}}{\alpha}
\biggr]
+\frac{3(12\alpha^3\beta-22\alpha\beta^2+\alpha^5)}{4\beta}
\biggr\}
\frac{1}{\Upsilon},
\label{eq:86}  \\[5mm] 
Z_1 &=& 
\frac{3\alpha^3}{64\beta^4}
\left[
3\alpha(2\beta-\alpha^2)
+\frac{7\alpha^6-5\alpha^2\beta^2-10\alpha^4\beta+8\beta^3}
{64(\alpha^2-\beta)^2} \frac{1}{\Upsilon^2}
\right],
\label{eq:87}  \\[5mm] 
Z_2 &=& 
\frac{\sqrt{3}\alpha^{5/2}}{1024\beta^4}
\left[
-\frac{3(12\alpha^3\beta-22\alpha\beta^2+\alpha^5)}{\alpha^2-\beta}
+ \frac{\alpha^2}{4} \frac{1}{\Upsilon^2}
\right]
\frac{1}{\Upsilon}.  
\label{eq:88}
\end{eqnarray}
Here, as stated above, we have considered the case 
$V_{0} \approx \Lambda \approx M^4$.  
Moreover, from Eq.\ (\ref{eq:74}) we have used 
the following relation:  
\begin{eqnarray}
M \approx \frac{H}{\Upsilon} 
=
\frac{\omega}{\Upsilon} \hspace{0.5mm} \frac{1}{t} 
\approx 
\frac{\alpha}{8\beta}\frac{1}{\Upsilon} \hspace{0.5mm} \frac{1}{t}, 
\label{eq:89}
\end{eqnarray}
where the last approximate equality follows from Eq.\ (\ref{eq:28}). 
\fi 

We now approximately evaluate $\mathrm{Tr} \hspace{0.5mm} {\cal M}^2$ and 
$\det {\cal M}^2$ by using Eqs.\ (\ref{eq:76})$-$(\ref{eq:82}) 
and then investigate the case in which the above three conditions can be 
satisfied.  
In order to confirm that the above three conditions can always be 
satisfied independently of the values of the sine functions, 
we investigate the case that the following three relations can be 
satisfied:  
\begin{eqnarray}
\tilde{\mathrm{Tr} \hspace{0.5mm} {\cal M}^2} &\equiv& 
(\mathrm{Tr} \hspace{0.5mm} {\cal M}^2)^{(0)} 
- \frac{\sqrt{X_1^2 +X_2^2}}{t^2}  > 0, 
\label{eq:90} \\[5mm]
\tilde{\det {\cal M}^2} &\equiv& 
(\det {\cal M}^2)^{(0)} 
- \frac{\sqrt{Y_1^2 +Y_2^2} + \sqrt{Z_1^2 + Z_2^2}}{t^4} > 0, 
\label{eq:91} \\[5mm]
\tilde{\Theta} &\equiv& \hspace{0mm}
4 \frac{(\det {\cal M}^2)^{(0)} 
+ 
\left( \sqrt{Y_1^2 +Y_2^2} + \sqrt{Z_1^2 + Z_2^2} \right) 
\hspace{0.5mm} t^{-4}
}
{
\left( \tilde{\mathrm{Tr} \hspace{0.5mm} {\cal M}^2} \right)^2
} \ll 1,  
\label{eq:92}
\end{eqnarray}
where $\tilde{\Theta}$ is positive.  
The first relation means that 
the minimum of $\mathrm{Tr} \hspace{0.5mm} {\cal M}^2$ 
is positive.  
Moreover, the second relation means that 
a value smaller than the minimum of $\det {\cal M}^2$ is positive.  
Furthermore, the third relation means that 
a value larger than the maximum of $\Theta$ is much smaller than unity.  
The reason we have adopted 
$\tilde{\det {\cal M}^2}$ as 
a value smaller than the minimum of $\det {\cal M}^2$ and 
$\tilde{\Theta}$ as a value larger than the maximum of $\Theta$ 
is that we cannot obtain analytical expressions of 
the minimum and 
maximum of $\det {\cal M}^2$, because, 
as seen from Eqs.\ (\ref{eq:79})$-$(\ref{eq:82}), 
$\det {\cal M}^2$ has two terms proportional to sine functions whose 
phases differ.  
\if
The first and second relations means that 
the minimum of $\mathrm{Tr} \hspace{0.5mm} {\cal M}^2$ 
and $\det {\cal M}^2$ are positive values.  Moreover, the third relation means 
that a larger value than the maximum of $\Theta$ is much smaller than unity.  
The reason why we have adopted $\tilde{\Theta}$ as a larger value than the maximum of $\Theta$ is that we could not obtain the analytical expression of 
the maximum of $\det {\cal M}^2$ because, 
as is seen in Eqs.\ (\ref{eq:81})$-$(\ref{eq:84}), 
$\det {\cal M}^2$ has two terms proportional to sine functions whose 
phases are different from each other.  
\fi

Moreover, as the effective values of $\lambda_{\pm}$ and the ratio 
$\lambda_{-}/\lambda_{+}$, 
we adopt the average values 
\begin{eqnarray}
\left< \lambda_{+} \right> &\equiv& 
(\mathrm{Tr} \hspace{0.5mm} {\cal M}^2)^{(0)}
\left[
1-\frac{(\det {\cal M}^2)^{(0)}}
{((\mathrm{Tr} \hspace{0.5mm} {\cal M}^2)^{(0)})^2 }
\right],
\label{eq:93} \\[3mm]
\left< \lambda_{-} \right> &\equiv& 
\frac{(\det {\cal M}^2)^{(0)}}{(\mathrm{Tr} \hspace{0.5mm} {\cal M}^2)^{(0)}},
\label{eq:94} \\[3mm]
\left< \frac{\lambda_{-}}{\lambda_{+}} \right> &\equiv& 
\frac{(\det {\cal M}^2)^{(0)}}
{((\mathrm{Tr} \hspace{0.5mm} {\cal M}^2)^{(0)})^2} 
\left[
1+\frac{(\det {\cal M}^2)^{(0)}}
{((\mathrm{Tr} \hspace{0.5mm} {\cal M}^2)^{(0)})^2 }
\right],  
\label{eq:95}
\end{eqnarray}
where `$\langle \hspace{1mm} \rangle$' denotes the average over 
many periods of the sine functions.  

Table I displays some examples of the values of the model parameters 
in the case in which an extremely large hierarchy of the eigenmasses of 
$f$ and $\tilde{f}$ can be realized, that is, 
the three relations (\ref{eq:90})$-$(\ref{eq:92}) can be satisfied.  
Here, 
from Eqs.\ (\ref{eq:28}), (\ref{eq:75}) and (45), 
we have approximately evaluated 
$\omega$, $\gamma$, and $t^{-1}$ 
by using 
$\omega \approx \alpha/(8\beta)$, 
$\gamma \approx 1/\sqrt{6\alpha \Upsilon^2}$, 
and 
$t^{-1} \approx (8\beta/\alpha) \Upsilon M$, 
respectively.  
Furthermore, 
Table II lists 
estimates of the quantities 
$\Xi_1 \equiv \tilde{\mathrm{Tr} \hspace{0.5mm} {\cal M}^2}/M^2$, 
$\Xi_2 \equiv \tilde{\det {\cal M}^2}/M^2$, 
and 
$\tilde{\Theta}$ in Eqs.\ (\ref{eq:90})$-$(\ref{eq:92}) 
and 
estimates of the effective values of the square of the eigenmasses 
$\left< \lambda_{+} \right>$ 
and 
$\left< \lambda_{-} \right>$, 
and the ratio 
$\left< \lambda_{-}/\lambda_{+} \right>$ in the cases displayed in Table I.  
Here we have estimated these quantities by using 
Eqs.\ (\ref{eq:77}), (\ref{eq:78}) and (\ref{eq:80})$-$(\ref{eq:82}).


\begin{table}[tbp]
\caption{
Examples of 
the values of the model parameters in the case that 
an extremely large hierarchy of 
the eigenmasses of $f$ and $\tilde{f}$ 
can be realized, that is, the three relations (\ref{eq:90})$-$(\ref{eq:92}) 
can be satisfied.  
Here we have approximately evaluated 
$\omega$, $\gamma$, and $t^{-1}$ 
by using 
$\omega \approx \alpha/(8\beta)$, 
$\gamma \approx 1/\sqrt{6\alpha \Upsilon^2}$, 
and 
$t^{-1} \approx (8\beta/\alpha) \Upsilon M$, respectively.
}
\begin{center}
\begin{tabular}
{ccccccc}
\hline
\hline
& $\epsilon_1$  
& $\epsilon_2$  
& $\Upsilon$ 
& $\omega$
& $\gamma$
& $t^{-1}/M$
\\[0mm]
\hline
(i)
& $4.1 \times 10^{-2}$
& $4.0 \times 10^{-2}$
& $3.5 \times 10^{1}$
& $3.1$
& $4.1 \times 10^{-2}$
& $1.1 \times 10^{1}$
\\[0mm]
(ii)
& $2.6 \times 10^{-2}$
& $2.5 \times 10^{-2}$
& $3.0 \times 10^{1}$
& $4.9$
& $6.0 \times 10^{-2}$
& $6.1$
\\[0mm]
(iii)
& $1.1 \times 10^{-2}$
& $1.0 \times 10^{-2}$
& $3.5 \times 10^{2}$
& $1.2 \times 10^{1}$
& $8.0 \times 10^{-3}$
& $2.9 \times 10^{1}$
\\[0mm]
(iv)
& $6.6 \times 10^{-3}$
& $6.5 \times 10^{-3}$
& $4.5 \times 10^{2}$
& $1.9 \times 10^{1}$
& $7.9 \times 10^{-3}$
& $2.4 \times 10^1$
\\[0mm]
(v)
& $3.1 \times 10^{-3}$
& $3.0 \times 10^{-3}$
& $1.5 \times 10^{4}$
& $4.1 \times 10^{1}$
& $3.5 \times 10^{-4}$
& $3.7 \times 10^2$
\\[0.5mm]
\hline
\end{tabular}
\end{center}
\label{table1}
\end{table}

\begin{table}[tbp]
\caption{
Estimates of the quantities 
in the three relations (\ref{eq:90})$-$(\ref{eq:92}) 
and of 
the effective values of the square of the eigenmasses of $f$ and $\tilde{f}$
and their ratio in the cases (i)$-$(v) displayed in Table I.  
Here we have defined 
$\Xi_1 \equiv \tilde{\mathrm{Tr} \hspace{0.5mm} {\cal M}^2}/M^2$ and 
$\Xi_2 \equiv \tilde{\det {\cal M}^2}/M^2$.  
Moreover, we have estimated these quantities by using 
Eqs.\ (\ref{eq:77}), (\ref{eq:78}) and (\ref{eq:80})$-$(\ref{eq:82}). 
}
\begin{center}
\begin{tabular}
{ccccccc}
\hline
\hline
& $\Xi_1$
& $\Xi_2$
& $\tilde{\Theta}$
& $\left< \lambda_{+} \right>/M^2$
& $\left< \lambda_{-} \right>/M^2$
& $\left< \lambda_{-}/\lambda_{+} \right>$
\\[0mm]
\hline
(i)
& $1.0 \times 10^{7}$
& $2.5 \times 10^{7}$
& $4.2 \times 10^{-5}$
& $1.0 \times 10^{7}$
& $5.4 \times 10^{1}$
& $5.4 \times 10^{-6}$
\\[0mm]
(ii)
& $1.0 \times 10^{6}$
& $5.0 \times 10^{6}$
& $3.0 \times 10^{-4}$
& $1.0 \times 10^{6}$
& $4.0 \times 10^{1}$
& $4.0 \times 10^{-5}$
\\[0mm]
(iii)
& $3.3 \times 10^{6}$
& $3.1 \times 10^{7}$
& $9.8 \times 10^{-4}$
& $3.4 \times 10^{6}$
& $4.1 \times 10^{2}$
& $1.2 \times 10^{-4}$
\\[0mm]
(iv)
& $8.2 \times 10^{7}$
& $2.1 \times 10^{9}$
& $2.7 \times 10^{-5}$
& $8.2 \times 10^{7}$
& $2.9 \times 10^{2}$
& $3.6 \times 10^{-6}$
\\[0mm]
(v)
& $4.1 \times 10^{9}$
& $5.8 \times 10^{11}$
& $3.8 \times 10^{-5}$
& $4.1 \times 10^{9}$
& $1.9 \times 10^{4}$
& $4.7 \times 10^{-6}$
\\[0.5mm]
\hline
\end{tabular}
\end{center}
\label{table2}
\end{table}


Figures 1, 2, and 3 depict 
estimates of 
$\Xi_1 = \tilde{\mathrm{Tr} \hspace{0.5mm} {\cal M}^2}/M^2$, 
$\Xi_2 = \tilde{\det {\cal M}^2}/M^2$, 
and 
$\tilde{\Theta}$, respectively.   
Moreover, Figs.\ 4 and 5 depict estimates of 
$\left< \lambda_{+} \right>/M^2$ and 
$\left< \lambda_{-} \right>/M^2$, 
and the ratio 
$\left< \lambda_{-}/\lambda_{+} \right>$.  
In all the figures, 
the solid lines represent the case $\Upsilon = 1.5 \times 10^4 $, 
and the dotted lines represent the case 
$\Upsilon = 3.5 \times 10^2 $.  
Moreover, the 
left panels correspond to the case 
$\epsilon_1 = \epsilon_2 + 1.0 \times 10^{-4},  
\epsilon_2 = 1.0 \times 10^{-3} + 1.0 \times 10^{-4}j \hspace{1mm} 
(0 \leq j \leq 89)$, where $j$ is an integer, 
while the right panels correspond to the case 
$\epsilon_1 = \epsilon_2 + 1.0 \times 10^{-3},  
\epsilon_2 = 1.0 \times 10^{-2} + 1.0 \times 10^{-3}j \hspace{1mm} 
(0 \leq j \leq 89)$.  
In Fig.\ 1, for all the cases, we have $\Xi_1 > 0$, and hence the relation 
(\ref{eq:90}) is satisfied.  
In Fig.\ 2, in the case $\Upsilon = 1.5 \times 10^4$, 
for $\epsilon_2 \leq 2.9 \times 10^{-3}$ and 
$\epsilon_2 \geq 4.5 \times 10^{-2}$, we have $\Xi_2 < 0$.  
Moreover, in the case $\Upsilon = 3.5 \times 10^2$, 
for $\epsilon_2 \leq 6.9 \times 10^{-3}$ and 
$\epsilon_2 \geq 4.5 \times 10^{-2}$, we have $\Xi_2 < 0$.  
Hence, in these cases, the relation (\ref{eq:91}) is not satisfied.  
In Fig.\ 3, 
in the case $\Upsilon = 1.5 \times 10^4$, 
for $\epsilon_2 \geq 4.5 \times 10^{-2}$, we have $\tilde{\Theta} < 0$.  
Moreover, in the case $\Upsilon = 3.5 \times 10^2$, 
for $\epsilon_2 \geq 4.6 \times 10^{-2}$, we have $\tilde{\Theta} < 0$.  
In the remaining cases, 
the relation 
$0 < \tilde{\Theta} \ll 1$ holds, and hence 
the relation (\ref{eq:92}) is satisfied.  
Consequently, from the above considerations, 
we see that 
all the relations (\ref{eq:90})$-$(\ref{eq:92}) can be 
satisfied; that is, an extremely large hierarchy of the eigenmasses of 
$f$ and $\tilde{f}$ can be realized 
in the case 
$\Upsilon = 1.5 \times 10^4$ for 
$\epsilon_1 = \epsilon_2 + 1.0 \times 10^{-4},  
\epsilon_2 = 1.0 \times 10^{-3} + 1.0 \times 10^{-4}j \hspace{1mm} 
(20 \leq j \leq 89)$, 
and 
$\epsilon_1 = \epsilon_2 + 1.0 \times 10^{-3},  
\epsilon_2 = 1.0 \times 10^{-2} + 1.0 \times 10^{-3}j \hspace{1mm} 
(0 \leq j \leq 34)$, 
and 
in the case 
$\Upsilon = 3.5 \times 10^2$ for 
$\epsilon_1 = \epsilon_2 + 1.0 \times 10^{-4},  
\epsilon_2 = 1.0 \times 10^{-3} + 1.0 \times 10^{-4}j \hspace{1mm} 
(60 \leq j \leq 89)$, 
and 
$\epsilon_1 = \epsilon_2 + 1.0 \times 10^{-3},  
\epsilon_2 = 1.0 \times 10^{-2} + 1.0 \times 10^{-3}j \hspace{1mm} 
(0 \leq j \leq 34)$.  
Furthermore, from Figs.\ 4 and 5 we see that in the above cases, 
both $\left< \lambda_{+} \right>/M^2$ and $\left< \lambda_{-} \right>/M^2$ 
are positive, and the ratio 
$\left< \lambda_{-}/\lambda_{+} \right>$ is much smaller than unity.  
From the above results, we see that in the case that 
the difference between the values of $\epsilon_1$ and $\epsilon_2$ is much 
smaller than $\epsilon_1$ and $\epsilon_2$ themselves, 
i.e., 
$|\epsilon_1-\epsilon_2| \ll O[\epsilon_i]$ [in other words, the 
two dilatons have an approximate $O(2)$ symmetric coupling], 
an extremely large hierarchy of the eigenmasses of 
$f$ and $\tilde{f}$ can be realized.  

Finally, we note that 
in all the above cases in which 
an extremely large hierarchy of the eigenmasses of 
$f$ and $\tilde{f}$ can be realized, 
the heavier eigenmass, $m_{+} \equiv \sqrt{\left< \lambda_{+} \right>}$, is 
comparable to or 
larger than the Hubble parameter in the inflationary stage, while 
the lighter eigenmass, $m_{-} \equiv \sqrt{\left< \lambda_{-} \right>}$, 
is much smaller than the Hubble parameter in the inflationary stage.  
For example, in the case (v) in Table I, we have 
$m_{+}/H \approx \sqrt{\left< \lambda_{+} \right>}/(\Upsilon M) = 4.3$ 
and 
$m_{-}/H \approx \sqrt{\left< \lambda_{-} \right>}/(\Upsilon M) = 9.2 \times 
10^{-3}$, where we have used Eq.\ (\ref{eq:74}).  
\if
For example, in the case (iii) in Table I, 
$m_{+}/H \approx \sqrt{\left< \lambda_{+} \right>}/(\Upsilon M) = 5.3$ 
and 
$m_{-}/H \approx \sqrt{\left< \lambda_{-} \right>}/(\Upsilon M) = 5.7 \times 
10^{-2}$, where we have used Eq.\ (\ref{eq:76}).  
\fi
Thus, it is expected that the scalar field with the lighter 
eigenmass, $m_{-}$, which is much smaller than 
the Hubble parameter in the inflationary stage, can 
be regarded as corresponding to the curvaton.  
\if
Thus it could be understood that the scalar field with 
the heavier eigenmass $m_{+}$ effectively corresponds to the inflaton, 
while the scalar field with the lighter eigenmass $m_{-}$, 
which is much smaller than 
the Hubble parameter in the inflationary stage, effectively corresponds to 
the curvaton.  
\fi

\subsection{Evolution of dilatons after inflation}
In this subsection, 
we discuss the evolution of dilatons after inflation, 
in particular, when and how the inflation ends and 
how reheating occurs in this model.  

As shown in the cases considered in Table I of the previous subsection, 
from this point, we mainly consider the following case 
$\varphi_{\mathrm{r}} = \sqrt{\varphi_1^2 + \varphi_2^2} 
\approx \gamma M$, where $\gamma (t) \ll 1$ during inflation.  
In this case, 
during the inflationary stage the dilatons $\varphi_i$ 
evolve as 
$\varphi_{\mathrm{r}} = Ct$ in the ansatz (\ref{eq:15}) around 
the origin of the potential 
$
V[\varphi_i] =V_{0} \cos \left( \varphi_{\mathrm{r}}/M \right) + \Lambda 
$ 
in Eq.\ (\ref{eq:3}).  
Although $\varphi_{\mathrm{r}}$ exists around 
the origin of the potential in the inflationary stage, 
after sufficient inflationary expansion,  
the amplitude of $\varphi_{\mathrm{r}}$ becomes large, and then 
$\varphi_{\mathrm{r}}$ approaches the minimum of the potential, i.e., 
$\varphi_{\mathrm{r}} = M\pi$, in which the value of the 
potential is $V = - V_{0} + \Lambda \approx 0$, where the last approximate 
equality follows from $V_{0} \approx \Lambda \approx M^4$.  
\if
(We here note that as the amplitude of $\varphi_{\mathrm{r}}$ increases, 
the approximation of replacing $V$ by $\Lambda$ used in \S 3 becomes invalid 
because the contribution of the part of cosine function in the dilaton 
potential $V$ in Eq.\ (\ref{eq:3}) increases.)  
\fi
As $\varphi_{\mathrm{r}}$ nears the 
potential minimum, the potential steepens, and the evolution of 
$\varphi_{\mathrm{r}}$ becomes more rapid; 
as $\varphi_{\mathrm{r}}$ approaches 
and then, inevitably, overshoots the minimum of its potential, 
it begins to 
oscillate about $\varphi_{\mathrm{r}} = M\pi$ on a time scale short 
compared to the Hubble time.  
Hence, after the oscillation epoch, $\varphi_{\mathrm{r}}$ does not 
evolve as $\varphi_{\mathrm{r}} = Ct$ but remains near the 
potential minimum, $\varphi_{\mathrm{r}} = M\pi$. 
An enormous vacuum energy of the dilatons then exists in the 
form of spatially coherent oscillations of the dilatons, corresponding 
to a condensate of zero-momentum $\varphi_i$ particles.  
Particle creations, or, equivalently, the decay of $\varphi_i$ particles 
into other, lighter fields to which it couples, will damp 
these oscillations.  Furthermore, as the decay products thermalize, the 
universe is reheated.  

\if
On the other hand, 
in another point of view, i.e., if we consider 
the evolution the quantities $f$ and $\tilde{f}$, 
we may interpret the evolutions of them as follows:  
As stated in the previous section, 
when we consider the eigenmasses of $f$ and $\tilde{f}$,  
the heavier eigenmass $m_{+} \equiv \sqrt{\left< \lambda_{+} \right>}$ is 
comparable to or 
larger than the Hubble parameter in the inflationary stage, while 
the lighter eigenmass $m_{-} \equiv \sqrt{\left< \lambda_{-} \right>}$ is much 
smaller than the Hubble parameter in the inflationary stage.  
Hence 
it is expected that 
the scalar field with the heavier 
eigenmass $m_{+}$ (comparable to or larger than $H$) can 
effectively correspond to the inflaton field, 
and that the scalar field with the lighter 
eigenmass $m_{-}$ ($\ll H$) can effectively correspond to the curvaton.  
After inflation, the inflaton with the heavier mass $m_{+}$ approaches its 
potential minimum and then starts oscillation 
with the mass $m_{+}$, and finally decays into radiation 
through the coupling to other lighter fields and 
then the Universe is reheated;  
on the other hand, the amplitude of the curvaton with the lighter 
eigenmass $m_{-}$ hardly change and hence the curvaton does not evolve 
during inflation.  
\fi

Finally, we estimate the number of \textit{e}-folds during inflation.  
This quantity is defined as \cite{Kolb}  
\begin{eqnarray}
N &\equiv& \int_{t_{\mathrm{i}}}^{t_{\mathrm{f}}} H dt  
\label{eq:4-1} \\[5mm]
&=& \omega \ln \left( \frac{t_{\mathrm{f}}}{t_{\mathrm{i}}} \right),
\label{eq:4-2}
\end{eqnarray}
where 
in deriving Eq.\ (\ref{eq:4-2}) 
we have used the relation $H=\omega/t$.  
Moreover, it follows from the considerations at the beginning of this 
subsection that 
$\varphi_{\mathrm{r}} (t_{\mathrm{i}}) \approx \gamma (t_{\mathrm{i}}) M$ 
and 
$\varphi_{\mathrm{r}} (t_{\mathrm{f}}) \approx M \pi$.  
On the other hand, from these relations and 
$\varphi_{\mathrm{r}} = Ct$, we obtain 
\begin{eqnarray}
t_{\mathrm{i}} &=& 
\frac{\varphi_{\mathrm{r}} (t_{\mathrm{i}})}{C} 
\approx \frac{\gamma (t_{\mathrm{i}})}{C} M, 
\label{eq:4-3} \\[5mm]
t_{\mathrm{f}} &=& 
\frac{\varphi_{\mathrm{r}} (t_{\mathrm{f}})}{C} 
\approx \frac{\pi}{C} M.
\label{eq:4-4}
\end{eqnarray}
Substituting Eqs.\ (\ref{eq:4-3}) and (\ref{eq:4-4}) into 
Eq.\ (\ref{eq:4-2}), we obtain 
\begin{eqnarray}
N \approx \omega \ln \left[ \frac{\pi}{\gamma(t_{\mathrm{i}}) } \right].
\label{eq:4-5}
\end{eqnarray}
For example, in the case (v) of Table I, 
it follows from the values 
$\omega = 4.1 \times 10^1$ and $\gamma(t_{\mathrm{i}}) \simeq 
3.5 \times 10^{-4}$ and Eq.\ (\ref{eq:4-5}) that 
$N = 3.7 \times 10^{2} > 70$.  
Thus, the expansion of the universe 
during the epoch of power-law inflation realized in this model is large 
enough to solve both the horizon and flatness problems.

\section{Curvaton scenario}
In the previous section, we showed that 
when two dilatons have an approximate $O(2)$ symmetric coupling, 
a scalar field corresponding to the curvaton exists in this theory.  
Considering this point, 
in this section, we investigate the case in which the curvaton scenario 
can be realized in this model.  
In particular, we consider a simple version of the curvaton scenario 
following the outline given in Refs.~\citen{Lyth} and \citen{Lyth2}.  

\subsection{Curvature perturbation} 
To begin with, we consider linear metric perturbations.  
(For a review of the theory of cosmological perturbations, see 
Ref.~\citen{C.P.T.}.)  
\if
To begin with, we consider linear metric perturbations incorporated into 
the spatially flat FRW background in the longitudinal gauge, 
\begin{eqnarray}
 {ds}^2 = (1+2\Phi){dt}^2 - a^2(t)(1-2\psi)d{\Vec{x}}^2,
\label{eq:96}
\end{eqnarray} 
where $\Phi$ and $\psi$ are gauge-invariant variables related to 
Bardeen's $\Phi_A$ and $\Phi_H$ as $\Phi = \Phi_A$ and $\psi = -\Phi_H$ 
\cite{Bardeen} 
(For a review of the theory of cosmological perturbations, see 
Ref.~\citen{C.P.T.}.)  
Moreover, in the case of no stress, we have $\Phi=\psi$.  
\fi
\if
We are mainly interested in 
the spatial curvature perturbation on comoving 
scales much larger than the Hubble scale, which are referred to as 
super-horizon scales.  
In order to define this quantity, it is necessary for us to specify 
a foliation of spacetime into space-like hypersurfaces (i.e., a slicing).  
The most convenient choices are the slicing of uniform energy density 
and the slicing orthogonal to comoving worldlines, which are practically 
the same on super-horizon scales.  
\fi
\if
The spatial curvature perturbation is of interest only on comoving 
scales much bigger than the Hubble scale, which is 
called super-horizon scales.  
To define it one has to specify a foliation of spacetime into 
spacelike hypersurfaces (slicing), and the most convenient choice
is the slicing of uniform energy density or the slicing 
orthogonal to comoving worldlines, which is practically the same
on super-horizon scales.  
\fi
The curvature perturbation on uniform-density hypersurfaces 
$\zeta$ 
is related to the gauge-dependent curvature perturbation $\psi$ 
on a generic slicing and the energy density perturbation $\delta\rho$ 
in that gauge through the following gauge-invariant formula:\cite{Wands}  
\begin{eqnarray}
\zeta
&=& -\psi - H\frac{\delta\rho}{\dot{\rho}}.  
\label{eq:97}
\end{eqnarray} 
Furthermore, the time dependence of $\zeta$ on super-horizon scales 
is given by \cite{Wands}
\begin{eqnarray}
\dot\zeta = -\frac{H}{\rho + P} \delta P_{\mathrm{nad}}, 
\label{eq:98}
\end{eqnarray} 
where 
$P$ is the pressure and 
$\delta P_{\mathrm{nad}} = \delta P - c_s^2\delta\rho$ 
is a non-adiabatic pressure perturbation.  Here, 
$c_s^2 \equiv \dot{P}/\dot\rho$ is the adiabatic speed of sound.  
In the usual inflation scenario, in which $\zeta$ is generated during 
inflation through the perturbation of a single-component inflaton field, 
it becomes practically time independent soon after horizon exit and 
remains so until the approach of horizon entry.  
By contrast, in the curvaton scenario, 
the curvature perturbation generated in the inflationary stage is 
negligible, 
and it is generated later through a non-adiabatic 
pressure perturbation associated with the curvaton perturbation.  
Finally, we note that from this point, we mainly consider the  
curvature perturbation on super-horizon scales.  

\subsection{Curvaton field} 
Next, we investigate the case in which the curvaton scenario 
can be realized in this model.  
As discussed in \S 3, power-law inflation in which 
the power-law exponent $\omega$ is much larger than unity 
can be realized in this model, and hence 
the Hubble parameter $H$ can be slowly varying:  
\begin{eqnarray}
\epsilon_H \equiv - \frac{\dot H}{H^2} = \frac{1}{\omega} \ll 1.  
\label{eq:99}
\end{eqnarray} 
The requirement for power-law inflation in the curvaton scenario is 
that 
the relation (\ref{eq:99}) be satisfied; that is, 
the power-law exponent $\omega$ should be much larger than unity.  
Furthermore, 
as noted in the previous section, 
in the framework of this theory, 
the scalar field with 
the lighter eigenmass, $m_{-}$, which is much smaller than 
the Hubble parameter in the inflationary stage, can 
be regarded as corresponding to the curvaton.  
From this point, we therefore regard this field as the curvaton 
$\sigma$ with mass $m_{\sigma} \equiv m_{-}$.  

\if
Furthermore, 
as noted in the previous section, 
in the framework of this theory  
the scalar field with 
the heavier eigenmass $m_{+}$ could 
effectively correspond to the inflaton 
and 
the scalar field with 
the lighter eigenmass $m_{-}$, which is much smaller than 
the Hubble parameter in the inflationary stage, could 
effectively correspond to the curvaton.  
From now on we therefore regard the former field as 
the inflaton and the latter field as the curvaton 
$\sigma$ with the mass $m_{\sigma} \equiv m_{-}$.  
\fi

Here we assume a spatially flat 
FRW spacetime with the metric (8), and we take the 
Lagrangian of the curvaton $\sigma$ to be given by 
\begin{eqnarray}
{\cal L}_\sigma &=& \frac{1}{2} (\partial \sigma)^2
- U[\sigma], 
\label{eq:100} 
\end{eqnarray} 
where $U[\sigma]$ is the potential of the curvaton.  
\if
The potential $U$ depends of course on all scalar fields
but we exhibit only the dependence on $\sigma$ which 
is assumed to have no significant coupling to the fields driving
inflation.  
\fi
Furthermore, we assume that the curvature perturbations is 
negligible during inflation.  
After the smallest cosmological scale leaves the horizon, the
curvature perturbation remains negligible until after the curvaton
starts to oscillate.  
Consequently, the curvaton evolves in unperturbed spacetime.  
In order to follow this evolution, we assume that the curvaton has 
no significant coupling to other fields, or, to be more precise, 
that the effect of any coupling can be integrated out to give a 
possibly time-dependent effective potential
$U$.  
\if
Furthermore, we assume that the curvature perturbations is 
negligible during inflation.  
After the smallest cosmological scale leaves the horizon, the
curvature perturbation remains negligible until after the curvaton
starts to oscillate.  
As a result, the curvaton 
evolves in unperturbed spacetime.  To follow the evolution it is
assumed that the curvaton has no significant coupling to other
fields or, to be more precise, that the effect of any coupling can be
integrated out to give a possibly time-dependent effective potential
$U$.  
\fi

We can now split the curvaton field as 
\begin{eqnarray}
\sigma(\Vec{x},t) = \sigma_0(t) + \delta \sigma(\Vec{x},t), 
\label{eq:101}
\end{eqnarray} 
where $\sigma_0$ is the classical field, while $\delta \sigma(\Vec{x},t)$ 
represents the quantum fluctuations around $\sigma_0$.  

The unperturbed curvaton field satisfies 
\begin{eqnarray}
\ddot{ \sigma }+ 3 H \dot{\sigma} + U_\sigma = 0, 
\label{eq:102}
\end{eqnarray} 
where the subscript $\sigma$ 
denotes partial differentiation with respect to $\sigma$.  

Expanding the perturbation of the curvaton $\delta \sigma$ 
in Fourier modes, 
\begin{eqnarray} 
 \delta \sigma(\Vec{x},t) = 
 \int \frac{d^3 k}{{(2\pi)}^{3/2}}
              e^{i \Vecs{k} \cdot \Vecs{x} } \delta \sigma_{\Vecs{k}}(t),  
\label{eq:103} 
\end{eqnarray}
where $\Vec{k}$ is the comoving wave number, and $k$ denotes 
its amplitude, $|\Vec{k}|$, 
we can write the equation for the perturbation as 
\begin{eqnarray} 
\ddot{\delta\sigma_{\Vecs{k}}} + 
3H\dot{\delta\sigma_{\Vecs{k}}} + 
\left( \frac{k^2}{a^2} + U_{\sigma\sigma} \right)
\delta\sigma_{\Vecs{k}} = 0.  
\label{eq:104} 
\end{eqnarray}
Here we assume that the curvaton potential 
is sufficiently flat during inflation, i.e., 
\begin{eqnarray} 
|U_{\sigma\sigma}|\ll H^2, 
\label{eq:105} 
\end{eqnarray}
and that on cosmological scales each Fourier component is in the
vacuum state well before horizon exit.  
Then the vacuum fluctuation causes a classical perturbation 
$\delta \sigma_{\Vecs{k}}$ well after
horizon exit, which satisfies Eq.\ (\ref{eq:104}), with a negligible 
gradient term, 
\begin{eqnarray} 
\ddot{\delta\sigma_{\Vecs{k}}} + 
3H\dot{\delta\sigma_{\Vecs{k}}} + 
U_{\sigma\sigma} 
\delta\sigma_{\Vecs{k}} = 0.  
\label{eq:106} 
\end{eqnarray}
\if
Here we assume that the curvaton potential 
is sufficiently flat during inflation, 
\begin{eqnarray} 
|U_{\sigma\sigma}|\ll H^2, 
\label{eq:105} 
\end{eqnarray}
and that on cosmological scales each Fourier component is in the
vacuum state well before horizon exit.  
The vacuum fluctuation then causes a classical perturbation 
$\delta \sigma_{\Vecs{k}}$ well after
horizon exit, which satisfies Eq.\ (\ref{eq:104}) with negligible
gradient term, 
\begin{eqnarray} 
\ddot{\delta\sigma_{\Vecs{k}}} + 
3H\dot{\delta\sigma_{\Vecs{k}}} + 
U_{\sigma\sigma} 
\delta\sigma_{\Vecs{k}} = 0.  
\label{eq:106} 
\end{eqnarray}
\fi
In the limit that the relation (\ref{eq:105}) 
is satisfied, the power 
spectrum of the perturbation of the 
curvaton on super-horizon scales is given by 
\begin{eqnarray} 
{\cal P}_{\delta \sigma}^{1/2} = \left( \frac{H_*}{2\pi} \right) 
\left( \frac{k}{a_*H_*} \right)^{\left( n_{\delta\sigma} - 1 \right)/2}, 
\label{eq:107} 
\end{eqnarray}
where 
the star denotes the epoch of horizon exit, $k=a_*H_*$. 
The spectral index 
specifying the slight scale dependence
is given by \cite{Lyth,Lyth2} 
\begin{eqnarray} 
n_{\delta\sigma} - 1 \equiv \frac{d \ln {\cal P}_{\delta\sigma}}{d \ln k} 
= 2 \eta_{\sigma\sigma} -2\epsilon_H, 
\label{eq:108} 
\end{eqnarray}
where $\eta_{\sigma\sigma}\equiv U_{\sigma\sigma}/(3H^2)$.  
\if
In the limit where the relation (\ref{eq:105})
is very well satisfied, the power 
spectrum of the perturbation of the 
curvaton on super-horizon scales is given by 
\begin{eqnarray} 
{\cal P}_{\delta \sigma}^{1/2} \approx \frac{H_*}{2\pi},
\label{eq:107} 
\end{eqnarray}
where 
the star denotes the epoch of horizon exit, $k=a_*H_*$. 
Here ${\cal P}_{\delta \sigma}$ is almost flat.  
To be more precise, the spectral index
specifying the slight scale-dependence
is given by \cite{Lyth,Lyth2} 
\begin{eqnarray} 
n_{\delta\sigma} - 1 \equiv \frac{d \ln {\cal P}_{\delta\sigma}}{d \ln k} 
= 2 \eta_{\sigma\sigma} -2\epsilon_H, 
\label{eq:108} 
\end{eqnarray}
where $\eta_{\sigma\sigma}\equiv U_{\sigma\sigma}/(3H^2)$.  
\fi

\if
Here we assume that the curvaton potential $U[\sigma]$ is 
given by the following simple double-well potential having 
a global minimum at $\sigma=v$:  
\begin{eqnarray} 
 U[\sigma] = \frac{\lambda}{4} \left( \sigma^2 - v^2 \right)^2, 
\label{eq:111} 
\end{eqnarray}
where $\lambda$ is a dimensionless constant. 
In the case $\sigma/v \ll 1$, 
the potential can be approximated as 
\begin{eqnarray}
U[\sigma] \simeq U_0 - \frac{1}{2} {m_{\sigma}}^2 \sigma^2, 
\label{eq:112}
\end{eqnarray} 
with $U_0=(\lambda/4)v^4$ and ${m_{\sigma}}^2=\lambda v^2$.  
\fi

Here we assume that the curvaton potential $U[\sigma]$ is 
given by the following quadratic form 
\begin{eqnarray} 
 U[\sigma] = \frac{1}{2} {m_{\sigma}}^2 \left( \sigma - v \right)^2, 
\label{eq:109} 
\end{eqnarray}
which has a global minimum at $\sigma=v$.  
The reason 
we have shifted the global minimum of the potential from the origin to 
$\sigma=v$ is as follows.  
In this model, as seen in \S 2 and \S 3,  
the dilatons $\varphi_i$ initially exist near the origin, 
along with the potential $V[\varphi_i]$ in Eq.\ (\ref{eq:3}), 
and then rapidly evolve in the inflationary stage.  
Hence, the initial field amplitude of 
the field variables $f$ and $\tilde{f}$ is also small, and then 
increases in proportion to $t^2$ in the inflationary stage.  
Because 
the curvaton considered here corresponds to the scalar field 
with the lighter eigenmass, $m_-$, 
it is conjectured that the initial amplitude of the curvaton is small 
and then increases.  
In this case, it follows from the discussion in \S 4.3 that  
$|U_{\sigma\sigma}|/H^2 \simeq {m_\sigma}^2/H^2 \ll 1$, 
and hence the relation (\ref{eq:105}) is satisfied.  
The spectral index of the primordial curvature perturbation 
is estimated as $0.99 \pm 0.04$ by using the first year WMAP data 
only \cite{Spergel}, where the uncertainty represents 
the 68\% confidence interval.  
For example, in the case (v) in Table I appearing in \S 4.3, 
$\epsilon_H = 1/\omega = 2.4 \times 10^{-2}$ and  
$\eta_{\sigma\sigma} \simeq {m_\sigma}^2/(3H^2) = 2.8 \times 10^{-5}$.  
Hence, it follows from Eq.\ (\ref{eq:108}) that 
$n_{\delta\sigma} = 1 + 
2 \eta_{\sigma\sigma} - 2 \epsilon_H = 0.98$, 
which is consistent with the above observational constraint of WMAP.  

The curvaton remains over damped until 
the Hubble parameter falls below the curvaton mass $m_\sigma$.  
The curvaton then starts to oscillate about its vacuum 
value, taken to be $\sigma=v$ in the curvaton potential $U[\sigma]$ in 
Eq.\ (\ref{eq:109}).  
\if
The curvaton remains over-damped until 
the Hubble parameter falls below the curvaton mass $m_\sigma$.  
The curvaton will then start to oscillate about its vacuum
value taken to be $\sigma=v$ in the curvaton potential $U[\sigma]$ in 
Eq.\ (\ref{eq:109}).  
\fi
Here we suppose that the oscillation 
starts during the radiation-dominated era after 
the reheating epoch following inflation.  
It follows from Eq.\ (\ref{eq:107}) that 
the spectrum of the fractional field perturbation at this stage is 
given by 
\begin{eqnarray} 
 {\cal P}^{1/2}_{\delta\sigma/\bar{\sigma}} 
= \frac {1}{2\pi} \frac{H_*}{{\bar{\sigma}}_*}, 
\label{eq:110} 
\end{eqnarray}
where we have defined $\bar{\sigma} \equiv |\sigma - v|$.  
\if
It follows from the Lagrangian of the curvaton in Eq.\ (\ref{eq:102}) 
that the energy density of the curvaton for the zero-mode is given 
by 
$
\sigma(\Vec{x},t)
\fi
Moreover, the energy density in the oscillating field is 
given by \cite{Lyth2} 
\begin{eqnarray} 
\rho_\sigma(\Vec{x},t) =
m_\sigma^2 \tilde{\bar{\sigma}}^2 (\Vec{x},t), 
\label{eq:111} 
\end{eqnarray}
where $\tilde{\bar{\sigma}}(\Vec{x},t)$ is the amplitude of the oscillation.
The perturbation in $\rho_\sigma$ depends on the curvaton field
perturbation through both a linear and a quadratic term.  
Assuming that the linear term dominates, however, we obtain 
\begin{eqnarray} 
\frac{\delta\rho_\sigma}{\rho_\sigma} \approx 
 2 \left( \frac{\delta\sigma}{\bar{\sigma}} \right)_*, 
\label{eq:112} 
\end{eqnarray}
where we have used $\delta \sigma = \delta \bar{\sigma}$.

\subsection{Generation of the curvature perturbation} 
In the previous subsection, we considered 
the epoch just after the Hubble parameter 
falls below the curvaton mass, and the curvaton oscillation starts.  
\if
So far we have reached the epoch just after the Hubble parameter 
falls below the curvaton mass, and the curvaton oscillation starts.  
\fi
Once the curvaton starts to oscillate, the energy density 
becomes a mixture of that of the curvaton and radiation.  
At this point, the pressure perturbation corresponding to this mixture 
becomes non-adiabatic, and hence, according to Eq.\ (\ref{eq:98}), 
the generation of the curvature perturbation begins.  
It ends when the pressure perturbation again becomes adiabatic, 
which is during 
the epoch of curvaton matter domination, or the epoch of 
curvaton decay, whichever is earlier.  
Moreover, 
at the stage at which the curvaton oscillation starts, 
it is assumed that the dominant portion of the energy density 
comes from the radiation.  
The curvaton, however, is assumed to be fairly 
long-lived, while decaying before nucleosynthesis.  
As long as the decay rate of the curvaton $\Gamma_\sigma$ 
is negligible, i.e., 
$\Gamma_\sigma \ll H$, we obtain 
$\rho_\sigma \propto a^{-3}(t)$ and 
$\rho_{\mathrm{r}} \propto a^{-4}(t)$, leading to
$\rho_\sigma/\rho_{\mathrm{r}} \propto a(t)$, where 
$\rho_\sigma$ is the energy density of the curvaton $\sigma$, and 
$\rho_{\mathrm{r}}$ is that of the radiation.  
It is this increase in the relative 
curvaton energy density which generates the curvature perturbation.  
\if
The curvaton, though, is supposed to be fairly 
long-lived, while decaying comfortably before nucleosynthesis.  
So long as the decay-rate of the curvaton $\Gamma_\sigma$ 
is negligible, i.e., 
$\Gamma_\sigma \ll H$, we obtain 
$\rho_\sigma \propto a^{-3}(t)$ and 
$\rho_{\mathrm{r}} \propto a^{-4}(t)$, leading to
$\rho_\sigma/\rho_{\mathrm{r}} \propto a(t)$, where 
$\rho_\sigma$ is the energy density of the curvaton $\sigma$ and 
$\rho_{\mathrm{r}}$ is that of the radiation.  
It is this increase in the relative 
curvaton energy density which generates the curvature perturbation.  
\fi

In order to analyze the generation of the curvature perturbation, 
it is convenient to separately consider the curvature 
perturbations $\zeta_{\mathrm{r}}$ and $\zeta_{\sigma}$ on, 
respectively, slices of uniform radiation density and 
curvaton density.  
These curvature perturbations on super-horizon scales 
are separately conserved \cite{Wands}, 
as the radiation and the curvaton are perfectly non-interacting fluids.  
The curvature perturbations are given by \cite{Wands}
\begin{eqnarray}
\zeta &=& -H\frac{\delta\rho}{\dot\rho} = 
-H\frac{\delta\rho_{\mathrm{r}} + \delta\rho_{\sigma}}
       {\dot{\rho_{\mathrm{r}}} + \dot{\rho_{\sigma}}}, 
\label{eq:113} \\[5mm] 
\zeta_{\mathrm{r}}  &=&  -H \frac{\delta\rho_{\mathrm{r}}}
{\dot\rho_{\mathrm{r}}}
=\frac{1}{4} \frac{\delta\rho_{\mathrm{r}}}{\rho_{\mathrm{r}}}, 
\label{eq:114} \\[5mm] 
\zeta_\sigma &=&  -H\frac{\delta\rho_\sigma}{\dot\rho_\sigma}
=\frac{1}{3} \frac{\delta\rho_\sigma}{\rho_\sigma},
\label{eq:115} 
\end{eqnarray} 
where the density perturbations are defined on the flat slicing of spacetime
($\psi=0$).  
It follows from Eqs.\ (\ref{eq:113})$-$(\ref{eq:115}) that 
the total curvature perturbation $\zeta$ can be written as 
\begin{eqnarray}
\zeta = (1-r) \zeta_{\mathrm{r}} + r \zeta_\sigma,  
\label{eq:116} 
\end{eqnarray} 
where the relative contribution of the curvaton to the total
curvature is given by 
\begin{eqnarray}
r = \frac{3\rho_\sigma}{4\rho_{\mathrm{r}} + 3\rho_\sigma}.  
\label{eq:117} 
\end{eqnarray} 
As is usually done in the curvaton scenario,   
here we assume that 
the curvature perturbation in 
the radiation produced at the end of inflation is negligible, i.e., 
$\zeta_{\mathrm{r}} \approx 0$.  
Furthermore, we assume that the curvaton decays instantaneously 
at the time that $H = H_{\mathrm{dec}} = \Gamma_\sigma$, where 
$H_{\mathrm{dec}}$ is the Hubble parameter in the decay epoch.  
In this case, it follows 
from Eq.\ (\ref{eq:116}) 
that the curvature perturbation during 
the decay epoch is given by 
\begin{eqnarray}
\zeta &\approx&  r_{\mathrm{dec}} \zeta_\sigma 
\label{eq:118} \\[5mm] 
&\approx& \frac{2}{3} r_{\mathrm{dec}} 
\left( \frac{\delta\sigma}{\bar{\sigma}} \right)_*,
\label{eq:119} 
\end{eqnarray} 
where the approximate equality in Eq.\ (\ref{eq:119}) 
follows from Eqs.\ (\ref{eq:112}) and (\ref{eq:115}).  
Here, 
$r_{\mathrm{dec}}$ is the relative contribution of the curvaton 
to the total curvature during the decay epoch, i.e., when 
$H = H_{\mathrm{dec}}
$.  
Consequently, it follows from Eqs.\ (\ref{eq:110}) and (\ref{eq:119}) 
that
the spectrum of the curvature perturbation in the curvaton scenario 
is given by  
\begin{eqnarray}
{{\cal P}_\zeta}^{1/2} &\approx& \frac{2}{3} r_{\mathrm{dec}}
{\cal P}^{1/2}_{\delta\sigma/\bar{\sigma}} 
\label{eq:120} \\[5mm]
&\approx&
\frac{r_{\mathrm{dec}}}{3\pi} 
\frac{H_*}{{\bar{\sigma}}_*}.  
\label{eq:121}
\end{eqnarray} 
The amplitude of the curvature perturbation 
required by the WMAP measurement \cite{Spergel} 
is estimated as 
\begin{eqnarray} 
{\cal P}_\zeta^{1/2} ({\mathrm{WMAP}}) \simeq 
5.2 \times 10^{-5}.  
\label{eq:122} 
\end{eqnarray} 
In the case that the curvaton dominates 
the energy density before it decays 
(that is, $r_{\mathrm{dec}} = 1$), 
it follows from Eq.\ (\ref{eq:120}) that 
the result in Eq.\ (\ref{eq:122}) 
implies the following 
amplitude of the spectrum of the fractional field perturbation of the 
curvaton: 
\begin{eqnarray} 
 {\cal P}^{1/2}_{\delta\sigma/\bar{\sigma}} \simeq 7.7 \times 10^{-5}.  
\label{eq:123} 
\end{eqnarray} 
Thus, in the case that the curvaton dominates 
the energy density before it decays, 
if the spectrum amplitude of 
the fractional field perturbation of the curvaton is approximately 
$7.7 \times 10^{-5}$ 
(i.e., 
it follows 
from Eq.\ (\ref{eq:110}) 
that 
$H_*/{\bar{\sigma}}_* \simeq 4.8 \times 10^{-4}$), 
then 
the curvature perturbation with 
the amplitude suggested by observations obtained from 
WMAP can be generated.  
\if
Finally we argue the possibility of the realization of the above 
relation $H_*/{\bar{\sigma}}_* \simeq 4.8 \times 10^{-4}$
\fi

Finally, we note that 
the arguments given in this section employ analysis in 
the Einstein frame.  It is not obvious whether the same property 
exists in the Jordan frame, which is the conformal frame considered in 
this model.  
It is shown in Appendix C, however, 
that 
the power spectrum of the curvature perturbation in the Einstein frame 
exactly coincides with that in the Jordan frame.  
\if
\footnote{
In Ref.~\citen{Tsujikawa1} the comoving curvature perturbation is 
introduced and then it is shown that 
the power spectrum of it in the Einstein frame exactly coincide with those 
in the Jordan frame.  On the other hand, 
in this section we consider 
the curvature perturbation on uniform-density hypersurfaces $\zeta$.  
The comoving curvature perturbation and 
the curvature perturbation on uniform-density hypersurfaces $\zeta$, 
however, are equal on superhorizon scales.\cite{Riotto1}  
} 
(The curvature perturbation in the Jordan frame is considered in 
detail in Ref.~\citen{Hwang1}).  
\fi
Thus, the results of the above arguments obtained from 
analysis in the Einstein frame, are identical to 
those in the Jordan frame.


\section{Conclusion}
In the present paper, we have shown that 
the curvaton scenario is realized 
in a theory 
with two dilatons coupled to the scalar curvature \cite{Yoshimura}.  
This theory has been considered in order to 
simultaneously solve the hierarchy problem between gravity and 
particle physics mass scales and the small cosmological constant 
or the dark energy problem.  
We have found that 
when both coupling constants 
between two dilatons and the scalar curvature 
are much smaller than unity, 
power-law inflation 
in which the power-law exponent is much larger than unity 
is realized in this model, and that 
when the difference between the values of 
the two coupling constants 
is much smaller than these values themselves 
[in other words, 
the two dilatons have an approximate $O(2)$ symmetric coupling], 
there exists 
a scalar field corresponding to the curvaton 
whose mass is much smaller than the Hubble parameter in the 
inflationary stage in this theory.  
This is realized 
without introducing any other scalar field that may 
play the role of the curvaton.  
Furthermore, we have investigated the simple version of the curvaton scenario 
in which the curvaton has a quadratic potential, 
and we demonstrated that 
a curvature perturbation with a sufficiently large amplitude and a nearly 
scale-invariant spectrum suggested by observations obtained 
from WMAP can be generated.  

\if
In the present paper we have discussed that 
the curvaton scenario could hold true in a theory 
with two dilatons coupled to the scalar curvature \cite{Yoshimura}.  
As a result, we have found that 
in the case the two dilatons have approximately O(2) symmetry 
there could be a scalar field corresponding to the curvaton 
in the framework of this theory 
without introducing any other scalar fields playing a role of the curvaton 
and thus the curvaton scenario could be realized.  
Furthermore, we have discussed the simple version of the curvaton scenario 
in which the curvaton has the quadratic potential.  
As a result, 
the curvature perturbation with the enough amplitude and nearly 
scale-invariant spectrum suggested by the observational results obtained 
from WMAP could be generated.  
\fi

\if
The serious problem of this model is that 
it is exceedingly difficult for the present model to be 
compatible with the scenario proposed in Ref.~\citen{Yoshimura}, 
which simultaneously solves the above two hierarchy problems.  
This reason is as follows:  In the present model,  
\fi

Finally, we note that 
in the present model, 
$\varphi_{\mathrm{r}} = 
\sqrt{\varphi_1^2 + \varphi_2^2} 
\approx \gamma (t) M$ exists around 
the origin of the potential $V$ 
in Eq.\ (\ref{eq:3}) 
in the inflationary stage; i.e., 
the case $\gamma (t) \ll 1$ during inflation is considered.  
After sufficient inflationary expansion,  
the amplitude of $\varphi_{\mathrm{r}}$ becomes large, and then 
$\varphi_{\mathrm{r}}$ approaches the first minimum of the potential.  
After the oscillation epoch, $\varphi_{\mathrm{r}}$ 
remains around the first potential minimum.  
Finally, the dilatons decay into radiation through coupling to other 
lighter fields, and then the universe is reheated.  
By contrast, 
in the scenario proposed in Ref.~\citen{Yoshimura}, 
the following case is considered: In the inflationary stage, 
$\varphi_{\mathrm{r}}$ goes over many local maxima of the potential and 
then changes for many periods; i.e., 
there exist the relations 
$\gamma (t) \gtrsim 1$ and 
$\gamma (t_{\mathrm{f}}) - \gamma (t_{\mathrm{i}}) \gg 1$, 
where $t_{\mathrm{f}}$ and $t_{\mathrm{i}}$ are the time at the end and 
beginning of inflation, respectively.  
Thus, the situation for the evolution of $\varphi_{\mathrm{r}}$ 
during inflation considered in the present model is different from 
that in the scenario proposed in Ref.~\citen{Yoshimura}.  
\if
This makes it difficult that 
both the present model and the scenario proposed in Ref.~\citen{Yoshimura} 
stand together. 
\fi

\if
the situation of the evolution of $\varphi_{\mathrm{r}}$ 
during inflation in the present model is different from that in the scenario 
in Ref.~\citen{Yoshimura}.  
The present model 
\fi

\section*{Acknowledgements}
The work of K.B.\ was partially supported by 
the Monbukagakusho 21st century COE Program 
``Towards a New Basic Science; Depth and Synthesis" 
and was also supported by a Grant-in-Aid provided by the 
Japan Society for the Promotion of Science.  
\if
The work of M.Y. was partially supported by the JSPS Grant-in-Aid for 
Scientific Research No.. 
\fi


\appendix
\section{Eigenmasses of $f$ and $\tilde{f}$}
In this appendix, 
following the discussion in \S 4, 
we first derive the fluctuation equation 
in terms of $f=\epsilon_1 \varphi_1^2 + \epsilon_2 \varphi_2^2$ and 
$\tilde{f}=\epsilon_1 \varphi_1^2 - \epsilon_2 \varphi_2^2$ 
in order to derive these mass values in the inflationary stage.  
Next, we investigate these eigenmasses 
in order to show that an extremely large hierarchy of 
the mass values of $f$ and $\tilde{f}$ in the inflationary 
stage can be realized.


\subsection{Background quantities} 
To begin with, 
as the preliminary stage for investigating 
the fluctuation equation in terms of $f$ and $\tilde{f}$, 
we consider the terms on the right-hand side of 
Eqs.\ (\ref{eq:10}) and (\ref{eq:11}).  
From $f = \epsilon_1 \varphi_1^2 + \epsilon_2 \varphi_2^2$ 
and ${\varphi_{\mathrm{r}}}^2 = \varphi_1^2 + \varphi_2^2$, we find 
\begin{eqnarray}
\varphi_1^2 = \frac{f - \epsilon_2 {\varphi_{\mathrm{r}}}^2}
{\epsilon_1 - \epsilon_2}, \hspace{5mm} 
\varphi_2^2 = \frac{-f + \epsilon_1 {\varphi_{\mathrm{r}}}^2}
{\epsilon_1 - \epsilon_2}.  
\label{eq:29}
\end{eqnarray}
Then, 
using Eq.\ (\ref{eq:29}), we obtain 
\begin{eqnarray}
\frac{\epsilon_1^2 \varphi_1^2 + \epsilon_2^2 \varphi_2^2}{f} 
&=&
(\epsilon_1 + \epsilon_2) - 
\epsilon_1 \epsilon_2 \frac{{\varphi_{\mathrm{r}}}^2}{f}, 
\label{eq:30} \\[5mm]
\frac{\varphi_1 \varphi_2}{f} 
&=&
\pm \frac{1}{\epsilon_1 - \epsilon_2} 
\sqrt{
\left(-1+ \epsilon_1 \frac{{\varphi_{\mathrm{r}}}^2}{f} \right) 
\left(1 - \epsilon_2 \frac{{\varphi_{\mathrm{r}}}^2}{f} \right)
}, 
\label{eq:31} \\[5mm]
\tilde{f} = \epsilon_1 \varphi_1^2 - \epsilon_2 \varphi_2^2
&=&
\frac{f}{\epsilon_1 - \epsilon_2}
\left[ (\epsilon_1 + \epsilon_2) - 
2 \epsilon_1 \epsilon_2 \frac{{\varphi_{\mathrm{r}}}^2}{f} \right]. 
\label{eq:32}
\end{eqnarray}
Furthermore, from 
$\dot{f} = 2 \left( \epsilon_1 \varphi_1 \dot{\varphi_1} 
+ \epsilon_2 \varphi_2 \dot{\varphi_2} \right)$ 
and 
$L = \varphi_1 \dot{\varphi_2}- \varphi_2 \dot{\varphi_1}$, we find 
\begin{eqnarray}
\dot{\varphi_1} = 
\frac{\varphi_1 \dot{f} - 2 \epsilon_2 \varphi_2 L}{2f}, \hspace{5mm} 
\dot{\varphi_2} = 
\frac{\varphi_2 \dot{f} + 2 \epsilon_1 \varphi_1 L}{2f}.  
\label{eq:33}
\end{eqnarray}
Using Eq.\ (\ref{eq:33}), we obtain 
\begin{eqnarray}
&& \hspace{-10mm}
\epsilon_1 \dot{\varphi_1}^2 + \epsilon_2 \dot{\varphi_2}^2
= 
\frac{1}{4f} \left( \dot{f}^2 + 4 \epsilon_1 \epsilon_2 L^2 \right),
\label{eq:34} \\[5mm]
&& \hspace{-10mm}
\epsilon_1 \dot{\varphi_1}^2 - \epsilon_2 \dot{\varphi_2}^2
=
\frac{1}{4f} \left[ 
\frac{\tilde{f}}{f} \left( \dot{f}^2 - 4 \epsilon_1 \epsilon_2 L^2  \right)
- 8 \epsilon_1 \epsilon_2 \dot{f} L 
\left( \frac{\varphi_1 \varphi_2}{f} \right) 
\right],
\label{eq:35} \\[5mm]
&& \hspace{-3.5mm}
\dot{\varphi_1}^2 + \dot{\varphi_2}^2 
= 
\dot{\varphi_{\mathrm{r}}}^2 + \frac{L^2}{\varphi_{\mathrm{r}}^2} 
\nonumber \\[3mm]
&& \hspace{14mm}
=
\frac{1}{4f^2} \left[ 
{\varphi_{\mathrm{r}}}^2 \dot{f}^2 + 
4 (\epsilon_1 - \epsilon_2) \varphi_1 \varphi_2 \dot{f} L + 
4 (\epsilon_1^2 \varphi_1^2 + \epsilon_2^2 \varphi_2^2) L^2
\right].  
\label{eq:36}
\end{eqnarray}
It follows from Eq.\ (\ref{eq:36}) that the quadric equation in terms of 
$L$ reads 
\begin{eqnarray}
&& \hspace{-15mm}
\left[ 
\left( \frac{\epsilon_1^2 \varphi_1^2 + \epsilon_2^2 \varphi_2^2}{f} 
\right) \left( \frac{{\varphi_{\mathrm{r}}}^2}{f} \right) -1
\right] L^2 
+ (\epsilon_1 - \epsilon_2) 
\left( \frac{{\varphi_{\mathrm{r}}}^2}{f} \right)
\left( \frac{\varphi_1 \varphi_2}{f} \right) \dot{f} L \nonumber \\[3mm]
&& \hspace{55mm} {}+
\frac{1}{4} \left( \frac{{\varphi_{\mathrm{r}}}^2}{f} \right)^2 \dot{f}^2 
-{\varphi_{\mathrm{r}}}^2 {\dot{\varphi_{\mathrm{r}}}}^2
=0.  
\label{eq:37}
\end{eqnarray}
The solution of Eq.\ (\ref{eq:37}) is given by 
\begin{eqnarray}
L = 
\left[
2 (\epsilon_1 - \epsilon_2) 
\left( \frac{\varphi_1 \varphi_2}{f} \right)
\right]^{-1}
\left[ - \left( \frac{{\varphi_{\mathrm{r}}}^2}{f} \right) \dot{f} 
\pm 2 \varphi_{\mathrm{r}} \dot{\varphi_{\mathrm{r}}}
\right], 
\label{eq:38}
\end{eqnarray}
where in deriving this expression we have used Eq.\ (\ref{eq:31}).  
Thus, from Eqs.\ (\ref{eq:30})$-$(\ref{eq:32}), 
(\ref{eq:34})$-$(\ref{eq:36}), and (\ref{eq:38}), 
we see that 
all the terms on the right-hand side of 
Eqs.\ (\ref{eq:10}) and (\ref{eq:11}) can 
be represented in terms of $f$ and $\varphi_{\mathrm{r}}$.  

Next, 
we consider approximate expressions of $A$, $B$, and $C$.  
We again assume $\varphi_1^2 \approx \varphi_2^2$ and 
$\dot{\varphi_1}^2 \approx \dot{\varphi_2}^2$, 
as in the case that 
we approximately determined $\omega$ in terms of $\epsilon_i$ 
in \S 3.   
Applying these approximate relations to Eq.\ (\ref{eq:20}) and 
using Eqs.\ (\ref{eq:24}), (\ref{eq:25}) and (\ref{eq:28}), we find 
\begin{eqnarray} 
A \approx \frac{32}{3} \left( \frac{\beta}{\alpha} \right)^2 \Lambda,  
\label{eq:39}
\end{eqnarray}
where in deriving this approximate expression 
we have employed the relation $\epsilon_i \ll 1$, 
that is, $\alpha/\beta \gg 1$, 
and hence $\omega \gg 1$.  
In the same way, from Eqs.\ (\ref{eq:21}) and (\ref{eq:39}) we find 
\begin{eqnarray} 
C^2 + \frac{B^2}{C^2} 
\approx 
\frac{32\beta}{3\alpha} \hspace{0.5mm} 
\Lambda.  
\label{eq:40}
\end{eqnarray}
Furthermore, applying one of the above approximate relations, 
$\varphi_1^2 \approx \varphi_2^2$, to ${\varphi_{\mathrm{r}}}^2/f$ and 
using the ansatz (\ref{eq:15}), we find 
\begin{eqnarray} 
\frac{{\varphi_{\mathrm{r}}}^2}{f} = \frac{C^2}{A} 
\approx \frac{2}{\alpha}.  
\label{eq:41}
\end{eqnarray}
Thus, from Eqs.\ (\ref{eq:39}) and (\ref{eq:41}), we find 
\begin{eqnarray} 
C \approx 
\sqrt{
\frac{64\beta^2}{3\alpha^3} 
\hspace{0.5mm} \Lambda}, 
\label{eq:42}
\end{eqnarray}
where we have taken the positive value of $C$. 

On the other hand, substituting the ansatz (\ref{eq:15}) into 
the solution of the quadric equation in terms of $L$, (\ref{eq:38}), 
we find 
\begin{eqnarray} 
B &=& \pm 2C^2
\left[
\left(-1+ \epsilon_1 \hspace{0.5mm} \frac{C^2}{A} \right) 
\left(1 - \epsilon_2 \hspace{0.5mm} \frac{C^2}{A} \right)
\right]^{-1/2}
\label{eq:43} \\[5mm]
&\approx&
\pm 
\frac{128\beta^2}{3\alpha^3} 
\sqrt{\frac{\alpha^2}{2\beta-\alpha^2}} \hspace{0.5mm} \Lambda, 
\label{eq:44}
\end{eqnarray} 
where the second approximate equality follows from 
Eqs.\ (\ref{eq:39}) and (\ref{eq:41}).

\subsection{Linearized equations} 
In this subsection, we discuss the fluctuation equation 
in terms of 
$f=\epsilon_1 \varphi_1^2 + \epsilon_2 \varphi_2^2$ and 
$\tilde{f}=\epsilon_1 \varphi_1^2 - \epsilon_2 \varphi_2^2$ 
in order to derive
these mass values in the inflationary stage.  
With the ansatz 
\begin{equation}
\begin{split}
f &= f_0 + \delta f, 
\hspace{5mm} \left| \delta f \right| \ll \left| f_0 \right|, \\[3mm] 
\tilde{f} &= {\tilde{f}}_0 + \delta \tilde{f}, 
\hspace{5mm} \left| \delta \tilde{f} \right| \ll \left| {\tilde{f}}_0 \right|,
\label{eq:45}
\end{split}
\end{equation}
where $f_0$ and ${\tilde{f}}_0$ are the zeroth-order quantities, which 
satisfy Eqs.\ (\ref{eq:10}) and (\ref{eq:11}), respectively, 
we obtain 
the linearized equations 
\begin{eqnarray}
\left(
\begin{array}{c}
  \ddot{\delta f} + 3H \dot{\delta f} \\
  \ddot{\delta \tilde{f}} + 3H \dot{\delta \tilde{f}}
\end{array}
\right)
&=&
-{\cal M}^2
\left(
\begin{array}{c}
 \delta f \\  \delta \tilde{f}
\end{array}
\right),  
\label{eq:46} 
\end{eqnarray}
where we have retained the terms up to first order in 
$\delta f$ and $\delta \tilde{f}$.  
Moreover, we have omitted the quantities in terms of 
$\dot{\delta f}$ and $\dot{\delta \tilde{f}}$ 
derived from the right-hand side 
of Eqs.\ (\ref{eq:10}) and (\ref{eq:11}).  
This is because, 
as we see from the ansatz (\ref{eq:15}) and Eq.\ (\ref{eq:32}), 
both $f$ and $\tilde{f}$ are proportional to $t^2$, and 
thus, in the large $t$ limit 
we have 
$\left| \dot{\delta f} \right| \ll \delta f$ and 
$\left| \dot{\delta \tilde{f}} \right| \ll \delta \tilde{f}$.  
Incidentally, the reason we have chosen the field variable 
$\tilde{f}$ as the partner of the coupling $f$ is that 
$\tilde{f}$ is a simple quantity satisfying the conditions that 
it have the same dimensions as $f$ and that it be expressed as a 
linear combination of $\varphi_1^2$ and $\varphi_2^2$.

In deriving the linearized equations (\ref{eq:46}), we have taken into 
account the following point:  
From Eqs.\ (\ref{eq:30})$-$(\ref{eq:32}), 
(\ref{eq:34})$-$(\ref{eq:36}), and (\ref{eq:38}), 
we see that all the terms on the right-hand side of 
Eqs.\ (\ref{eq:10}) and (\ref{eq:11}) can 
be represented in terms of $f$ and $\varphi_{\mathrm{r}}$, 
as noted in the previous subsection.  
Furthermore, 
from Eq.\ (\ref{eq:32}), we find 
$
{\varphi_{\mathrm{r}}}^2 
=
\left[ (\epsilon_1 + \epsilon_2)f 
-(\epsilon_1 - \epsilon_2) \tilde{f}
\right]/ (2 \epsilon_1 \epsilon_2). 
$
Hence, the fluctuation $\delta \varphi_{\mathrm{r}}$ 
induced by $\delta f$ and $\delta \tilde{f}$ 
can be represented as follows: 
\begin{eqnarray}
\delta \varphi_{\mathrm{r}} 
=
\frac{1}{4\epsilon_1 \epsilon_2 \varphi_{\mathrm{r}}} 
\left[ (\epsilon_1 + \epsilon_2) \delta f
-(\epsilon_1 - \epsilon_2) \delta \tilde{f}
\right].  
\label{eq:47}
\end{eqnarray}
Thus, all the fluctuations derived from the right-hand side of 
Eqs.\ (\ref{eq:10}) and (\ref{eq:11}) can be represented 
in terms of $\delta f$ and $\delta \tilde{f}$ 
by using Eqs.\ (\ref{eq:30})$-$(\ref{eq:32}), 
(\ref{eq:34})$-$(\ref{eq:36}), (\ref{eq:38}) and (\ref{eq:47}).  

From the above, we find that ${\cal M}^2$ is given by 
\begin{eqnarray}
{\cal M}^2 \equiv 
\left(
\begin{array}{cc}
  {{\cal M}^2}_{11} & {{\cal M}^2}_{12} \\
  {{\cal M}^2}_{21} & {{\cal M}^2}_{22} 
\end{array}
\right),   
\label{eq:48}
\end{eqnarray}
with
\begin{eqnarray}
{{\cal M}^2}_{11} &=& 
-4 F^{-2} \left( 2F-1 \right) Q 
\hspace{0.5mm} 
\frac{1}{f}
\hspace{0.5mm} 
V
\nonumber \\[3mm]
&& \hspace{0mm}
{}-2 F^{-1} 
\left[ 
-1 + \frac{\epsilon_1 + \epsilon_2}{\epsilon_1 \epsilon_2} 
\left( \frac{1}{4} \frac{f}{{\varphi_{\mathrm{r}}}^2} + U \right)
-\frac{6Q}{F}
\right] 
\frac{1}{\varphi_{\mathrm{r}}}
\hspace{0.5mm} 
V^{\prime} 
\nonumber \\[3mm]
&& \hspace{0mm}
{}+F^{-1} 
\hspace{0.5mm}
\frac{\epsilon_1 + \epsilon_2}{2 \epsilon_1 \epsilon_2} 
\hspace{0.5mm}
\frac{f}{{\varphi_{\mathrm{r}}}^2} 
\hspace{0.5mm} 
V^{\prime \prime} 
\nonumber \\[3mm]
&& \hspace{0mm}
{}-F^{-1} 
\Biggl\{
\frac{Q}{F} \hspace{0.5mm}
\frac{\dot{\varphi_1}^2 + \dot{\varphi_2}^2}{f} 
+ \frac{\epsilon_1 + \epsilon_2}{\epsilon_1 \epsilon_2} 
\hspace{0.5mm} 
U 
\hspace{0.5mm} 
\frac{L^2}{{\varphi_{\mathrm{r}}}^4}
-\frac{1 + 6(\epsilon_1 + \epsilon_2)}{2F} J 
\nonumber \\[3mm]
&& \hspace{14mm}
{}+
\frac{Q}{\epsilon_1 \epsilon_2} 
\hspace{0.5mm} 
\frac{I}{{\varphi_{\mathrm{r}}}^2} 
\left[ 
Q^2 \frac{I}{f} - Q \frac{\dot{f}}{f} 
\pm (\epsilon_1 + \epsilon_2) 
\frac{\dot{\varphi_{\mathrm{r}}}}{\varphi_{\mathrm{r}}}
\right] 
\Biggr\} 
\label{eq:49} \\[5mm]
&\approx& 
-\frac{3\alpha(2\beta-\alpha^2)}{8\beta^2}
\frac{V}{\Lambda t^2}
+\frac{3\sqrt{3}\alpha^{3/2}(\alpha^2-4\beta)}{16\beta(\alpha^2-\beta)}
\frac{V^{\prime}}{\sqrt{\Lambda} t}
+\frac{\alpha^2}{2(\alpha^2-\beta)} 
V^{\prime \prime}
\nonumber \\[3mm]
&& \hspace{0mm}
{}
+\frac{C_{11}}{\beta(\alpha^2-\beta)(2\beta-\alpha^2)}
\hspace{0.5mm}
\frac{1}{t^2},  
\label{eq:50} \\[5mm]
{{\cal M}^2}_{12} &=& 
-4 F^{-2} (\epsilon_1 - \epsilon_2)
\hspace{0.5mm}
\frac{1}{f} \hspace{0.5mm} 
V
\nonumber \\[3mm] 
&& \hspace{0mm}
{}-2 F^{-1} (\epsilon_1 - \epsilon_2) 
\left[ 
-\frac{1}{\epsilon_1 \epsilon_2} 
\left( \frac{1}{4} \frac{f}{{\varphi_{\mathrm{r}}}^2} + U \right)
+ \frac{6}{F}
\right] 
\frac{1}{\varphi_{\mathrm{r}}}
\hspace{0.5mm} 
V^{\prime} 
\nonumber \\[3mm]
&& \hspace{0mm}
{}-F^{-1} 
\hspace{0.5mm}
\frac{\epsilon_1 - \epsilon_2}{2 \epsilon_1 \epsilon_2} 
\hspace{0.5mm}
\frac{f}{{\varphi_{\mathrm{r}}}^2} 
\hspace{0.5mm} 
V^{\prime \prime} 
\nonumber \\[3mm]
&& \hspace{0mm}
{}+F^{-1} 
\hspace{0.5mm}
(\epsilon_1 - \epsilon_2) 
\Biggl[
\frac{1}{F} \hspace{0.5mm}
\frac{\dot{\varphi_1}^2 + \dot{\varphi_2}^2}{f} 
+ \frac{U}{\epsilon_1 \epsilon_2} 
\hspace{0.5mm}
\frac{L^2}{{\varphi_{\mathrm{r}}}^4}
+\frac{3}{F} J
\nonumber \\[3mm]
&& \hspace{28mm}
{}+\frac{Q}{\epsilon_1 \epsilon_2} 
\hspace{0.5mm} 
\frac{I}{{\varphi_{\mathrm{r}}}^2} 
\left( 
Q \frac{I}{f} - \frac{\dot{f}}{f} 
\pm \frac{\dot{\varphi_{\mathrm{r}}}}{\varphi_{\mathrm{r}}}
\right)
\Biggr]
\label{eq:51} \\[5mm]
&\approx& 
\pm \sqrt{2\beta-\alpha^2}
\biggl\{
-\frac{3\alpha^2}{8\beta^2}
\frac{V}{\Lambda t^2}
+\frac{\sqrt{3}\alpha^{3/2}(\alpha^2+8\beta)}{16\alpha\beta(\alpha^2-\beta)}
\frac{V^{\prime}}{\sqrt{\Lambda} t}
-\frac{\alpha}{2(\alpha^2-\beta)} 
V^{\prime \prime}
\nonumber \\[3mm]
&& \hspace{22mm}
{}
+\frac{C_{12}}{\beta(\alpha^2-\beta)(2\beta-\alpha^2)}
\hspace{0.5mm}
\frac{1}{t^2}
\biggr\},
\label{eq:52} \\[5mm]
{{\cal M}^2}_{21} &=& 
4 F^{-2} 
\hspace{0.5mm}
\frac{Q}{\epsilon_1 - \epsilon_2} 
\hspace{0.5mm}
\left[ (\epsilon_1 + \epsilon_2) + 24 \epsilon_1 \epsilon_2 \right]
\hspace{0.5mm} 
\frac{1}{f}
\hspace{0.5mm} 
V
\nonumber \\[3mm]
&& \hspace{0mm}
{}-\frac{F^{-1}}{\epsilon_1 - \epsilon_2} 
\biggl\{ 
Q
\left(
-\frac{12Q}{F} 
+ \frac{\epsilon_1 + \epsilon_2}{2 \epsilon_1 \epsilon_2} 
\hspace{0.5mm} 
\frac{f}{{\varphi_{\mathrm{r}}}^2} 
\right) 
+ 48 \epsilon_1 \epsilon_2 
\left[ 
\left( 1 + \frac{6Q}{F} \right) P 
- Q \hspace{0.5mm} \frac{{\varphi_{\mathrm{r}}}^2}{f}
\right]
\nonumber \\[3mm]
&& \hspace{19mm}
{}+12(\epsilon_1 + \epsilon_2) 
\left( 
2 Q - P \hspace{0.5mm} \frac{f}{{\varphi_{\mathrm{r}}}^2}
\right)
+\frac{(\epsilon_1 + \epsilon_2)(\epsilon_1 - \epsilon_2)}
{\epsilon_1 \epsilon_2}
\hspace{0.5mm} 
W
\biggr\}
\hspace{0.5mm}
\frac{1}{\varphi_{\mathrm{r}}}
\hspace{0.5mm} 
V^{\prime} 
\nonumber \\[3mm]
&& \hspace{0mm}
{}-F^{-1} 
\hspace{0.5mm}
\frac{\epsilon_1 + \epsilon_2}{\epsilon_1 - \epsilon_2} 
\left( 
-\frac{Q}{2 \epsilon_1 \epsilon_2} + 12P
\right)
\frac{f}{{\varphi_{\mathrm{r}}}^2} 
\hspace{0.5mm} 
V^{\prime \prime} 
\nonumber \\[3mm]
&& \hspace{0mm}
{}+\frac{1}{\epsilon_1 - \epsilon_2} 
\hspace{0.5mm} 
\frac{1}{f^2}
\left[ 
Q \left( \dot{f}^2 - 4 \epsilon_1 \epsilon_2 L^2 \right) 
+ \left( 4 \epsilon_1 \epsilon_2 P - Q^2 \right) \dot{f} I 
\right]
\nonumber \\[3mm]
&& \hspace{0mm}
{}-F^{-1}
\biggl\{
\frac{Q}{(\epsilon_1 - \epsilon_2)F}
\hspace{0.5mm}
\left[ (\epsilon_1 + \epsilon_2) + 24 \epsilon_1 \epsilon_2 \right]
\left( \frac{\dot{\varphi_1}^2 + \dot{\varphi_2}^2}{f} + 3J \right)
\nonumber \\[3mm]
&& \hspace{14mm}
{}
+W 
\left( 
3J + \frac{\epsilon_1 + \epsilon_2}{2 \epsilon_1 \epsilon_2} 
\hspace{0.5mm} 
\frac{L^2}{{\varphi_{\mathrm{r}}}^4}
\right)
\biggr\}
\nonumber \\[3mm]
&& \hspace{0mm}
{}+
\frac{1}{\epsilon_1 - \epsilon_2} 
\left\{
\frac{\dot{f}}{f} 
-
\left[
Q + 
\frac{\epsilon_1 - \epsilon_2}{2 \epsilon_1 \epsilon_2}
\hspace{0.5mm} 
\frac{W}{F}
\left( 
\frac{f}{{\varphi_{\mathrm{r}}}^2} + 12 \epsilon_1 \epsilon_2
\right)
\right]
\frac{I}{f}
\right\}
\nonumber \\[3mm]
&& \hspace{55mm}
{}\times 
\left[ Q^2 \frac{I}{f} - Q \frac{\dot{f}}{f} 
\pm (\epsilon_1 + \epsilon_2) \frac{\dot{\varphi_{\mathrm{r}}}}
{\varphi_{\mathrm{r}}}
\right]
\label{eq:53} \\[5mm]
&\approx& 
\pm \sqrt{2\beta-\alpha^2}
\left\{
\frac{3\alpha^2}{8\beta^2}
\frac{V}{\Lambda t^2}
-\frac{9\sqrt{3}\alpha^{5/2}}{16\beta(\alpha^2-\beta)}
\frac{V^{\prime}}{\sqrt{\Lambda} t}
+\frac{\alpha}{2(\alpha^2-\beta)} 
V^{\prime \prime}
\right\}
\nonumber \\[3mm]
&& \hspace{22mm}
{}
+\frac{C_{21}}
{(\pm \sqrt{2\beta-\alpha^2})\beta(\alpha^2-\beta)}
\hspace{0.5mm}
\frac{1}{t^2},
\label{eq:54} \\[5mm] 
{{\cal M}^2}_{22} &=& 
-4 F^{-1} 
\left[ 
(\epsilon_1 + \epsilon_2) - 6(\epsilon_1 - \epsilon_2)
\hspace{0.5mm}
\frac{W}{F}
\right]
\frac{1}{f}
\hspace{0.5mm} 
V
\nonumber \\[3mm]
&& \hspace{0mm} 
{}+F^{-1} 
\biggl\{
2
\left[ 
1-\frac{6Q}{F} 
+ 6\left( 2Q - P \hspace{0.5mm} \frac{f}{{\varphi_{\mathrm{r}}}^2} \right) 
+ 144 \epsilon_1 \epsilon_2 \hspace{0.5mm} \frac{P}{F} 
\right] 
\nonumber \\[3mm]
&& \hspace{13mm}
{}+
\frac{1}{2 \epsilon_1 \epsilon_2} 
\left[ 
2(\epsilon_1 - \epsilon_2) W 
+ Q \hspace{0.5mm} \frac{f}{{\varphi_{\mathrm{r}}}^2} 
\right]
\biggr\}
\hspace{0.5mm}
\frac{1}{\varphi_{\mathrm{r}}}
\hspace{0.5mm} 
V^{\prime} 
\nonumber \\[3mm]
&& \hspace{0mm}
{}-F^{-1} 
\left( \frac{Q}{2 \epsilon_1 \epsilon_2} - 12P \right)
\frac{f}{{\varphi_{\mathrm{r}}}^2} 
\hspace{0.5mm} 
V^{\prime \prime}
\nonumber \\[3mm]
&& \hspace{0mm}
{}+\frac{1}{2f^2} 
\left( 
-\dot{f}^2 + 4 \epsilon_1 \epsilon_2 L^2 + 2 Q \dot{f} I 
\right)
\nonumber \\[3mm]
&& \hspace{0mm}
{}+F^{-1} 
\left\{
\left[ 
(\epsilon_1 + \epsilon_2) - 6(\epsilon_1 - \epsilon_2)
\hspace{0.5mm}
\frac{W}{F}
\right]
\left( 
\frac{\dot{\varphi_1}^2 + \dot{\varphi_2}^2}{f} + 3J
\right)
+ \frac{\epsilon_1 - \epsilon_2}{2 \epsilon_1 \epsilon_2}
\hspace{0.5mm}
W
\hspace{0.5mm}
\frac{L^2}{{\varphi_{\mathrm{r}}}^4}
\right\}
\nonumber \\[3mm]
&& \hspace{0mm}
{}-
\left\{
\frac{\dot{f}}{f}
- 
\left[ 
Q + \frac{\epsilon_1 - \epsilon_2}{2 \epsilon_1 \epsilon_2}
\hspace{0.5mm} 
\frac{W}{F}
\left( 
\frac{f}{{\varphi_{\mathrm{r}}}^2} + 12 \epsilon_1 \epsilon_2 
\right)
\right]
\frac{I}{f} 
\right\} 
\left( 
Q \frac{I}{f} - \frac{\dot{f}}{f} 
\pm \frac{\dot{\varphi_{\mathrm{r}}}}{\varphi_{\mathrm{r}}}
\right)
\label{eq:55} \\[5mm]
&\approx&
-\frac{3\alpha^3}{8\beta^2}
\frac{V}{\Lambda t^2}
+\frac{\sqrt{3}\alpha^{3/2}(14\beta-5\alpha^2)}{16\beta(\alpha^2-\beta)}
\frac{V^{\prime}}{\sqrt{\Lambda} t}
-\frac{2\beta-\alpha^2}{2(\alpha^2-\beta)} 
V^{\prime \prime}
\nonumber \\[3mm]
&& \hspace{0mm}
{}
+\frac{C_{22}}
{\beta(\alpha^2-\beta)(2\beta-\alpha^2)}
\hspace{0.5mm}
\frac{1}{t^2},
\label{eq:56} 
\end{eqnarray}
where 
\begin{eqnarray}
F &=& 
1 + 12 
\left[ 
(\epsilon_1 + \epsilon_2) 
- \epsilon_1 \epsilon_2 \frac{{\varphi_{\mathrm{r}}}^2}{f} 
\right] = 1+12U 
\hspace{0.5mm} \approx \hspace{0.5mm}
1+ 12 \frac{\beta}{\alpha}, 
\label{eq:57} \\[3mm]
Q &=& (\epsilon_1 + \epsilon_2) 
- 2 \epsilon_1 \epsilon_2 \frac{{\varphi_{\mathrm{r}}}^2}{f}
\hspace{0.5mm} \approx \hspace{0.5mm}
\frac{2\beta - \alpha^2}{\alpha}, 
\label{eq:58} \\[3mm]
U &=& (\epsilon_1 + \epsilon_2) 
- \epsilon_1 \epsilon_2 \frac{{\varphi_{\mathrm{r}}}^2}{f}
\hspace{0.5mm} \approx \hspace{0.5mm}
\frac{\beta}{\alpha}, 
\label{eq:59} \\[3mm]
J &=& 
\frac{1}{f^2} 
\left( \dot{f}^2 + 4 \epsilon_1 \epsilon_2 L^2 \right)
\hspace{0.5mm} \approx \hspace{0.5mm}
4 \left( \frac{7\alpha^2 - 6\beta}{2\beta - \alpha^2} \right) 
\frac{1}{t^2},
\label{eq:60} \\[3mm] 
I &=& -\frac{1}{2P} 
\left[ - \left( \frac{{\varphi_{\mathrm{r}}}^2}{f} \right) \dot{f} 
\pm 2 \varphi_{\mathrm{r}} \dot{\varphi_{\mathrm{r}}}
\right]
\hspace{0.5mm} \approx \hspace{0.5mm}
\frac{128}{3} \hspace{0.5mm} \frac{\beta^2}{\alpha (2\beta - \alpha^2)} 
\hspace{0.5mm} \Lambda \hspace{0.5mm} t, 
\label{eq:61} \\[3mm] 
P &=& - 
\left(1- \epsilon_1 \frac{{\varphi_{\mathrm{r}}}^2}{f} \right) 
\left(1 - \epsilon_2 \frac{{\varphi_{\mathrm{r}}}^2}{f} \right)
\hspace{0.5mm} \approx \hspace{0.5mm}
\frac{2\beta - \alpha^2}{\alpha^2}, 
\label{eq:62} \\[3mm] 
W &=& (\epsilon_1 - \epsilon_2) + 
\frac{\epsilon_1 + \epsilon_2}{\epsilon_1 - \epsilon_2} 
\hspace{0.5mm}  
Q
\hspace{0.5mm} \approx \hspace{0.5mm}
\pm 2 \sqrt{2\beta - \alpha^2},  
\label{eq:63} \\[3mm]
C_{11} &=& 
7\alpha^4\beta-16\beta^3-2\alpha^2\beta^2-\alpha^6
-\frac{96\beta(\alpha^2-\beta)^2(2\beta-\alpha^2)}{\alpha},
\label{eq:64}  \\[5mm]
C_{12} &=& 
14\alpha\beta^2-\alpha^3\beta-\alpha^5 
+\frac{48\beta(\alpha^2-\beta)
\left[
2\alpha(\alpha^2-\beta)-3\beta(7\alpha^2-6\beta)
\right]
}
{\alpha},
\label{eq:65}  \\[5mm] 
C_{21} &=& 
\alpha^5-8\alpha\beta^2-5\alpha^3\beta 
+12\beta(-7\alpha^4+\alpha^2\beta+10\beta^2),
\label{eq:66}  \\[5mm] 
C_{22} &=& 
\alpha
\left[
\alpha(14\beta^2-\alpha^4-\alpha^2\beta)
+12\beta(7\alpha^4-9\alpha^2\beta-2\beta^2)
\right].  
\label{eq:67}   
\end{eqnarray}
Here, the expression for $F$ given in (\ref{eq:57}) follows from 
Eqs.\ (\ref{eq:12}) and (\ref{eq:30}).  

In deriving the last approximate equality in each of the equations
(\ref{eq:50}), (\ref{eq:52}), (\ref{eq:54}), and 
(\ref{eq:56})$-$(\ref{eq:63}), 
we have used Eqs.\ (\ref{eq:24}), (\ref{eq:25}), 
(\ref{eq:39}), (\ref{eq:40}), (\ref{eq:42}), and (\ref{eq:44}).  
Moreover, 
in the last approximate equality in each of the equations 
(\ref{eq:50}), (\ref{eq:52}), (\ref{eq:54}), and (\ref{eq:56}),
we have retained all the leading-order terms and 
some of the sub-leading order terms in $\epsilon_i$ and 
omitted the rest of the sub-leading order terms, which are 
explicitly unimportant.  
Furthermore, 
in deriving the last approximate equality in 
(\ref{eq:61}), we have taken the negative sign, so that 
$I$ has a finite value.  
If we take the positive sign, 
under the approximation in (\ref{eq:41}), we have $I \approx 0$.  
Since 
it follows from Eqs.\ (\ref{eq:38}), (\ref{eq:61}), and (\ref{eq:62}) 
that the angular momentum is given by $L = - \left( \pm P^{1/2} \right) I$, 
and because here we consider the case $L\not= 0$, 
we have taken the negative sign so that the angular momentum can 
have a finite value.  
Hence, in deriving the approximate expressions 
(\ref{eq:50}), (\ref{eq:52}), (\ref{eq:54}) and (\ref{eq:56}), 
we have taken the negative sign 
in the last term on the right-hand side of each of 
Eqs.\ (\ref{eq:49}), (\ref{eq:51}), (\ref{eq:53}) and (\ref{eq:55}), 
because the sign of these terms is determined by the angular 
momentum $L$, and $L$ is proportional to $I$, as shown above.  

We finally emphasize that all the terms in each of the 
components of ${\cal M}^2$ 
can be expressed in terms of the background quantities 
$f=At^2$, $L=Bt$ and $\varphi_{\mathrm{r}}=Ct$ in the inflationary stage, 
and hence can be approximately represented in terms of 
$\alpha$ and $\beta$ by using the expressions 
(\ref{eq:39}), (\ref{eq:40}), (\ref{eq:42}) and (\ref{eq:44}), 
as shown in expressions 
(\ref{eq:50}), (\ref{eq:52}), (\ref{eq:54}) and (\ref{eq:56}).

\subsection{Mass diagonalization} 
Next, in order to show that an extremely large hierarchy of 
the mass values of $f$ and $\tilde{f}$ in the inflationary 
stage can be realized, we investigate these eigenmasses by 
solving the 
characteristic 
equation of ${\cal M}^2$,  
\begin{eqnarray} 
\det \left({\cal M}^2 - \lambda E  \right)  = 0 
\hspace{1mm}
\Longleftrightarrow 
\hspace{1mm}
\lambda^2 - \mathrm{Tr} \hspace{0.5mm} {\cal M}^2  \hspace{0.5mm} \lambda 
+ \det {\cal M}^2  = 0, 
\label{eq:68}
\end{eqnarray}
where $\lambda$ is the eigenvalue of ${\cal M}^2$ and 
$E$ is the unit matrix.  The solutions of Eq.\ (\ref{eq:68}) 
are equivalent to the diagonal components of the diagonalized 
form of ${\cal M}^2$, namely, the square of the eigenmasses of $f$ and 
$\tilde{f}$.  
The solution of Eq.\ (\ref{eq:68}) is given by 
\begin{eqnarray} 
\lambda_{\pm} = \frac{\mathrm{Tr} \hspace{0.5mm} {\cal M}^2}{2} 
\left[
1 \pm 
\sqrt{1-
\frac{4\det {\cal M}^2}{(\mathrm{Tr} \hspace{0.5mm} {\cal M}^2)^2}
}
\hspace{0.5mm}
\right].
\label{eq:69}
\end{eqnarray}
In this subsection, 
in order to evaluate the solutions (\ref{eq:69}), 
we derive approximate expressions of 
$\mathrm{Tr} \hspace{0.5mm} {\cal M}^2 $ and $\det {\cal M}^2$.  

From Eq.\ (\ref{eq:3}) and the expressions 
(\ref{eq:50}), (\ref{eq:52}), (\ref{eq:54}), and (\ref{eq:56}), 
we can obtain approximate expressions for 
$\mathrm{Tr} \hspace{0.5mm} {\cal M}^2 $ and $\det {\cal M}^2$.  
For convenience in describing these expressions, 
we define 
$
\mathrm{Tr} \hspace{0.5mm} {\cal M}^2 \equiv
(\mathrm{Tr} \hspace{0.5mm} {\cal M}^2)^{(0)} 
+ (\mathrm{Tr} \hspace{0.5mm} {\cal M}^2)^{(1)} 
$
and
$
\det {\cal M}^2 \equiv 
(\det {\cal M}^2)^{(0)} 
+ (\det {\cal M}^2)^{(1)}
+ (\det {\cal M}^2)^{(2)},
$
where 
$(\mathrm{Tr} \hspace{0.5mm} {\cal M}^2)^{(0)}$ 
and $(\det {\cal M}^2)^{(0)}$ are the parts independent of 
the sine functions, 
while 
$(\mathrm{Tr} \hspace{0.5mm} {\cal M}^2)^{(1)}$ 
and $(\det {\cal M}^2)^{(1)}$ are the parts proportional to 
$\sin \left[ \varphi_{\mathrm{r}}/M + 
\arctan \left( X_1/X_2 \right) \right]$ and 
$\sin \left[ \varphi_{\mathrm{r}}/M + 
\arctan \left( Y_1/Y_2 \right) \right]$, respectively, 
and $(\det {\cal M}^2)^{(2)}$ is the part proportional to 
$\sin \left[ 2\varphi_{\mathrm{r}}/M + 
\arctan \left( Z_1/Z_2 \right) \right]$.  
[$X_1$, $X_2$, $Y_1$, $Y_2$, $Z_1$ and $Z_2$ are 
given in Eqs.\ (\ref{eq:83})$-$(\ref{eq:88}).] 
We then obtain 
the approximate expressions for 
$\mathrm{Tr} \hspace{0.5mm} {\cal M}^2 $ and $\det {\cal M}^2$ 
as 
\if
It follows from Eq.\ (\ref{eq:3}) and the expressions 
(\ref{eq:52}), (\ref{eq:54}), (\ref{eq:56}), and (\ref{eq:58}) 
that the approximate expressions of 
$\mathrm{Tr} \hspace{0.5mm} {\cal M}^2 $ and $\det {\cal M}^2$ 
are given by  
\fi
\begin{eqnarray}
\mathrm{Tr} \hspace{0.5mm} {\cal M}^2 &=& 
{{\cal M}^2}_{11} + {{\cal M}^2}_{22} 
\nonumber \\[5mm]
&\equiv& 
(\mathrm{Tr} \hspace{0.5mm} {\cal M}^2)^{(0)} 
+ (\mathrm{Tr} \hspace{0.5mm} {\cal M}^2)^{(1)}, 
\label{eq:76} \\[5mm]
(\mathrm{Tr} \hspace{0.5mm} {\cal M}^2)^{(0)} 
&\approx& \hspace{0mm}
\left[
-\frac{3\alpha}{4\beta} 
+ \frac{C_{11}+C_{22}}{\beta(\alpha^2-\beta)(2\beta-\alpha^2)}
\right] 
\hspace{0mm}
\frac{1}{t^2},
\label{eq:77} \\[5mm]
(\mathrm{Tr} \hspace{0.5mm} {\cal M}^2)^{(1)}
&\approx& \hspace{0mm}
\sqrt{X_1^2 +X_2^2} 
\sin \left( \frac{\varphi_{\mathrm{r}}}{M} + 
\arctan \frac{X_1}{X_2} \right)
\frac{1}{t^2},
\label{eq:78}  
\end{eqnarray}
and 
\begin{eqnarray}
\det {\cal M}^2
&=& 
{{\cal M}^2}_{11} {{\cal M}^2}_{22} - {{\cal M}^2}_{12} {{\cal M}^2}_{21}
\nonumber \\[5mm]
&\equiv& 
(\det {\cal M}^2)^{(0)} 
+ (\det {\cal M}^2)^{(1)}
+ (\det {\cal M}^2)^{(2)},
\label{eq:79} \\[5mm]
(\det {\cal M}^2)^{(0)} 
&\approx& \hspace{0mm}
\biggl\{
\frac{3\alpha}{8\beta^3(\alpha^2-\beta)}
\left[
-\frac{\alpha^2}{2\beta-\alpha^2}C_{11} -C_{22}
+ \alpha\left( C_{21} - C_{12} \right)
\right]
\nonumber \\[3mm]
&& \hspace{-5mm}
{}
+ \frac{C_{11}C_{22} - (2\beta-\alpha^2)C_{12}C_{21}}
{\beta^2(\alpha^2-\beta)^2(2\beta-\alpha^2)^2}
\nonumber \\[3mm]
&& \hspace{-10mm}
{}
+\frac{3\alpha^3}{64\beta^4}
\left[
9\alpha(2\beta-\alpha^2)
+\frac{\alpha^6-11\alpha^2\beta^2+2\alpha^4\beta+8\beta^3}
{64(\alpha^2-\beta)^2} \frac{1}{\Upsilon^2}
\right]
\biggr\} \hspace{0.5mm}
\frac{1}{t^4},
\label{eq:80} \\[5mm]
(\det {\cal M}^2)^{(1)}
&\approx& \hspace{0mm}
\sqrt{Y_1^2 +Y_2^2} 
\sin \left( \frac{\varphi_{\mathrm{r}}}{M} + 
\arctan \frac{Y_1}{Y_2} \right) 
\hspace{0mm}
\frac{1}{t^4},
\label{eq:81} \\[5mm]
(\det {\cal M}^2)^{(2)}
&\approx& \hspace{0mm}
\sqrt{Z_1^2 +Z_2^2} 
\sin \left( 2\frac{\varphi_{\mathrm{r}}}{M} + 
\arctan \frac{Z_1}{Z_2} \right)
\hspace{0mm}
\frac{1}{t^4},
\label{eq:82}  
\end{eqnarray}
where
\begin{eqnarray}
X_1 &=& 
-\frac{3\alpha}{4\beta}
\left(1+\frac{\alpha}{48\beta}\frac{1}{\Upsilon^2} \right),
\label{eq:83}  \\[5mm]
X_2 &=& 
\frac{\sqrt{3} \alpha^{5/2}}{64\beta^2} 
\frac{1}{\Upsilon},
\label{eq:84}  \\[5mm]
Y_1 &=& 
\frac{3\alpha}{8\beta^3(\alpha^2-\beta)}
\left[
-\frac{\alpha^2}{2\beta-\alpha^2}C_{11} -C_{22}
+ \alpha\left( C_{21} - C_{12} \right)
\right]
+\frac{9\alpha^4(2\beta-\alpha^2)}{16\beta^4}
\nonumber \\[3mm]
&& \hspace{0mm}
{}
-\frac{\alpha^2}{128\beta^3(\alpha^2-\beta)}
\biggl\{
\frac{1}{\alpha^2-\beta}
\left[
-C_{11}+\frac{\alpha^2}{2\beta-\alpha^2}C_{22} 
+ \alpha\left( C_{21} - C_{12} \right)
\right]
\nonumber \\[3mm]
&& \hspace{31.5mm}
{}
+\frac{3\alpha(2\beta^2-\alpha^4)}{4\beta}
\biggr\}
\frac{1}{\Upsilon^2},
\label{eq:85}  \\[5mm] 
Y_2 &=& 
-\frac{\sqrt{3}\alpha^{5/2}}{128\beta^3(\alpha^2-\beta)}
\biggl\{
\frac{1}{\alpha^2-\beta}
\biggl[ 
\frac{(14\beta-5\alpha^2)C_{11} + 3(\alpha^2-4\beta)C_{22}}
{2\beta-\alpha^2}
+9 \alpha C_{12}
\nonumber \\[3mm]
&& \hspace{28.5mm}
{}
-\frac{(\alpha^2+8\beta)C_{21}}{\alpha}
\biggr]
+\frac{3(12\alpha^3\beta-22\alpha\beta^2+\alpha^5)}{4\beta}
\biggr\}
\frac{1}{\Upsilon},
\label{eq:86}  \\[5mm] 
Z_1 &=& 
\frac{3\alpha^3}{64\beta^4}
\left[
3\alpha(2\beta-\alpha^2)
+\frac{7\alpha^6-5\alpha^2\beta^2-10\alpha^4\beta+8\beta^3}
{64(\alpha^2-\beta)^2} \frac{1}{\Upsilon^2}
\right],
\label{eq:87}  \\[5mm] 
Z_2 &=& 
\frac{\sqrt{3}\alpha^{5/2}}{1024\beta^4}
\left[
-\frac{3(12\alpha^3\beta-22\alpha\beta^2+\alpha^5)}{\alpha^2-\beta}
+ \frac{\alpha^2}{4} \frac{1}{\Upsilon^2}
\right]
\frac{1}{\Upsilon}.  
\label{eq:88}
\end{eqnarray}
Here 
we have considered the case 
$V_{0} \approx \Lambda \approx M^4$.  
Moreover, from Eq.\ (\ref{eq:74}), we have used 
the relation 
\begin{eqnarray}
M \approx \frac{H}{\Upsilon} 
=
\frac{\omega}{\Upsilon} \hspace{0.5mm} \frac{1}{t} 
\approx 
\frac{\alpha}{8\beta}\frac{1}{\Upsilon} \hspace{0.5mm} \frac{1}{t}, 
\label{eq:89}
\end{eqnarray}
where the last approximate equality follows from Eq.\ (\ref{eq:28}).


\section{Power-Law Inflation in the Einstein Frame}
In this appendix, we show that power-law inflation can be 
realized in this model not only in the Jordan frame, 
as shown in \S 3, 
but also in the Einstein frame.  

\subsection{Action} 
To begin with, we make the following conformal transformation of 
the action in Eq.\ (\ref{eq:1}):  
\begin{eqnarray}
g_{\mu \nu} \hspace{0.5mm} \rightarrow \hspace{0.5mm}
\hat{g}_{\mu \nu} = \Omega^2 g_{\mu \nu},
\label{eq:B.1}
\end{eqnarray}
with 
\begin{eqnarray}
\Omega = \sqrt{2f}.  
\label{eq:B.2}
\end{eqnarray}
Here, the hat denotes quantities in the Einstein frame.  
Moreover, 
$f(\varphi_i) = 
\epsilon_1 \varphi_1^2 + \epsilon_2 \varphi_2^2$ in Eq.\ (\ref{eq:2}) 
is the coupling between the dilaton and the scalar curvature.  
The action in the Einstein frame is then given by \cite{F-M,Tsujikawa1} 
\begin{eqnarray}
S_{\mathrm{E}} &=& 
\int d^{4} \hat{x} \sqrt{-\hat{g}} 
\Biggl\{
-\frac{1}{2} \hat{R} 
+ \frac{1}{2} \sum_i 
\left[ \frac{3}{2} \left( \frac{f_{\varphi_i}}{f} \right)^2 + \frac{1}{2f}
\right]
\hat{g}^{\mu \nu} {\partial}_{\mu}{\varphi_i}{\partial}_{\nu}{\varphi_i} 
\nonumber \\[3mm]
&& \hspace{21.5mm}
{}+6 \epsilon_1 \epsilon_2 \frac{\varphi_1 \varphi_2}{f^2} 
\hat{g}^{\mu \nu} {\partial}_{\mu}{\varphi_1}{\partial}_{\nu}{\varphi_2} 
- \hat{V}[\varphi_i] + 
{\hat{{\cal L}}}_{\mathrm{m}}
\Biggr\}, 
\label{eq:B.3} 
\end{eqnarray}
with 
\begin{eqnarray}
\hat{V}[\varphi_i] = \frac{V[\varphi_i]}{4f^2}.  
\label{eq:B.4}
\end{eqnarray}
Here, the subscript $\varphi_i$ denotes 
partial differentiation with respect to $\varphi_i$.  
Moreover, in deriving the expression for the third term on the 
right-hand side of Eq.\ (\ref{eq:B.3}), namely, the cross term, 
we have used the symmetry between $\mu$ and $\nu$.  
Because there exists this cross term in the action in Eq.\ (\ref{eq:B.3}),
we introduce new scalar fields defined as 
\begin{eqnarray}
{\partial}_{\mu} \tilde{\varphi}_1 &\equiv& 
\frac{1}{\sqrt{\epsilon_1^2 \varphi_1^2 + \epsilon_2^2 \varphi_2^2}} 
\left( \epsilon_1 \varphi_1 {\partial}_{\mu} \varphi_1 + 
\epsilon_2 \varphi_2 {\partial}_{\mu} \varphi_2
\right),
\label{eq:B.5}  \\[5mm]
{\partial}_{\mu} \tilde{\varphi}_2 &\equiv& 
\frac{1}{\sqrt{\epsilon_1^2 \varphi_1^2 + \epsilon_2^2 \varphi_2^2}} 
\left( -\epsilon_2 \varphi_2 {\partial}_{\mu} \varphi_1 + 
\epsilon_1 \varphi_1 {\partial}_{\mu} \varphi_2
\right),
\label{eq:B.6}  
\end{eqnarray}
so that the cross term can be removed.  Then the action 
in Eq.\ (\ref{eq:B.3}) can be rewritten in the form 
\begin{eqnarray}
S_{\mathrm{E}} &=& 
\int d^{4} \hat{x} \sqrt{-\hat{g}} 
\Biggl\{
-\frac{1}{2} \hat{R} 
+ \frac{1}{2} \sum_i 
\left[ \frac{3}{2} \left( \frac{f_{\varphi_i}}{f} \right)^2 + \frac{1}{2f} 
+ (-1)^{1+i} \hspace{0.5mm} 
6 \left( \frac{\epsilon_2 \varphi_2}{f} \right)^2
\right]
\hat{g}^{\mu \nu} 
{\partial}_{\mu}{\tilde{\varphi}_i}{\partial}_{\nu}{\tilde{\varphi}_i} 
\nonumber \\[3mm]
&& \hspace{21.5mm}
{} 
- \hat{V}[\varphi_i] + 
{\hat{{\cal L}}}_{\mathrm{m}}
\Biggr\}, 
\label{eq:B.7} 
\end{eqnarray}
where $i$ in the third term in the brackets on the right-hand side 
of Eq.\ (\ref{eq:B.7}) corresponds to the subscript of 
$\varphi_i \hspace{1mm} (i=1,2)$.  

\if
Here we have derived the expressions of $\tilde{\varphi}_1$ in (\ref{eq:B.5-2}) and $\tilde{\varphi}_2$ in (\ref{eq:B.6-2}) 
by integrating both the sides of expressions 
(\ref{eq:B.5}) and (\ref{eq:B.6}), respectively.  
In deriving expressions (\ref{eq:B.5-2}) and (\ref{eq:B.6-2}) 
we have regarded $\varphi_1$ and $\varphi_2$ 
as to be independent of each other.  Moreover, we have chosen 
integration constants so that both $\tilde{\varphi}_1$ and $\tilde{\varphi}_2$ 
can be equal to zero in the case $\varphi_1 = 0$ and $\varphi_2 = 0$.  
\fi

Furthermore, we introduce new scalar fields defined as 
\begin{eqnarray}
\frac{d {\hat{\tilde{\varphi}}}_i}{d \tilde{\varphi}_i} \equiv 
\sqrt{{D}_{\mathrm{E}} \left( \varphi_i \right)},
\label{eq:B.8}
\end{eqnarray}
with
\begin{eqnarray} 
{D}_{\mathrm{E}} \left( \varphi_i \right) 
= \frac{3}{2} \left( \frac{f_{\varphi_i}}{f} \right)^2 + \frac{1}{2f} 
+ (-1)^{1+i} \hspace{0.5mm} 
6 \left( \frac{\epsilon_2 \varphi_2}{f} \right)^2, 
\hspace{3mm} ( \hspace{0.5mm} > 0 \hspace{0.5mm}  ) 
\label{eq:B.9}
\end{eqnarray}
so that the action in Eq.\ (\ref{eq:B.7}) can be rewritten in 
the following canonical form:  
\begin{eqnarray}
S_{\mathrm{E}} = 
\int d^{4} \hat{x} \sqrt{-\hat{g}} 
\left[
-\frac{1}{2} \hat{R} 
+ \frac{1}{2} 
\hat{g}^{\mu \nu} {\partial}_{\mu} {\hat{\tilde{\varphi}}}_i 
{\partial}_{\nu} {\hat{\tilde{\varphi}}}_i
- \hat{V}[\varphi_i] + 
{\hat{{\cal L}}}_{\mathrm{m}}
\right].  
\label{eq:B.10} 
\end{eqnarray}
It is conjectured that 
if the potential $\hat{V}$ is written in terms of 
${\hat{\tilde{\varphi}}}_1$ and ${\hat{\tilde{\varphi}}}_2$, 
the form is complicated.

\subsection{Field equations} 
Field equations can be derived by taking variations of the 
action in Eq.\ (\ref{eq:B.10}) with respect to the 
metric $\hat{g}_{\mu\nu}$ and the dilatons ${\hat{\varphi}}_i$ as follows: 
\begin{eqnarray}
\hat{R}_{\mu \nu} - \frac{1}{2}\hat{g}_{\mu \nu} \hat{R} = 
{\hat{T}}^{(\mathrm{m})}_{\mu \nu} + 
{\hat{T}}^{( \varphi_i )}_{\mu \nu}, 
\label{eq:B.11}
\end{eqnarray}
with 
\begin{eqnarray}
{\hat{T}}^{(\varphi_i)}_{\mu \nu}
= {\partial}_{\mu} {\hat{\tilde{\varphi}}}_i 
{\partial}_{\nu} {\hat{\tilde{\varphi}}}_i
- {\hat{g}}_{\mu\nu} 
\left[ \frac{1}{2} 
\hat{g}^{\mu \nu} {\partial}_{\mu} {\hat{\tilde{\varphi}}}_i {\partial}_{\nu} 
{\hat{\tilde{\varphi}}}_i  
- \hat{V}[\varphi_i] 
\right], 
\label{eq:B.12}
\end{eqnarray}
and 
\begin{eqnarray}
\hat{\Box} {\hat{\tilde{\varphi}}}_i  
=
-\frac{\partial \hat{V}}{\partial {\hat{\tilde{\varphi}}}_i},
\label{eq:B.13}
\end{eqnarray}
where 
$\hat{\Box} \equiv \hat{g}^{\mu \nu} {\hat{\nabla}}_{\mu} 
{\hat{\nabla}}_{\nu}$.  

We now assume spatially flat 
Friedmann-Robertson-Walker (FRW) spacetime with the metric
\begin{eqnarray}
{ds}^2 = g_{\mu\nu}dx^{\mu}dx^{\nu} =  {dt}^2 - a^2(t)d{\Vec{x}}^2,
\label{eq:B.14}
\end{eqnarray}
to which we apply the conformal transformation in Eq.\ (\ref{eq:B.1}) as 
follows:  
\begin{eqnarray}
{d\hat{s}}^2 = \hat{g}_{\mu\nu}dx^{\mu}dx^{\nu} = 
\Omega^2 {ds}^2 = 
\Omega^2{dt}^2 - \Omega^2 a^2(t)d{\Vec{x}}^2.  
\label{eq:B.15}
\end{eqnarray}
We write the above line element in the form 
\begin{eqnarray}
{d\hat{s}}^2 = {d\hat{t}}^2 - {\hat{a}}^2 \left( \hat{t} \right) d{\Vec{x}}^2,
\label{eq:B.16}
\end{eqnarray}
through which $\hat{t}$ is defined as the cosmic time in the Einstein frame, 
and $\hat{a} \left( \hat{t} \right)$ as the scale factor.  
From Eqs.\ (\ref{eq:B.15}) and (\ref{eq:B.16}), we find 
\begin{eqnarray}
d\hat{t} &=& \Omega dt,
\label{eq:B.17} \\[3mm]
\hat{a} &=& \Omega a.  
\label{eq:B.18} 
\end{eqnarray}
In the FRW metric (\ref{eq:B.16}), 
the background gravitational field equations read 
\begin{eqnarray}
\hat{H}^2 &=&  
\left( \frac{1}{\hat{a}} \frac{d\hat{a}}{d\hat{t}} \right)^2 = 
\frac{1}{3} 
\left\{ 
{\hat{T}}^{(\mathrm{m})}_{00} + 
\frac{1}{2} \left[
\left( \frac{d{\hat{\tilde{\varphi}}}_1}{d\hat{t}} \right)^2 + 
\left( \frac{d{\hat{\tilde{\varphi}}}_2}{d\hat{t}} \right)^2
\right]
+ \hat{V}
\right\},
\label{eq:B.19} \\[5mm] 
\frac{d\hat{H}}{d\hat{t}} &=& 
-\frac{1}{6}
\left\{- {\hat{T}}^{(\mathrm{m})} + 4 {\hat{T}}^{(\mathrm{m})}_{00} 
+ 3 \left[
\left( \frac{d{\hat{\tilde{\varphi}}}_1}{d\hat{t}} \right)^2 + 
\left( \frac{d{\hat{\tilde{\varphi}}}_2}{d\hat{t}} \right)^2
\right]
\right\}  
\label{eq:B.20}.  
\end{eqnarray} 
Moreover, 
the equations of motion for the background homogeneous scalar fields read 
\begin{eqnarray}
\frac{d^2 {\hat{\tilde{\varphi}}}_i}{d{\hat{t}}^2} 
+ 3 \hat{H} \frac{d {\hat{\tilde{\varphi}}}_i}{d\hat{t}} 
= 
-\frac{\partial \hat{V}}{\partial {\hat{\tilde{\varphi}}}_i}.  
\label{eq:B.21}
\end{eqnarray}

\subsection{Power-law inflation} 
Next, we show that power-law inflation can be 
realized in the Einstein frame.  
Substituting Eq.\ (\ref{eq:B.2}) with 
$f=At^2$ in the ansatz (\ref{eq:15}) in the Jordan frame
into Eq.\ (\ref{eq:B.17}), we find 
\begin{eqnarray}
d\hat{t} = \sqrt{2A} t dt.  
\label{eq:B.22} 
\end{eqnarray}
Integrating Eq.\ (\ref{eq:B.22}), we obtain 
\begin{eqnarray}
\hat{t} = \sqrt{\frac{A}{2}} t^2, 
\label{eq:B.23} 
\end{eqnarray}
where we have chosen the constant of integration so that 
$\hat{t}=0$ at $t=0$.  

Here we seek solutions with the following ansatz:  
\begin{eqnarray}
\hat{a} \propto {\hat{t}}^{\hat{\omega}}.  
\label{eq:B.24} 
\end{eqnarray}
We now assume that 
in the inflationary stage, the cosmic energy density is 
dominated by the dilatons, and hence 
${\hat{T}}^{(\mathrm{m})}_{\mu \nu}$ is negligible.  
Substituting the ansatz (\ref{eq:B.24}) into 
Eqs.\ (\ref{eq:B.19}) and (\ref{eq:B.20}) and 
using Eq.\ (\ref{eq:B.4}) with $f=At^2$, we obtain 
\begin{eqnarray}
\frac{\hat{\omega}^2}{\hat{t}^2} &=& 
\frac{1}{6}
\left[
\left( \frac{d{\hat{\tilde{\varphi}}}_1}{d\hat{t}} \right)^2 + 
\left( \frac{d{\hat{\tilde{\varphi}}}_2}{d\hat{t}} \right)^2 
\right]
+ \frac{V}{24A} \frac{1}{\hat{t}^2},  
\label{eq:B.25} \\[5mm]
-\frac{\hat{\omega}}{\hat{t}^2} &=& 
-\frac{1}{2}
\left[
\left( \frac{d{\hat{\tilde{\varphi}}}_1}{d\hat{t}} \right)^2 + 
\left( \frac{d{\hat{\tilde{\varphi}}}_2}{d\hat{t}} \right)^2
\right].  
\label{eq:B.26}
\end{eqnarray}
Eliminating 
$
\left[ d{\hat{\tilde{\varphi}}}_1/ \left( d\hat{t} \right) \right]^2 + 
\left[ d{\hat{\tilde{\varphi}}}_2/ \left( d\hat{t} \right) \right]^2
$
from Eqs.\ (\ref{eq:B.25}) and (\ref{eq:B.26}), we obtain 
\begin{eqnarray}
\hat{\omega}^2 - \frac{1}{3} \hat{\omega} - \frac{V}{24A} = 0.  
\label{eq:B.27}
\end{eqnarray}
The solution of Eq.\ (\ref{eq:B.27}) is given by 
\begin{eqnarray}
\hat{\omega} &=& \frac{1}{6} 
\left( 1 + \sqrt{1+ \frac{3}{2} \frac{V}{A} } \right) 
\label{eq:B.28} \\[5mm]
&\approx& \frac{1}{6} 
\left( 1 + \sqrt{1 + 9 \omega^2 } \right) 
\label{eq:B.29}, 
\end{eqnarray}
where we have taken the positive solution of $\hat{\omega}$.  
Moreover, in deriving the approximate equality in Eq.\ (\ref{eq:B.29}), 
we have ignored variation of the potential, and hence 
replaced $V$ by its average value $\Lambda$ (as done in \S 3 
in deriving the solutions of power-law inflation in the Jordan frame) 
and used the relation 
$
\Lambda / A \approx 3/(32) \left( \alpha/\beta \right)^2 
\approx 6 \omega^2, 
$
where the first approximate equality follows from Eq.\ (\ref{eq:39}), and 
in deriving the last approximate equality we have used Eq.\ (\ref{eq:28}).  
Hence, it follows from Eq.\ (\ref{eq:B.29}) that 
if the power-law exponent $\omega$ in the Jordan frame is 
much larger than unity, 
the power-law exponent $\hat{\omega}$ in the Einstein frame 
can also be much larger than unity.  
Because power-law inflation with $\omega \gg 1$ can be realized in 
the Jordan frame, as shown in \S 3 and \S 4.2, 
power-law inflation 
can also be realized in the Einstein frame.

Finally, we estimate the number of \textit{e}-folds during inflation 
in the Einstein frame.  As in the Jordan frame, 
this number is defined as \cite{Kolb}  
\begin{eqnarray}
\hat{N} &\equiv& \int_{\hat{t}_{\mathrm{i}}}^{\hat{t}_{\mathrm{f}}} 
\hat{H} d\hat{t}  
\label{eq:B.30} \\[5mm]
&=& \hat{\omega} \ln \left( 
\frac{\hat{t}_{\mathrm{f}}}{\hat{t}_{\mathrm{i}}} \right),
\label{eq:B.31}
\end{eqnarray}
where 
in deriving Eq.\ (\ref{eq:B.31}), 
we have used $\hat{H}=\hat{\omega}/\hat{t}$. 
Applying Eq.\ (\ref{eq:B.23}) to Eq.\ (\ref{eq:B.31}), we obtain 
\begin{eqnarray}
\hat{N} &=& 
2 \hat{\omega} \ln \left( \frac{t_{\mathrm{f}}}{t_{\mathrm{i}}} \right) 
\label{eq:B.32} \\[5mm]
&\approx& 
2 \hat{\omega} 
\ln \left[ \frac{\pi}{\gamma(t_{\mathrm{i}}) } \right].  
\label{eq:B.33}
\end{eqnarray}
Here, in deriving Eq.\ (\ref{eq:B.33}), 
we have used Eqs.\ (\ref{eq:4-3}) and (\ref{eq:4-4}).  
For example, in the case (v) in Table I, 
it follows from 
$\omega = 4.1 \times 10^1$ and Eq.\ (\ref{eq:B.29}) 
that $\hat{\omega} = 2.1 \times 10^1$.  
Furthermore, it follows from this value of $\hat{\omega}$, the relation 
$\gamma(t_{\mathrm{i}}) \simeq 3.5 \times 10^{-4}$ and 
Eq.\ (\ref{eq:B.33}) that 
$\hat{N} = 3.7 \times 10^2 > 70$.  Thus, the expansion of the universe 
during the power-law inflation in the Einstein frame 
is large enough to solve both the horizon and flatness problems.  
Thus our result in the Einstein frame is the same as that 
in the Jordan frame, considered in \S 4.3.


\section{Equivalence of the Power Spectra of the Curvature Perturbations 
in the Jordan and Einstein Frames}
In this appendix, we show that 
the power spectra of the curvature perturbations 
in the Jordan and Einstein frames are the same, 
following the outline given in Ref.~\citen{Tsujikawa1}.  
Here we note that 
in this appendix we consider the case in which there exists a scalar field 
coupled to the scalar curvature, as shown in Eq.\ (\ref{eq:C.1}).  

\subsection{Curvature perturbations in the Jordan frame} 
We first consider the power spectrum of curvature 
perturbations in the Jordan frame.  
To begin with, we consider the action 
\begin{eqnarray}
{\cal S}_{\mathrm{J}}=\int d^4 x \sqrt{-g} 
\left[ -\frac{1}{2} {\cal F}(\phi)R
+\frac{1}{2}  (\nabla \phi)^2- {\cal V}[\phi] \right],
\label{eq:C.1}
\end{eqnarray}
where ${\cal F}(\phi)$ is the coupling between the scalar field 
$\phi$ and the scalar curvature $R$, and 
${\cal V}[\phi]$ is the potential of the scalar field;  
both ${\cal F}(\phi)$ and ${\cal V}[\phi]$ are 
arbitrary functions of the scalar field $\phi$.

We consider a general perturbed metric for scalar perturbations 
\begin{eqnarray}
\hspace{-5mm}
ds^2 &=& \left( 1+2\Phi \right)dt^2 - 2a(t){B_{\mathrm{p}}}_{,i}dx^idt
-a^2(t)\left[ \left( 1-2\psi \right) \delta_{ij} 
+2{E_{\mathrm{p}}}_{,i,j}\right]dx^i dx^j,
\end{eqnarray}
where a comma denotes a flat-space coordinate 
derivative.  
Here, it is convenient to introduce the 
comoving curvature perturbation ${\cal R}$ defined as 
\begin{eqnarray}
{\cal R} \equiv - \psi -
\frac{H}{\dot{\phi}}\delta \phi, 
\label{eq:C.3}
\end{eqnarray}
where $\delta \phi$ is the perturbation of the field $\phi$.  
It follows from the action in Eq.\ (\ref{eq:C.1}) that 
the equation of motion is given by \cite{Hwang1}
\begin{eqnarray}
\frac{1}{a^3Q_{\mathrm{S}}} \left( a^3 Q_{\mathrm{S}} 
\dot{\cal R} \right)^{\bullet}
+ \frac{k^2}{a^2}{\cal R}=0,
\label{eq:C.4}
\end{eqnarray}
with 
\begin{eqnarray}
Q_{\mathrm{S}} = \frac{\dot{\phi}^2 + 3 \dot{{\cal F}}^2/ 
\left( 2 {\cal F} \right) }
{\left[ H + \dot{{\cal F}}/ \left( 2 {\cal F} \right) \right]^2}.  
\label{eq:C.5}
\end{eqnarray}
If we ignore the contribution of the decaying mode, the curvature 
perturbation is conserved in the large-scale limit ($k \to 0$).  
We here introduce the 
new variables $z=a\sqrt{Q_{\mathrm{S}}}$ and $u=a{\cal R}$, 
so that Eq.\ (\ref{eq:C.4}) can be rewritten in the form 
\begin{eqnarray}
u^{\prime \prime} + \left(k^2 - \frac{z^{\prime \prime}}{z} \right) u = 0,
\label{eq:C.6}
\end{eqnarray}
where the prime denotes differentiation 
with respect to the conformal time $\eta=\int dt /a(t)$.  
Here, the gravitational term 
$z^{\prime \prime}/z$ can be written as 
\begin{eqnarray}
\frac{z^{\prime \prime}}{z}=\left( aH \right)^2 
\left[(1+\delta_{\mathrm{S}})(2 + \delta_{\mathrm{S}}
+\varepsilon)
+\frac{\delta_{\mathrm{S}}^{\prime}}{aH} \right],
\label{eq:C.7}
\end{eqnarray}
where 
\begin{eqnarray}
\varepsilon &=& \frac{\dot{H}}{H^2},
\label{eq:C.8} \\[3mm]
\delta_{\mathrm{S}} &=& \frac{\dot{Q}_{\mathrm{S}}}{2HQ_{\mathrm{S}}}.
\label{eq:C.9}
\end{eqnarray}
In the context of slow-roll inflation, it is a good approximation 
to ignore the variations of $\varepsilon$ and $\delta_{\mathrm{S}}$.
In the case of power-law inflation, the conformal time is given by 
$\eta=-1/[(1+\varepsilon)aH]$.  From this expression of 
$\eta$ and Eq.\ (\ref{eq:C.7}), we obtain 
\begin{eqnarray}
\frac{z^{\prime \prime}}{z} = \frac{\gamma_{\mathrm{S}}}{\eta^2},
\label{eq:C.10}
\end{eqnarray}
with
\begin{eqnarray}
\gamma_{\mathrm{S}} = \frac{\left( 1 + \delta_{\mathrm{S}} \right)
\left( 2 + \delta_{\mathrm{S}} + \varepsilon \right)}
{\left( 1 + \varepsilon \right)^2}.
\label{eq:C.11}
\end{eqnarray}
The solution of Eq.\ (\ref{eq:C.6}) is then given by 
\begin{eqnarray}
u= \frac{\sqrt{\pi |\eta|}}{2} 
\left[ c_1 H_{\nu_{\mathrm{S}}}^{(1)}(k|\eta|) + 
c_2 H_{\nu_{\mathrm{S}}}^{(2)}(k|\eta|) \right],
\label{eq:C.12}
\end{eqnarray}
where 
\begin{eqnarray}
\nu_{\mathrm{S}} \equiv \sqrt{\gamma_{\mathrm{S}} + \frac{1}{4}}.  
\label{eq:C.13}
\end{eqnarray}
Here $H_{\nu_{\mathrm{S}}}^{(1)}$ and $H_{\nu_{\mathrm{S}}}^{(2)}$ are 
$\nu_{\mathrm{S}}$th-order Hankel functions of types 1 and 2, 
respectively.  
We choose the coefficients to be $c_1=0$ and $c_2=1$, 
so that positive frequency solutions in the Minkowski
vacuum can be recovered in the asymptotic past.  
Then, using the relation $H_{\nu_{\mathrm{S}}}^{(2)} 
(k|\eta|) \to (i/\pi) 
\Gamma(\nu_{\mathrm{S}}) \left( k|\eta|/2 \right)^{-\nu_{\mathrm{S}}}$ 
for long wavelength perturbations ($ k \to 0$), we find that 
the power spectrum of the comoving curvature perturbation 
${\cal P}_{\cal R} \equiv \left[ k^3 / \left(2\pi^2 \right) \right]
\left| {\cal R} \right|^2$ is given by 
\begin{eqnarray}
{\cal P}_{\cal R} = \frac{1}{Q_{\mathrm{S}}}
\left(\frac{H}{2\pi}\right)^2
\left(\frac{1}{aH|\eta|}\right)^2
\left(\frac{\Gamma(\nu_{\mathrm{S}})}{\Gamma(3/2)}\right)^2
\left(\frac{k|\eta|}{2}\right)^{3-2\nu_{\mathrm{S}}}.  
\label{eq:C.14}
\end{eqnarray}
Furthermore, the spectral index 
$n_{\mathrm{S}} \equiv 1 + d\ln{\cal P}_{\cal R}/ \left( d \ln k \right)$ 
is given by 
\begin{eqnarray}
n_{\mathrm{S}} - 1 = 3 - 2\nu_{\mathrm{S}}=3-\sqrt{4\gamma_{\mathrm{S}}+1},
\label{eq:C.15}
\end{eqnarray}
where the last equality follows from the relation (\ref{eq:C.13}).

\subsection{Curvature perturbations in the Einstein frame} 
Next, in order to show that 
the power spectrum of the curvature perturbations in the Jordan frame is 
identical to that in the Einstein frame, 
we consider the power spectrum of curvature 
perturbations in the Einstein frame.  

To begin with, we apply the following conformal transformation 
to the action in Eq.\ (\ref{eq:C.1}): 
\begin{eqnarray}
g_{\mu \nu} \hspace{0.5mm} \rightarrow \hspace{0.5mm} 
\hat{g}_{\mu \nu} = \Omega_{\mathrm{S}} g_{\mu \nu},
\label{eq:C.16}
\end{eqnarray}
with 
\begin{eqnarray}
\Omega_{\mathrm{S}} =  {\cal F}.
\label{eq:C.17}
\end{eqnarray}
The action in the Einstein frame is then given by \cite{F-M,Tsujikawa1} 
\begin{eqnarray}
{\cal S}_{\mathrm{E}} &=& 
\int d^{4} \hat{x} \sqrt{-\hat{g}} 
\left\{
-\frac{1}{2} \hat{R} 
+ \frac{1}{2} 
\left[ \frac{3}{2} 
\left( \frac{{\cal F}_{\phi}}{{\cal F}} \right)^2 + \frac{1}{{\cal F}}
\right]
\left( \hat{\nabla} \phi \right)^2
- \hat{V}[\phi]
\right\}, 
\label{eq:C.18} 
\end{eqnarray}
with 
\begin{eqnarray}
\hat{V}[\phi] = \frac{V[\phi]}{{\cal F}^2}.  
\label{eq:C.19}
\end{eqnarray} 
Here, the subscript $\phi$ denotes 
partial differentiation with respect to $\phi$.  
Furthermore, we introduce the new scalar fields 
\begin{eqnarray}
\frac{d {\hat{\phi}}}{d \phi} \equiv 
\sqrt{{\cal D}_{\mathrm{E}} \left( \phi \right)},
\label{eq:C.20}
\end{eqnarray}
with
\begin{eqnarray} 
{\cal D}_{\mathrm{E}} \left( \phi \right) 
= \frac{3}{2} 
\left( \frac{{\cal F}_{\phi}}{{\cal F}} \right)^2 
+ \frac{1}{{\cal F}}, 
\hspace{3mm} ( \hspace{0.5mm} > 0 \hspace{0.5mm}  )
\label{eq:C.21}
\end{eqnarray}
so that the action in Eq.\ (\ref{eq:C.18}) can be rewritten in 
the following canonical form:  
\begin{eqnarray}
{\cal S}_{\mathrm{E}} = 
\int d^{4} \hat{x} \sqrt{-\hat{g}} 
\left[
-\frac{1}{2} \hat{R} 
+ \frac{1}{2} \left( \hat{\nabla} \hat{\phi} \right)^2
- \hat{V}[\phi] 
\right].  
\label{eq:C.22} 
\end{eqnarray}

Now we consider a perturbed metric in the Einstein frame, 
\begin{eqnarray}
d\hat{s}^2 &=& \Omega_{\mathrm{S}} ds^2 
\label{eq:C.23} \\[3mm] 
&=& 
\left( 1+2 \hat{\Phi} \right)d\hat{t}^2 
- 2\hat{a}(\hat{t}){\hat{B_{\mathrm{p}}}}_{,i}d\hat{x}^id\hat{t}
\nonumber \\[1.5mm]
&& 
- \hat{a}^2(\hat{t})\left[ \left( 1-2 \hat{\psi} \right) \delta_{ij} 
+2{\hat{E_{\mathrm{p}}}}_{,i,j} \right]d\hat{x}^i d\hat{x}^j,
\label{eq:C.24} 
\end{eqnarray}
and decompose the conformal factor into background and perturbed parts as 
\begin{eqnarray}
\Omega_{\mathrm{S}} (\Vec{x}, t) = {\bar{\Omega}}_{\mathrm{S}}(t) 
\left(1+\frac{\delta \Omega_{\mathrm{S}}(\Vec{x}, t)}
{{\bar{\Omega}}_{\mathrm{S}}(t)}\right).
\label{eq:C.25} 
\end{eqnarray}
In what follows we drop the bar on 
${\bar{\Omega}}_{\mathrm{S}}(t)$.  
From the above, we obtain the following relations:  
\begin{eqnarray}
\hat{a} &=& a\sqrt{\Omega_{\mathrm{S}}}, \hspace{2mm}
d\hat{t}=\sqrt{\Omega_{\mathrm{S}}}dt, \hspace{2mm}
\hat{H}=\frac{1}{\sqrt{\Omega_{\mathrm{S}}}} 
\left(H+\frac{{\dot{\Omega}}_{\mathrm{S}}}{2\Omega_{\mathrm{S}}} \right),
\nonumber \\[3mm]
\hat{\Phi} &=& \Phi + \frac{\delta \Omega_{\mathrm{S}}}
{2\Omega_{\mathrm{S}}}, \hspace{2mm}
\hat{\psi}=\psi+\frac{\delta \Omega_{\mathrm{S}}}{2\Omega_{\mathrm{S}}}.
\label{eq:C.26} 
\end{eqnarray}
Using these relations, it is shown that 
curvature perturbations in the Einstein frame exactly 
coincide with those in the Jordan frame as follow:  
\begin{eqnarray}
\hat{\cal R} &\equiv& -\hat{\psi}
-\frac{\hat{H}}
{d \hat{\phi}/ \left( d \hat{t} \right)} \delta \hat{\phi} 
\label{eq:C.27} \\[5mm] 
&=& -\psi-\frac{H}{\dot{\phi}} \delta \phi = {\cal R}.
\label{eq:C.28} 
\end{eqnarray}
Hence, from Eq.\ (\ref{eq:C.28}) we find 
\begin{eqnarray}
\hat{\cal P}_{\cal R}={\cal P}_{\cal R}. 
\label{eq:C.29}  
\end{eqnarray}

Furthermore, we introduce the following quantities:  
\begin{eqnarray}
\hat{\gamma}_{\mathrm{S}} &=& 
\frac{ \left( 1+\hat{\delta}_{\mathrm{S}} \right)
\left( 2+\hat{\delta}_{\mathrm{S}} + \hat{\varepsilon} \right)}
{\left( 1+\hat{\varepsilon} \right)^2},
\hspace{2mm}
\hat{\varepsilon}=\frac{d\hat{H}/ \left( d \hat{t} \right) }{\hat{H}^2},
\nonumber \\[3mm]
\hat{\delta}_{\mathrm{S}} &=& 
\frac{d\hat{Q}_{\mathrm{S}}/\left( d \hat{t} \right)}
{2\hat{H}\hat{Q}_{\mathrm{S}}},
\hspace{2mm}
\hat{Q}_{\mathrm{S}} = \left[ \frac{ d \hat{\phi} / \left( d \hat{t} \right)}
{\hat{H}} \right]^2 = \frac{Q_{\mathrm{S}}}{{\cal F}}.  
\label{eq:C.30}  
\end{eqnarray}
The spectral index of scalar perturbations in the 
Einstein frame is then given by 
\begin{eqnarray}
\hat{n}_{\mathrm{S}}-1=3-\sqrt{4\hat{\gamma}_{\mathrm{S}}+1}.
\label{eq:C.31}  
\end{eqnarray}
Moreover, 
the two quantities $\hat{\varepsilon}$ and  $\hat{\delta}_{\mathrm{S}}$ 
can be expressed as follows:  
\begin{eqnarray}
\hat{\varepsilon} &=& \frac{\varepsilon-\vartheta}{1+\vartheta} +
\frac{\dot{\vartheta}}{H \left( 1+\vartheta \right)^2}, 
\label{eq:C.32} \\[5mm]  
\hat{\delta}_{\mathrm{S}} &=& \frac{\delta-\vartheta}{1+\vartheta}, 
\label{eq:C.33}
\end{eqnarray}
with
\begin{eqnarray}
\vartheta=\frac{\dot{{\cal F}}}{2H{\cal F}}.
\label{eq:C.34}
\end{eqnarray}
When the variation of $\vartheta$ is negligible 
($\dot{\vartheta} \simeq 0$), which is the case in the context of slow-roll 
inflation, it can be shown that 
$\hat{\gamma}_{\mathrm{S}} = \gamma_{\mathrm{S}}$.  
Hence, the spectral index in Eq.\ (\ref{eq:C.31}) in the Einstein frame 
coincides with that in the Jordan frame: 
$\hat{n}_{\mathrm{S}}=n_{\mathrm{S}}$.   

Finally, we note that in the above discussion, we have considered 
comoving curvature perturbations ${\cal R}$ and then shown that 
the power spectrum of such perturbations in the Einstein frame exactly 
coincides with that in the Jordan frame.  
Contrastingly, 
in \S 5 we considered 
curvature perturbations on uniform-density hypersurfaces $\zeta$.  
Comoving curvature perturbations and 
curvature perturbations on uniform-density hypersurfaces $\zeta$, 
however, are identical on superhorizon scales.\cite{Riotto1}  
Thus, the results of \S 5 obtained from analysis in the Einstein frame are 
identical to those in the Jordan frame.

\newpage

\begin{figure}[tbp]
\begin{center}
  \begin{minipage}{65mm}
  \begin{center}
  \unitlength=1mm
\resizebox{!}{6.5cm}{
   \includegraphics{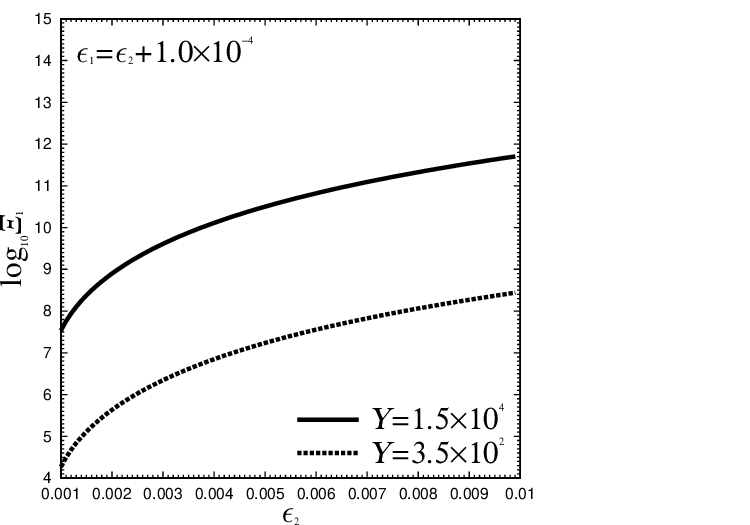}
                  }
  \end{center}
  \end{minipage}
 \hspace{1mm}
  \begin{minipage}{65mm}
  \begin{center}
  \unitlength=1mm
\resizebox{!}{6.5cm}{
   \includegraphics{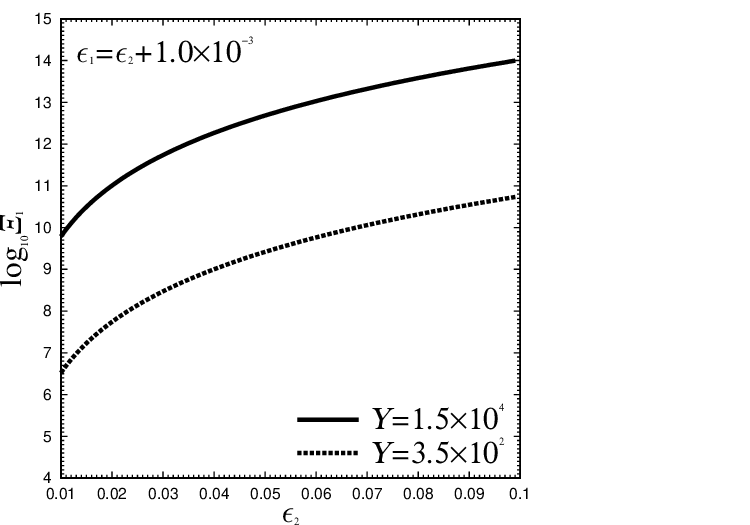}
                  }
  \end{center}
  \end{minipage}
\caption{ 
Estimates of 
$\Xi_1 = \tilde{\mathrm{Tr} \hspace{0.5mm} {\cal M}^2}/M^2$ 
in the relation (\ref{eq:90}).  
The solid curve is for the case $\Upsilon = 1.5 \times 10^4 $, 
and the dotted curve is for the case 
$\Upsilon = 3.5 \times 10^2 $.  
The left panel describes 
the case 
$\epsilon_1 = \epsilon_2 + 1.0 \times 10^{-4},  
\epsilon_2 = 1.0 \times 10^{-3} + 1.0 \times 10^{-4}j \hspace{1mm} 
(0 \leq j \leq 89)$, where $j$ is an integer, 
while the right panel describes the case 
$\epsilon_1 = \epsilon_2 + 1.0 \times 10^{-3},  
\epsilon_2 = 1.0 \times 10^{-2} + 1.0 \times 10^{-3}j \hspace{1mm} 
(0 \leq j \leq 89)$. 
}
\end{center}
\end{figure}

\begin{figure}[tbp]
\begin{center}
  \begin{minipage}{65mm}
  \begin{center}
  \unitlength=1mm
\resizebox{!}{6.5cm}{
   \includegraphics{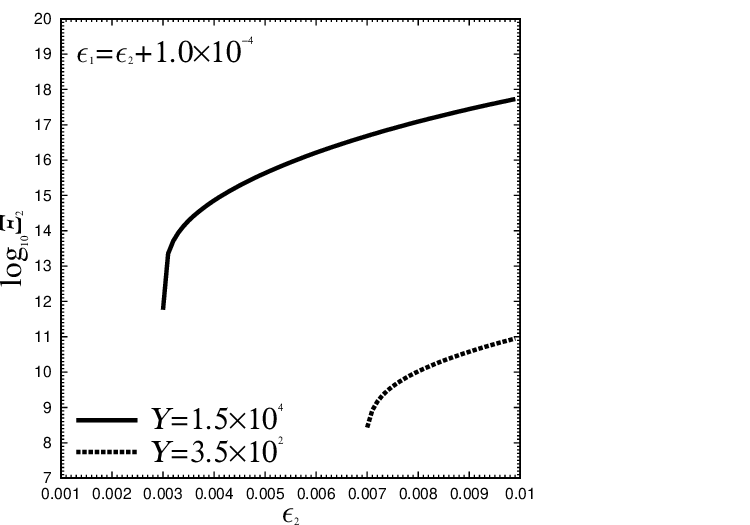}
                  }
  \end{center}
  \end{minipage}
 \hspace{1mm}
  \begin{minipage}{65mm}
  \begin{center}
  \unitlength=1mm
\resizebox{!}{6.5cm}{
   \includegraphics{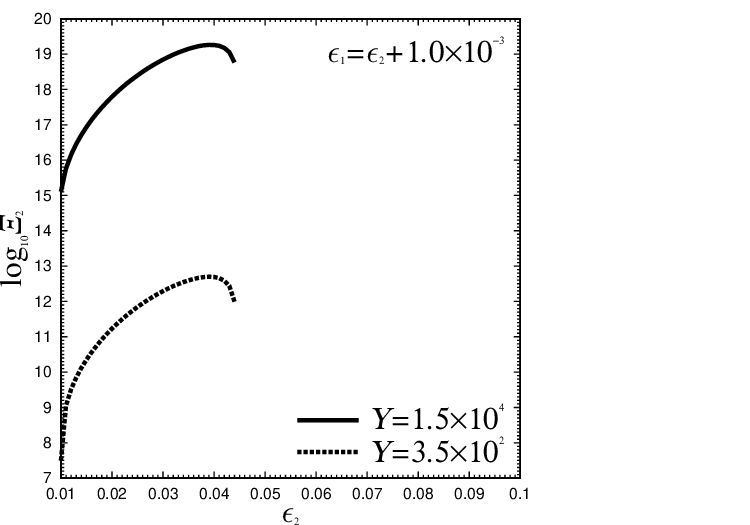}
                  }
  \end{center}
  \end{minipage}
\caption{
Estimates of  
$\Xi_2 = \tilde{\det {\cal M}^2}/M^2$ 
in the relation (\ref{eq:91}).  
The solid curve is for the case $\Upsilon = 1.5 \times 10^4$, 
and the dotted curve is for the case 
$\Upsilon = 3.5 \times 10^2 $.  
The left panel describes the case 
$\epsilon_1 = \epsilon_2 + 1.0 \times 10^{-4},  
\epsilon_2 = 1.0 \times 10^{-3} + 1.0 \times 10^{-4}j \hspace{1mm} 
(0 \leq j \leq 89)$, where $j$ is an integer, 
while the right panel describes the case 
$\epsilon_1 = \epsilon_2 + 1.0 \times 10^{-3},  
\epsilon_2 = 1.0 \times 10^{-2} + 1.0 \times 10^{-3}j \hspace{1mm} 
(0 \leq j \leq 89)$. 
In the case $\Upsilon = 1.5 \times 10^4 $, 
for $\epsilon_2 \leq 2.9 \times 10^{-3}$ and 
$\epsilon_2 \geq 4.5 \times 10^{-2}$, we have $\Xi_2 < 0$.  
Similarly, in the case $\Upsilon = 3.5 \times 10^2 $, 
for $\epsilon_2 \leq 6.9 \times 10^{-3}$ and 
$\epsilon_2 \geq 4.5 \times 10^{-2}$, we have $\Xi_2 < 0$.  
}
\end{center}
\end{figure}

\begin{figure}[tbp]
\begin{center}
  \begin{minipage}{65mm}
  \begin{center}
  \unitlength=1mm
\resizebox{!}{6.5cm}{
   \includegraphics{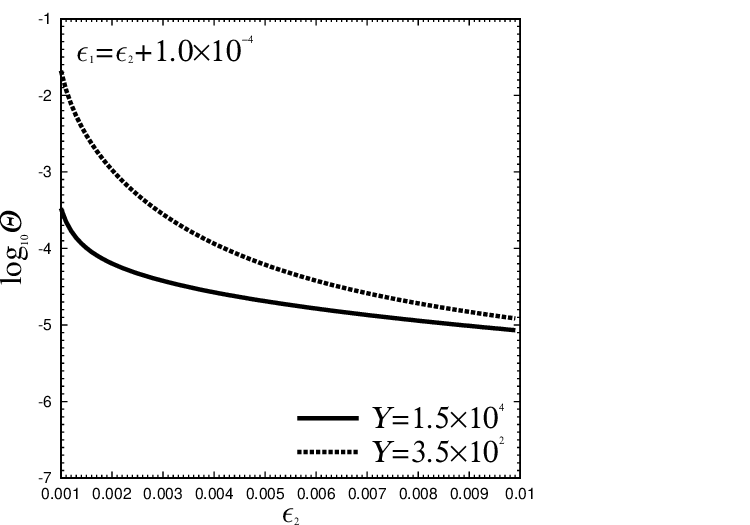}
                  }
  \end{center}
  \end{minipage}
 \hspace{1mm}
  \begin{minipage}{65mm}
  \begin{center}
  \unitlength=1mm
\resizebox{!}{6.5cm}{
   \includegraphics{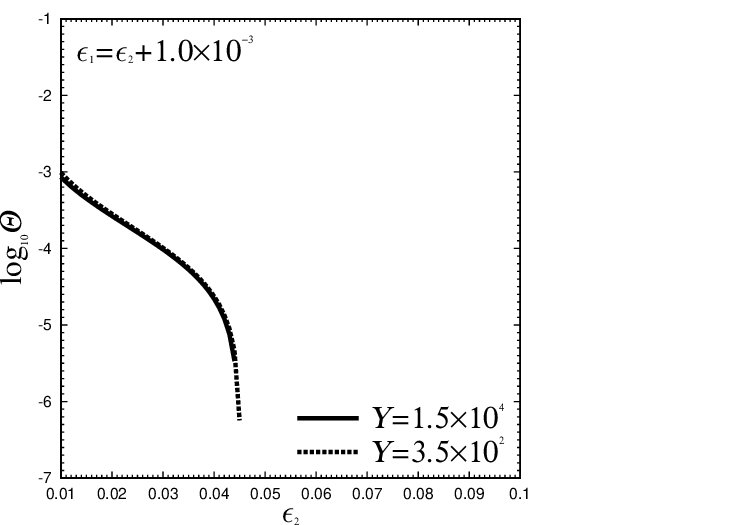}
                  }
  \end{center}
  \end{minipage}
\caption{
Estimates of 
$\tilde{\Theta}$ 
in the relation (\ref{eq:92}).  
The solid curve is for the case $\Upsilon = 1.5 \times 10^4$, 
and the dotted curve is for the case 
$\Upsilon = 3.5 \times 10^2 $.  
The left panel describes the case 
$\epsilon_1 = \epsilon_2 + 1.0 \times 10^{-4},  
\epsilon_2 = 1.0 \times 10^{-3} + 1.0 \times 10^{-4}j \hspace{1mm} 
(0 \leq j \leq 89)$, where $j$ is an integer, 
while the right panel describes the case 
$\epsilon_1 = \epsilon_2 + 1.0 \times 10^{-3},  
\epsilon_2 = 1.0 \times 10^{-2} + 1.0 \times 10^{-3}j \hspace{1mm} 
(0 \leq j \leq 89)$. 
In the case $\Upsilon = 1.5 \times 10^4$, 
for $\epsilon_2 \geq 4.5 \times 10^{-2}$, we have $\tilde{\Theta} < 0$.  
Similarly, in the case $\Upsilon = 3.5 \times 10^2 $, 
for $\epsilon_2 \geq 4.6 \times 10^{-2}$, we have $\tilde{\Theta} < 0$.
}
\end{center}
\end{figure}

\begin{figure}[tbp]
\begin{center}
  \begin{minipage}{65mm}
  \begin{center}
  \unitlength=1mm
\resizebox{!}{6.5cm}{
   \includegraphics{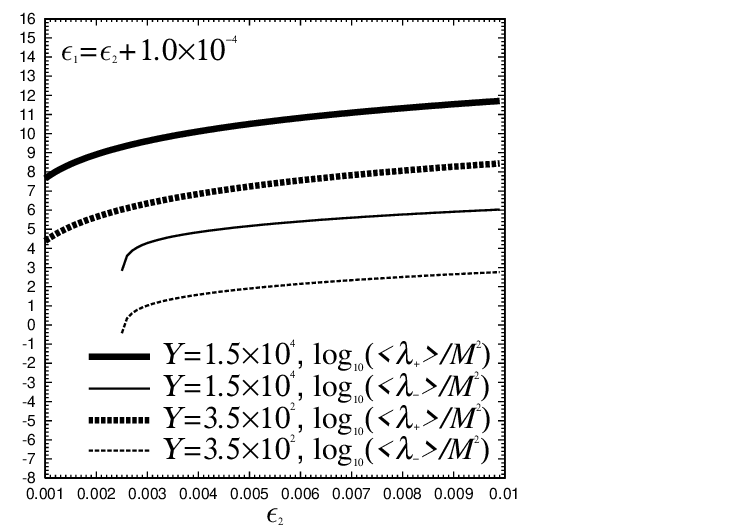}
                  }
  \end{center}
  \end{minipage}
 \hspace{1mm}
  \begin{minipage}{65mm}
  \begin{center}
  \unitlength=1mm
\resizebox{!}{6.5cm}{
   \includegraphics{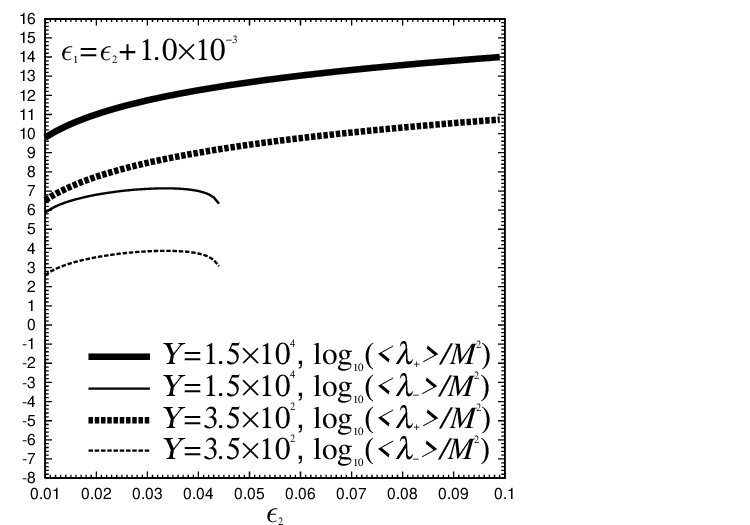}
                  }
  \end{center}
  \end{minipage}
\caption{ 
Estimates of  
$\left< \lambda_{+} \right>/M^2$ and 
$\left< \lambda_{-} \right>/M^2$ 
in Eqs.\ (\ref{eq:93}) and (\ref{eq:94}).  
The solid curves are for the case $\Upsilon = 1.5 \times 10^4$, 
and the dotted curves are for the case 
$\Upsilon = 3.5 \times 10^2 $.  
Thick curves 
represent $\left< \lambda_{+} \right>/M^2$, and the 
thin curves represent $\left< \lambda_{-} \right>/M^2$.  
The left panel describes the case 
$\epsilon_1 = \epsilon_2 + 1.0 \times 10^{-4},  
\epsilon_2 = 1.0 \times 10^{-3} + 1.0 \times 10^{-4}j \hspace{1mm} 
(0 \leq j \leq 89)$, where $j$ is an integer, 
while the right panel describes the case 
$\epsilon_1 = \epsilon_2 + 1.0 \times 10^{-3},  
\epsilon_2 = 1.0 \times 10^{-2} + 1.0 \times 10^{-3}j \hspace{1mm} 
(0 \leq j \leq 89)$. 
In both the cases $\Upsilon = 1.5 \times 10^4 $ 
and $\Upsilon = 3.5 \times 10^2 $, 
for $\epsilon_2 \leq 2.4 \times 10^{-3}$ and 
$\epsilon_2 \geq 4.5 \times 10^{-2}$, we have 
$\left< \lambda_{-} \right> /M^2 < 0$.  
}
\end{center}
\end{figure}

\begin{figure}[tbp]
\begin{center}
  \begin{minipage}{65mm}
  \begin{center}
  \unitlength=1mm
\resizebox{!}{6.5cm}{
   \includegraphics{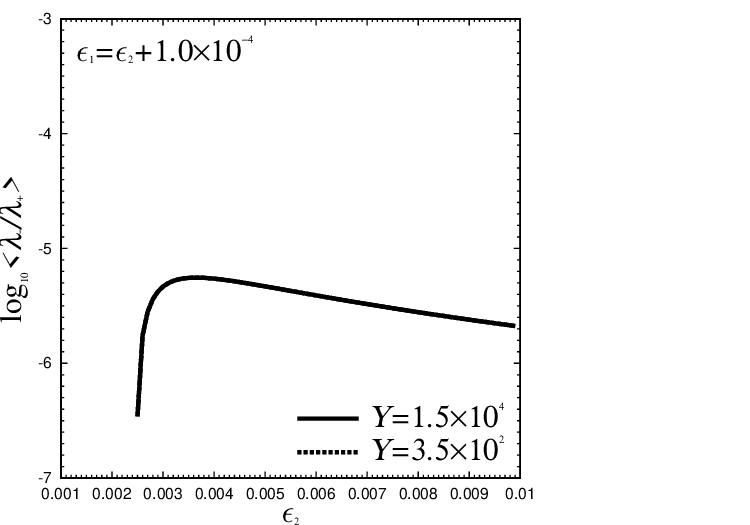}
                  }
  \end{center}
  \end{minipage}
 \hspace{1mm}
  \begin{minipage}{65mm}
  \begin{center}
  \unitlength=1mm
\resizebox{!}{6.5cm}{
   \includegraphics{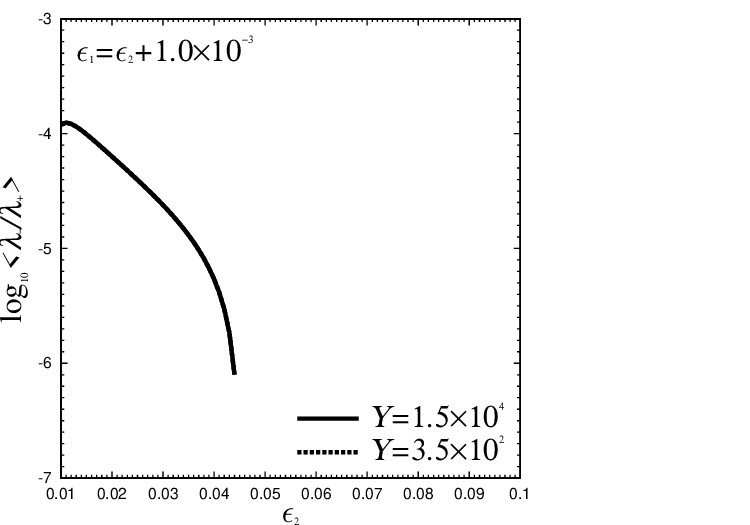}
                  }
  \end{center}
  \end{minipage}
\caption{ 
Estimates of 
$\left< \lambda_{-}/\lambda_{+} \right>$ 
in Eq.\ (\ref{eq:95}).  
The solid curve is for the case $\Upsilon = 1.5 \times 10^4$, 
and the dotted curve is for the case 
$\Upsilon = 3.5 \times 10^2 $.  
In this figure, the former and latter coincide. 
The left panel describes the case 
$\epsilon_1 = \epsilon_2 + 1.0 \times 10^{-4},  
\epsilon_2 = 1.0 \times 10^{-3} + 1.0 \times 10^{-4}j \hspace{1mm} 
(0 \leq j \leq 89)$, where $j$ is an integer, 
while the right panel describes the case 
$\epsilon_1 = \epsilon_2 + 1.0 \times 10^{-3},  
\epsilon_2 = 1.0 \times 10^{-2} + 1.0 \times 10^{-3}j \hspace{1mm} 
(0 \leq j \leq 89)$. 
\if
Moreover, 
the left panel is 
for the case 
$\epsilon_1 = \epsilon_2 + 1.0 \times 10^{-4},  
\epsilon_2 = 1.0 \times 10^{-3} + 1.0 \times 10^{-4}j \hspace{1mm} 
(j =0, 1, 2, ..., 89)$, 
while the right panel is for the case 
$\epsilon_1 = \epsilon_2 + 1.0 \times 10^{-3},  
\epsilon_2 = 1.0 \times 10^{-2} + 1.0 \times 10^{-3}j \hspace{1mm} 
(j =0, 1, 2, ..., 89)$.  
In the left panel, we have plotted the value of 
$\left< \lambda_{-}/\lambda_{+} \right>$ 
for the case 
$\epsilon_1 = \epsilon_2 + 1.0 \times 10^{-4},  
\epsilon_2 = 1.0 \times 10^{-3} + 1.0 \times 10^{-4}j \hspace{1mm} 
(j =0, 1, 2, ..., 89)$, 
while in the right panel, for the case 
$\epsilon_1 = \epsilon_2 + 1.0 \times 10^{-3},  
\epsilon_2 = 1.0 \times 10^{-2} + 1.0 \times 10^{-3}j \hspace{1mm} 
(j =0, 1, 2, ..., 89)$.  
In the left panel, $\epsilon_1 = \epsilon_2 + 1.0 \times 10^{-4} 
(1.0 \times 10^{-3} \leq \epsilon_2 \leq 9.9 \times 10^{-3} )$, while 
in the right panel $\epsilon_1 = \epsilon_2 + 1.0 \times 10^{-3} 
(1.0 \times 10^{-2} \leq \epsilon_2 \leq 9.9 \times 10^{-2})$.  
\fi
In both the cases $\Upsilon = 1.5 \times 10^4 $ 
and $\Upsilon = 3.5 \times 10^2$,  
for $\epsilon_2 \leq 2.4 \times 10^{-3}$ and 
$\epsilon_2 \geq 4.5 \times 10^{-2}$, we have 
$\left< \lambda_{-}/\lambda_{+} \right> < 0$.  
}
\end{center}
\end{figure}

\end{document}